\documentclass{svjour3}                     

\usepackage{graphicx}
\usepackage[]{natbib}
\usepackage{xcolor}
\usepackage{amsmath, amssymb}
\usepackage{hyperref}

\def\msun{\ifmmode M_\odot \else $M_\odot$ \fi}

\newcommand{\mc}{M$_c$}

\newcommand{\msunyr}{M$_\odot$yr$^{-1}$}
\newcommand{\dsfr}{M$_\odot$yr$^{-1}$kpc$^{-2}$}

\newcommand{\tcr}{\tau_{\rm cr}}
\newcommand{\reff}{R_{\rm eff}}
\newcommand{\pc}{{\rm pc}}
\newcommand{\yr}{{\rm yr}}
\newcommand{\myr}{{\rm Myr}}

\begin{document}

\title{Star clusters near and far; \\tracing star formation across cosmic time
}

\titlerunning{Cluster formation near and far}        

\author{Angela Adamo, Peter Zeidler, J.~M.~Diederik Kruijssen, M\'{e}lanie Chevance, Mark Gieles, Daniela Calzetti, Corinne Charbonnel, Hans Zinnecker, Martin G.~H. Krause}

\authorrunning{Adamo, Zeidler, Kruijssen et al.} 

\institute{Angela Adamo \at Department of Astronomy, Oskar Klein Centre, Stockholm University, AlbaNova University Centre, SE-106 91 Stockholm, Sweden\\
\email{adamo@astro.su.se}          
           \and
             Peter Zeidler \at Department of Physics and Astronomy, Johns Hopkins University, Baltimore, MD 21218, USA\\
             Space Telescope Science Institute, 3700 San Martin Drive, Baltimore, MD-21218, USA\\
              \email{zeidler@stsci.edu} 
            \and
            J.~M.~Diederik Kruijssen \at Astronomisches Rechen-Institut,
              Zentrum f\"ur Astronomie der Universit\"at Heidelberg,
              M\"onchhofstra\ss{}e 12-14, 69120 Heidelberg, Germany\\
              \email{kruijssen@uni-heidelberg.de}
             \and
             M\'{e}lanie Chevance \at
              Astronomisches Rechen-Institut, Zentrum f\"ur Astronomie der Universit\"at Heidelberg, M\"onchhofstra\ss e 12-14, 69120 Heidelberg, Germany 
              \and
                 Mark Gieles \at
             Institut de Ci\`{e}ncies del Cosmos (ICCUB-IEEC), Universitat de Barcelona, Mart\'{i} i Franqu\`{e}s 1, 08028 Barcelona, Spain\\
             ICREA, Pg. Lluis Companys 23, 08010 Barcelona, Spain       
             \and
             Daniela Calzetti  \at
                  Departement of Astronomy, University of Massachusetts, LGRT-B 619J, 710 North Pleasant Street, Amherst, U.S.A.
                                   \and
            Corinne Charbonnel \at
            Department of Astronomy, University of Geneva, Chemin de Pegase 51, 1290 Versoix, Switzerland\\
            IRAP, CNRS \& Univ. of Toulouse, 14, av.E.Belin, 31400 Toulouse, France
                       \and
              Hans Zinnecker \at Universidad Autonoma de Chile
                    (Nucleo de Astroquimica y Astrofisica) 
                Avda Pedro de Valdivia 425,
                Providencia,
                Santiago de Chile, Chile
              \and
              Martin G. H. Krause \at
              Centre for Astrophysics Research, 
              School of Physics, Astronomy and Mathematics, 
              University of Hertfordshire, College Lane,
              Hatfield, Hertfordshire AL10 9AB, UK}
            
\date{Received: 31 Jan 2020 / Accepted: 08 May 2020}

\maketitle

\newpage
\begin{abstract}
Star clusters are fundamental units of stellar feedback and unique tracers of their host galactic properties.  In this review, we will first focus on their constituents, i.e.\ detailed insight into their stellar populations and their surrounding ionised, warm, neutral, and molecular gas. We, then, move beyond the Local Group to review star cluster populations at various evolutionary stages, and in diverse galactic environmental conditions accessible in the local Universe. At high redshift, where conditions for cluster formation and evolution are more extreme, we are only able to observe the integrated light of a handful of objects that we believe will become globular clusters. We therefore discuss how numerical and analytical methods, informed by the observed properties of cluster populations in the local Universe, are used to develop sophisticated simulations potentially capable of disentangling the genetic map of galaxy formation and assembly that is carried by globular cluster populations.

\end{abstract}

\tableofcontents

\clearpage
\section{Introduction}
\label{sec:intro}

\noindent Deciding if a grouping of stars is a gravitationally bound star cluster is very challenging when the kinematics of the single stars cannot be traced. To use star clusters as tracers of galaxy evolution, it requires the understanding of the formation and evolution of gravitationally bound stellar systems; i.e. star clusters that are likely to survive for a certain time span within their host systems. A definition of ``star cluster'' valid at any position in space and time in our Universe is very much a challenge. We will discuss the dynamical conditions of clusters when we can access their stellar kinematic information, i.e. properties of star cluster populations resolved in their stellar components (young star clusters, YSCs, and globular clusters, GCs) in our own Galaxy and local neighbours (Section 2). We point out the interested reader that a detailed description of the physics of cluster formation and evolution as single entity can be found in \citet{Krause_20}, another review in this series.

The focus of this work is to review cluster formation and evolution in the framework of galaxy assembly. To achieve this goal it requires to move beyond the Local Group. At increasing distances we lose information about the detailed components of star clusters (we will provide a detailed description of the assumptions made) but we acquire a statistical view of the entire cluster population and wider ranges of galactic environments and diverse physical conditions for star and cluster formation.  Throughout Section~\ref{sec:local_universe_clusters} to Section~\ref{yscsurvival}, we will describe the most recent results we have acquired on cluster populations from studies of local galaxies and the lesson we have learned so far about cluster formation and evolution.  However we can not restrict a review on cluster populations to our local Universe. In recent years, we have made huge progresses in probing under which conditions star formation operates in increasingly younger galaxies where the bulk of their stellar mass is assembled. GC populations are witnesses of these key phases of galaxy evolution. We need a way to read the information encoded in their properties as single entities and overall populations. In Section~\ref{sec:redshift_formation_GC} and Section~\ref{sec:simulations} we will discuss how we can link GC formation and evolution to the galaxy assembly history of our Universe. Analytical implementations (based on the lesson we learn from studying YSC populations in the local Universe) of cluster populations or self-consistent formation of clusters into cosmological simulations of our Universe will be summarised in light of their capabilities to successfully reproduce well established GC properties (mass function, blue tilt, age-metallicity relation,  multi-peaked metallicity distributions, etc.). Simulations are not the only accessible way to probe cluster formation at redshift beyond zero. In recent years, thanks to the help of magnification lenses, we have been able to directly detect the formation sites of GCs. These detections are paving the way to new fundamental questions such as what conditions are required for GCs to form, at which redshift GCs start to form, did GCs contribute to reionise our Universe at redshift beyond 6. In the last Section (Section~\ref{sec:outlook}) we will summarise the current observational results, predictions for the redshift formation of GCs and how next generation facilities can open our horizon into the formation of  gravitationally bound stellar structures at any redshift.

\subsection{Definition of a bound star cluster}

We all agree what a star cluster is when we see one, but coming up with a quantitative  definition is hard. Most people agree that a star cluster is a group of stars, but so are galaxies, so not all stellar groupings are star clusters. Any additional specification brings a risk of over-simplifying things and chopping up continuous distributions in discrete samples. For example, we could try and exclude galaxies from the definition and define a star cluster as a stellar grouping that formed in a single burst, from a well-mixed molecular cloud, implying that all stars have the same age and chemical composition. However,  all  stellar systems will have an age spread to some degree ($\sim10^{5-6}~\yr$), and light-element abundance spreads (HeCNONaMgAl) are the norm in old, massive GCs, leading \citet{Carretta_10} to define a bona fide GC by the presence of a Na-O anti-correlation (more on this in \S~\ref{multipop}). Also, nuclear star clusters display clear spreads in age (Gyrs) and [Fe/H] \citep[$\sim1~$dex, see][for a recent review]{seth2019}. Alternatively, we could argue that star clusters are dark matter (DM) free, which  removes galaxies. However, the first star clusters to form in the Universe may have formed in small DM halos ($\sim10^9~\msun$) which have since been stripped \citep[e.g.,][]{trenti15}. 

Despite the risks that are associated with defining what a star cluster is, there is a need for a working definition that separates star clusters from  associations, where the former are assumed to be gravitationally bound and the latter are usually assumed to be unbound. This is particularly important in extra-galactic samples, where large stellar groupings ($\gtrsim10~\pc$) are only  marginally spatially resolved and therefore resemble  compact star clusters. To test whether a system is bound, one needs kinematics to compare to an independent photometric mass estimate. This is not available for most samples. \citet{Gieles_11}  propose a  simple proxy, namely an estimate of the dynamical age of the stellar group, defined as the ratio of the age over the crossing timescale: $\Pi \equiv {\rm age}/\tcr$, where $\tcr \propto \sqrt{\reff^{3}/(GM)}$, with $\reff$  the half-light radius, $G\simeq0.0045~\pc^3\msun^{-1}\myr^{-2}$  the gravitational constant and $M$  the cluster mass. 
Stellar groups with $\Pi\gtrsim1$ have evolved for more than a crossing time, suggestive of them being bound. 
For groups with $\Pi\lesssim1$ it is not possible to say whether the group is bound, hence by applying this cut at young ages, some bound systems might be excluded. 
However, because unbound groups expand,  they quickly evolve to $\Pi\lesssim1$, hence at older ages a separation at $\Pi=1$ is safe.
We note that a more accurate definition of the crossing time involves the velocity dispersion: $\tcr\propto \reff/\sigma$. For a system in virial equilibrium the two definitions can be used interchangeably, but for unbound systems that are  flying apart with high velocities $\reff/\sigma<<\sqrt{\reff^{3}/(GM)}$. Hence using the crossing time based on kinematics -- albeit more accurate -- could lead to an erroneous classification as bound ($\Pi>>1$) if virial equilibrium is assumed. Defining $\tcr$ in terms of $M$ and $\reff$ avoids this, and has as the additional advantage that it is straightforward to determine this ratio for large numbers of stellar groups in extra-galactic samples using photometry only. A final word of caution is in place: the distribution of $\Pi$ for young stellar groupings in the Milky Way and nearby galaxies is continuous \citep{Gieles_11}, making a classification of objects with $\Pi\simeq1$ arbitrary and care should be exercised when discussing individual objects near the boundary. The simplicity of the definition and the ease of determining it for a large (extra-galactic) sample of stellar groups has made it a popular tool to attempt to separate bound star clusters from unbound associations \citep[e.g.,][]{bastian2012, johnson2016, ryon2017}. 


\subsection{The rich chemical footprints of GCs and viable channels to their origin}
\label{multipop}

\noindent As discussed in the previous section, the dynamical state of a stellar grouping can be used as a classification of star clusters and associations. Apart from the clarity it provides when discussing observational samples, this classification might also provide insight into open questions regarding their formation mechanism. Do bound star clusters  form in a fundamentally different way than unbound associations and fields stars? Or are star clusters simply the high density tail of the density distribution of the interstellar gas from which they form? 

In this section we discuss the additional clues that come from the stellar populations. The {(very small, or even absent)} star-to-star  spread in [Fe/H] and $\alpha$- and s-process elements in open clusters in the Milky Way is consistent with the measurement uncertainty of high-resolution spectra \citep[$\lesssim 0.05~$dex, e.g.,][]{2007AJ....133.1161D}, suggesting that all stars formed from a chemically homogeneous (i.e. well-mixed) molecular cloud. Somewhat surprisingly, the same level of homogeneity was found for unbound moving groups \citep{2007AJ....133..694D}. These  moving groups tends to be older ($\sim50~$Myr) than associations ($\sim10~$Myr). If they are the same objects in different evolutionary phases, it suggests that both bound clusters and unbound associations form in a similar way.  

The picture is completely different for old Galactic clusters: nearly all GCs in the Milky Way ($\gtrsim10~$Gyr) display star-to-star light-element  abundance variations (He, C, N, O, Na, Mg, Al). In fact, GC stars that have field-like chemical composition (usually called first population, 1P) are only a minority (10 to 30\,\%, depending on the GC), while the chemically anomalous stars (or second population, 2P) dominate \citep[e.g.,][]{bastianlardo2018}. The abundance variations display particular  anti-correlations, that are the tell-tale of hot-hydrogen burning \citep*{1990SvAL...16..275D,2007A&A...470..179P,2017A&A...608A..28}:  C--N (CNO-cycle, $\gtrsim20\,$MK), O--Na (NeNa-chain, $\gtrsim45\,$MK) and Mg--Al (MgAl-chain, $\gtrsim70\,$MK) anti-correlations. Additional support for the hot-hydrogen burning hypothesis comes from the fact that some GCs display broadened or split main sequences in the colour magnitude diagram \citep{Anderson_02,Bedin_etal04,Piotto_etal2007,Piotto_etal2012,Piotto_etal2015,Han_etal2009}, which has been attributed to He spreads  \citep{2004ApJ...612L..25N,2005ApJ...631..868D, 2016EAS....80..177C,Lagioia_etal19}. 
More recently, the so-called chromosome map was introduced, which is based on specific combinations of {\it HST} filters that are sensitive to He and N \citep{Milone_etal15}; it turns out to be a very powerful photometric tool to detect the presence of multiple stellar populations in Galactic and extra-galactic GCs \citep{Milone_etal17,Zennaro_etal19,Saracino_etal19}. 
In the last few years, evidence for N enhancement has been found in younger clusters ($\gtrsim2~$Gyr) in the Magellanic Clouds  \citep[e.g.,][]{2017MNRAS.465L..39H, 2018MNRAS.473.2688M,2018MNRAS.477.4696M}. Although a N spread is not necessarily pointing at the same origin as the variations in ONaMgAl seen  in the older halo GCs, it is at the moment something that can not be explained by (single) stellar evolution models. 
As mentioned earlier, the O--Na anti-correlation is so ubiquitous that it has been suggested to be the unique identifying property of a genuine GC \citep{Carretta_10}. The ubiquity does not mean that all GCs are similar; quite the contrary, the details of the multiple populations are different in every GC \citep{Milone_etal17}. Some note-worthy trends with GC properties have been identified, that might provide clues to the origin of these multiple populations. For example, the Mg--Al anti-correlation is only found in the most massive and metal-poor GCs \citep{Carretta_09, 2017MNRAS.467..412P,Masseron_etal2019,2019MNRAS.tmp.3134M}. In these clusters, also the minimum(maximum) O(Na) abundance is lower(higher) \citep{2009A&A...505..117C}. Both findings are expected if these anti-correlations are the result of hot-H burning and the temperature of the polluter was higher  in more massive and metal-poor GCs. In addition, both the He spread inferred from the main sequence broadening and the fraction of  stars with anomalous abundances
 correlate with GC mass \citep[][]{2014ApJ...785...21M, Milone_etal17,Milone_etal18}, implying that more polluted material is produced per unit of cluster mass in massive GCs. More details on the observational signatures can be found in \citet{bastianlardo2018} and \citet{2019A&ARv..27....8G}.
 
 It is broadly accepted that anomalous (2P) GC stars formed out of original proto-cluster gas mixed with the H-burning yields ejected by massive (M$>5$ \msun) and short-lived 1P stellar polluters  \citep[e.g.,][and references therein]{2017A&A...608A..28}. 
 However, there is no consensus on the nature of the polluter(s) and the pollution mechanism  \citep[e.g.,][]{Renzini_etal15,2016EAS....80..177C,bastianlardo2018}. 
 The vast majority of GCs show no spread in iron abundance, which suggests that (self-)enrichment by supernovae plays no role \citep[e.g.,][and references therein]{Simmerer_etal13,Marino_etal15,Marino_etal18}. GCs with clear [Fe/H] spreads, such as Omega Centauri ($\omega$ Cen) and M54,  are among the most massive clusters and are (former?) nuclear clusters. Several possible polluters have been proposed, which all reach the required high temperatures at some stage of their evolution: Asymptotic Giant Branch (AGB) stars
\citep[$\sim5-6.5\,\msun$,][]{2001ApJ...550L..65V}, massive stars \citep[$\gtrsim20\,\msun$,
][]{2006A&A...448L..37M,2006A&A...458..135P,2007A&A...475..859D,2007A&A...464.1029D}, massive binaries \citep[$\gtrsim10\,\msun$,][] {2009A&A...507L...1D,Bastian_etal13}, and
supermassive stars \citep[SMSs, $\gtrsim10^3\,\msun$,][]{2014MNRAS.437L..21D,2018MNRAS.478.2461G}. 

The different aspects and the pros and cons of the corresponding scenarios are described in \citet{Krause_20}, a review of this series. Here we would like to conclude on the fact that observations of 
proto-GC at high redshift will help discriminate between the different options in the near future. As we present in Section\ref{sec:redshift_formation_GC}, proto-GC candidates have already been detected with the aid of gravitational lensing. Their number detections will significantly increase in the \textit{James Webb} Space Telescope (JWST) and the European Extremely Large Telescope (E-ELT) era and spectroscopic sampling of their light will certainly help to confirm the nature of the stellar populations hosted in these systems.
 On the theoretical side, evolutionary synthesis models are developed to predict the characteristics of proto-GCs in the early Universe \citep{Renzini2017,pfeffer2019a,Pozzetti_etal19}. In this context, \cite{Martins_etal2020} developed the first synthetic models of proto-GCs hosting multiple stellar populations and a SMS. They compute theoretical combined spectra and
 synthetic photometry in UV, optical, and near-IR bands
 over a wide range of redshift (1 to 10), and predict the detectability of cool SMS in proto-GCs through deep imaging with JWST NIRCAM camera. 


\section{Young star cluster populations within the Local Group}
\label{sec:cluster_stats}

\noindent The properties of YSCs in the local, resolvable Universe span a wide parameter space. Their masses range from low-mass associations, such as the Orion Nebular Cloud \citep[ONC, e.g.,][]{O'Dell_08a,Muench_08,Robberto_13}, the Upper Scorpius association \citep[e.g.,][]{Preibisch_08}, or the Pleiades, to the massive super star clusters (SSCs, defined in this review as clusters with stellar masses above $10^5$ \msun) like R136 \citep{Hunter_95,Crowther_16} in the Large Magellanic Cloud (LMC) or in other galaxies of the Local group \citep[e.g.,][]{hunter2001, sabbi2008}. While it is still possible to resolve the most luminous stars in nearby galaxies \citep{Sabbi_18, Sacchi_18}, only star clusters located in the Milky Way or the Magellanic Clouds are close enough that a major fraction of their stellar population can be resolved with existing telescopes, such the \textit{Hubble} Space Telescope (HST) or from the ground using adoptive optic (AO) systems to correct for atmospheric turbulence (e.g., the Very Large Telescope (VLT), Keck, or Gemini).

Observing the low-mass stellar populations is crucial to understand cluster formation and evolution. While most of the energy is emitted by the most massive stars, the majority of the mass budget is bound in the low-mass stars, which influence the gravitational potential of their host systems' long-term evolution. 

New instruments like large integral field units (IFUs), such as the Multi Object Spectrographic Explorer \citep[MUSE,][]{Bacon_10} mounted at the VLT, the Gaia satellite \citep{Gaia_16,Gaia_18}, and long baseline photometric observations allow us, for the first time, to study the detailed 3D dynamics of the majority of stars in these resolved star clusters, including the dynamics of the gas \citep[e.g.,][]{Kamann_13,McLeod_15,Zeidler_18,Lennon_18,wright18,ward18,ward19,Getman_19,Zari_19}. This provides insights into the star cluster formation modes: Do star clusters form hierarchically, following the structure of the giant molecular cloud (GMC) \citep[e.g.,][]{kruijssen12b,Parker_14,Longmore_14,Dale_15,walker15,walker16,barnes19,Krumholz_19,ward19}, or do they form in monolithic, central starburst-like events \citep[e.g.,][]{Lada_84a, Bastian_06,Banerjee_15}? Future missions and telescopes, such as JWST, the E-ELT, or the Thirty Meter Telescope (TMT), will provide the necessary angular resolution and wavelength ranges to further investigate the low-mass end of the initial mass function (IMF)\footnote{We will refer to the stellar mass function as IMF throughout the review, although because of stellar evolution, the high-mass end of the stellar mass will disappear within a few Myr timescales} and the embedded objects in the surrounding HII regions. This will lead to a better understanding of the star formation and feedback processes in these HII regions under the influence of a large central population of massive, luminous OB stars, eventually shedding light on the formation process of the stellar populations within GCs.

The scope of this Section is not to describe the detailed, individual parameters of each star-forming region but rather give a more general overview over the observed parameter space provided by star clusters that are close enough to be resolved into their components. When looking at YSCs in local galaxies outside the Local group, the rich information contained in each of these star-forming regions and very young clusters will be collapsed into a handful amount of pixels. We give up on their  single physical components and look at them in a statistical meanighful way.

\subsection{Young massive star clusters in the Milky Way and Magellanic Clouds}
\label{sec:MW_MCl_clusters}

\noindent Compared to other, more distant galaxies \citep[e.g.,][]{Gascoigne_52,Hodge_61b,Hodge_61a}, the Milky Way host relatively few massive YSCs, none of which are expected to survive a significant fraction of a Hubble time \citep[][and references therein]{Krumholz_19}. Yet, together with the Magellanic Clouds, these are the only places where individual stars can be resolved down to the hydrogen burning limit or below, even in the dense star clusters \citep[e.g.,][]{Stolte_06,Sabbi_07,Zeidler_15}. Milky Way and Magellanic Cloud star clusters are crucial to understand the first few million years, during which star formation is on-going and a significant amount of gas is still present. Stellar and gas dynamics and interactions, feedback processes, and possible secondary triggered star formation in the surrounding HII regions is still poorly understood due to the lack of sufficient, large scale observations. To trace star cluster evolution over a longer time scale sophisticated simulations are necessary, yet these are only as good as their initial conditions. To understand cluster formation in a more distant Universe, unresolved with current telescopes, the local star cluster observations must suffice to deepen the knowledge about these cluster initial conditions.

With $\sim5\times10^4\,{\rm M}_\odot$ \citep[][and references therein]{Gennaro_17}, Westerlund 1 (Wd1) is the most massive YSC in the Milky Way. With an age of $\sim5$\,Myr it has already undergone several supernova explosions and is dynamically more evolved than other, younger massive Milky Way star clusters ($m>10^4\,M_\odot$), such as Westerlund~2 \citep[Wd2,][]{Westerlund_61} or NGC~3603. The Milky Way also hosts two YSCs in extreme environments, the Arches and Quintuplet Cluster \citep[e.g.,][]{Figer_99,Stolte_10,Stolte_15}. Being only $\sim60$~pc away from the Galactic center \citep{kruijssen15b}, the star-forming regions in this environment are characterised by high stellar and gas densities \citep{walker15}, highly compressive tidal fields \citep{kruijssen2019c}, turbulent motion \citep{oka01,henshaw16}, and they are located in a very steep gravitational potential. Their observations however are challenging because they are not only one of the densest and most efficient star-forming regions but are affected by a visual extinction exceeding 20\,mag. The LMC hosts the only nearby SSC. With an estimated age of 1--2\,Myr, R136, located in 30~Doradus (30~Dor) or the Tarantula Nebula, hosts the most massive and luminous stars known with masses up to $\sim300\,M_\odot$ and a spectral types of O2-3V \citep{Crowther_16}.

While the young Milky Way star clusters mainly have Solar metallicity, star clusters in the LMC \citep[distance: 50\,kpc, $A_V \approx 0.3$,][]{Schaefer_08,Imara_07} and the Small Magellanic Cloud \citep[SMC; distance: 62\,kpc][]{Hilditch_05} are located in a more metal poor environment, corresponding to the properties at higher redshifts ($z$). The typical metallicity in the LMC is $0.5\,{\rm Z}_\odot$ with a dust-to-gas ratio of $\sim 1/3$ of the Milky Way and the SMC has even lower values \citep[$0.25\,\rm Z_\odot$, dust-to-gas ratio: $\sim 1/6$ of the Milky Way, e.g.,][]{Russell_92,Rolleston_99,Lee_05,Roman-Duval_14}. This allows star and star cluster formation and evolution studies in lower metallicity environments, where effective stellar temperatures and luminosities are higher, resulting in faster stellar evolution and lower mass-loss rates \citep{Kudritzki_00}. An increased stellar temperature leads to higher far-ultra-violet (FUV) fluxes emitted by the most massive O- and B-stars.  Additionally, the low extinction toward the Magellanic Clouds allows detailed observations in the ultra-violet (UV) and FUV.

\subsection{Observing young star-forming regions}

Because of the large dynamical range of the physical processes in star clusters, observations across the full electromagnetic spectrum are necessary to fully understand these systems.

Stars and star clusters form due to the gravitational collapse of (parts of) GMCs \citep[][]{kennicuttevans2012,krumholz_araa19}. These GMCs mostly contain neutral (HI) and molecular (H$_2$) hydrogen at temperatures of 10--50\,K. These temperatures make it necessary to observe star formation sites with radio telescopes, such as the Atacama Large Millimeter Array (ALMA) to understand the GMC dynamics by observing various CO transition lines of the cold interstellar medium \citep[ISM, e.g.,][]{Yonekura_05,Furukawa_09,heyer09,sun18,Tsuge_19}. The unprecedented high spatial resolution of radio telescopes also allows for the direct observation of protoplanetary and debris disks around young stars in the later stages of the star formation process \citep[e.g.,][]{ALMA_15,van_der_Marel_18}.

\begin{figure}
  \includegraphics[width = \linewidth]{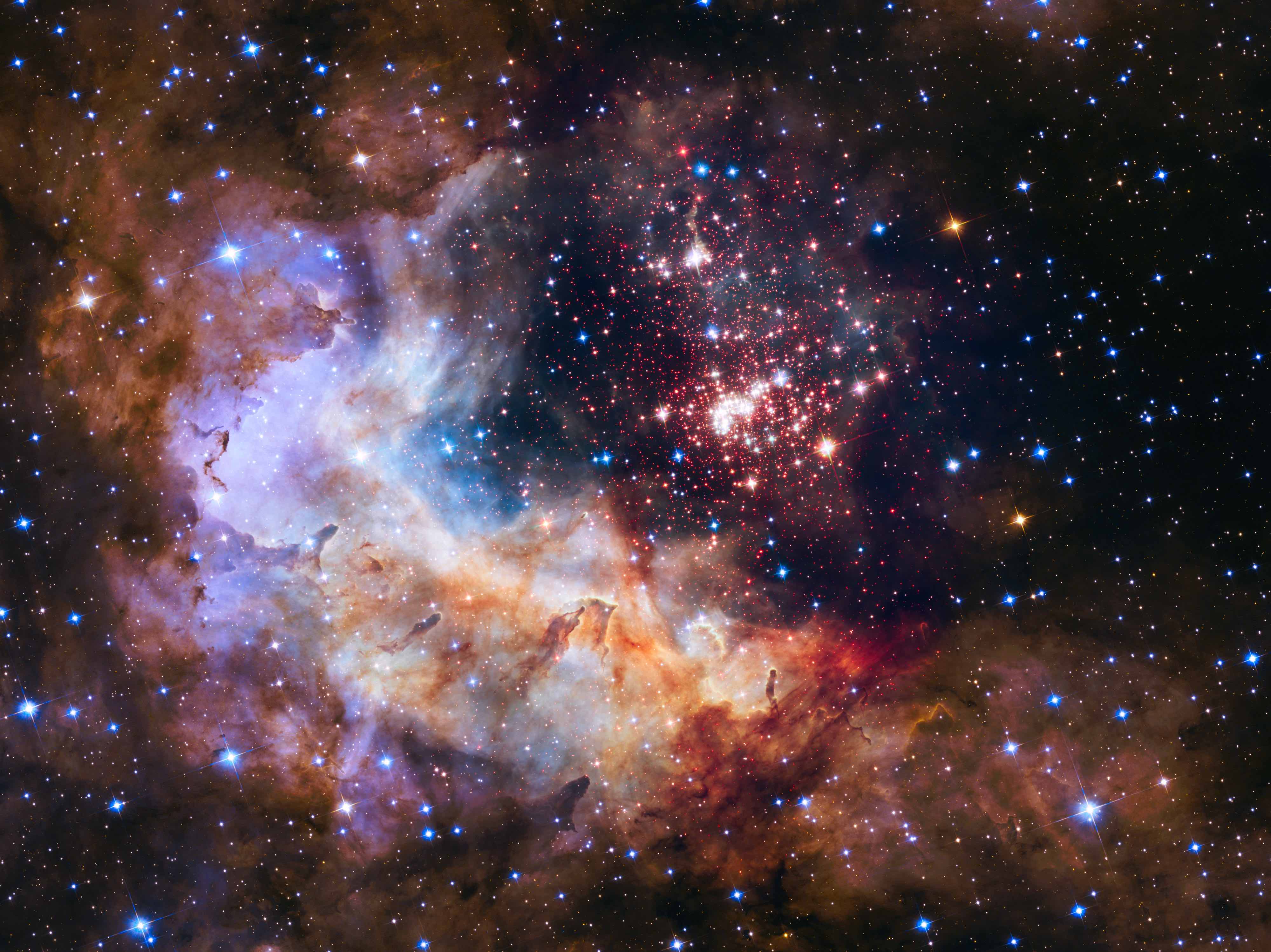}
\caption{The Galactic young massive star cluster Westerlund 2 in the center of the HII region RCW~49 as it is seen by HST in the optical and near-infrared. This image shows the cluster's very massive O-star population with up to 80~$M_\odot$ stars but also very low-mass stars with only 0.1~$M_\odot$ or even below. \textit{Credit: NASA, ESA, the Hubble Heritage Team (STScI/AURA), A. Nota (ESA/STScI), and the Westerlund 2 Science Team.}}
\label{fig:Wd2HST}       
\end{figure}

Observations from the infrared (far to near) to the FUV are necessary to observe stellar IMF and the resulting wide range of stellar masses (from 80--100\,$M_\odot$ OB stars to the faint 0.1\,$M_\odot$ dwarf stars, see Fig.~\ref{fig:Wd2HST}), high (differential) extinction, still deeply embedded objects, YSOs, and disk-bearing objects. Class 0-III YSOs and active star formation may still be present in remaining gas and dust clouds \citep{Carlson_07}. Mid and far-infrared space telescopes, such as \textit{Spitzer}, the Stratospheric Observatory for Infrared Astronomy (SOFIA), and \textit{Herschel}, are able to look more deeply into the dense gas clouds and observe and classify these objects \citep[e.g.,][]{Gaczkowski_13}.

Combining various optical and NIR broad-band filters, such as $UBVIJHK$-filters allow the construction of various two-color diagrams and color-magnitude diagrams (CMDs), which can be used together with model isochrone fitting to determine age, distance, and extinction of the stellar population as well as the individual stellar masses. This method is widely applied to very young and open clusters in the Milky Way and in general to star clusters in both the Magellanic Clouds and M31 \citep[e.g.,][]{zinnecker2007, Glatt2010, johnson2015}. 

Stars that are still in their pre-main sequence phase show strong excess emission in the NIR due to their circumstellar disks, Balmer line emission in the optical, and X-Ray emission due to large hot stellar coronae, magnetic coupling of the disks to the stellar surface, and flaring of the central stars. While X-ray observations require space missions (e.g., Chandra or XMM-Newton), the optical and NIR observations can also be obtained from the ground, and with extreme AO even at similar spatial resolutions as HST provides. Combining these broad-band observations with narrow-band observations such as the H$\alpha$ or Pa$\beta$ filter allows to detect pre-main sequence stars with active mass accretion \citep[e.g.,][]{deMarchi_11,Beccari_15,Zeidler_16b,Kalari_19} revealing protoplanetary disks. NUV and FUV photometry and spectroscopy from space is necessary to study and classify the most massive OB-stars. Their spectral energy distribution (SED) peaks in the (F)UV and most of their parameters are degenerate in optical CMDs. With these data stellar winds and the FUV flux budget can be measured \citep{Crowther_16}, which is responsible for accelerated disk dispersal of the lower mass stars in the close vicinity \citep{Clarke_07}, creating photo-dissociated regions (PDRs) in the surrounding gas cloud, as well as triggering secondary generation star formation. NUV photometry also allows to directly probe the UV extinction curve via the stellar color excess. 

\begin{figure}
  \includegraphics[width = \linewidth]{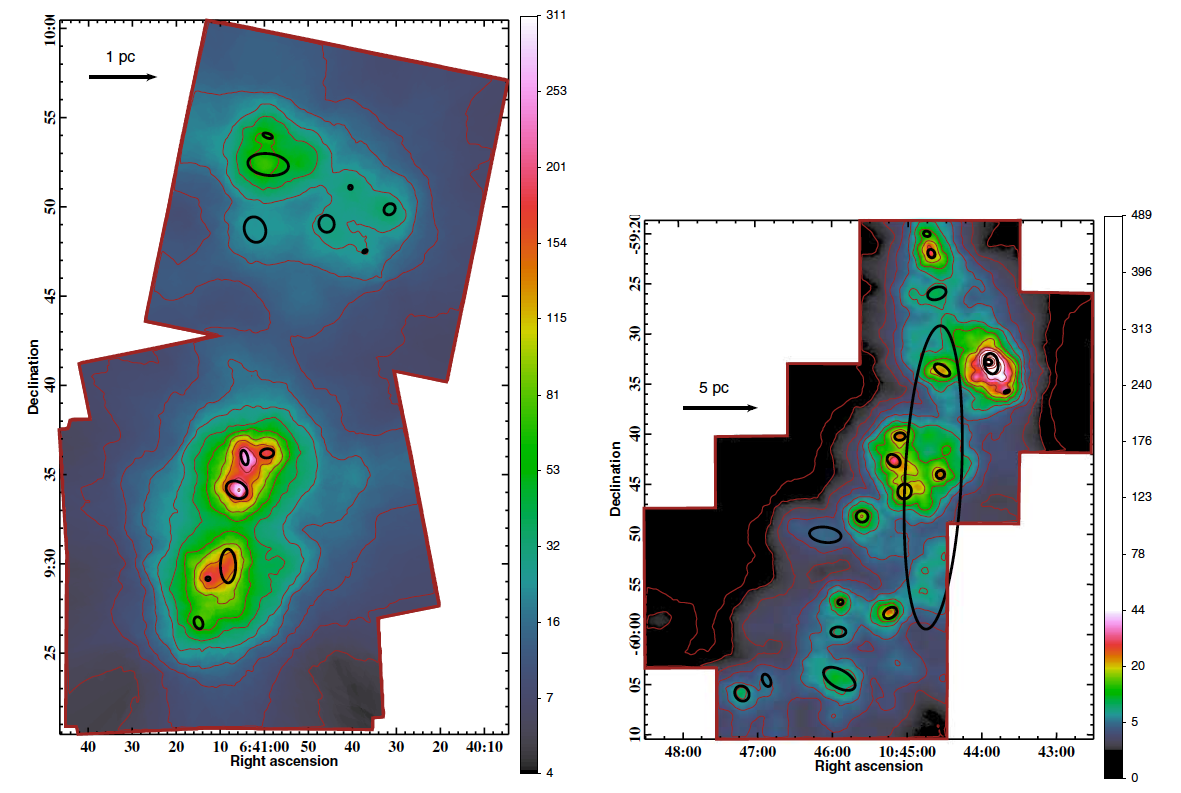}
\caption{The smoothed projected stellar surface density from the Massive Young Star-Forming Complex Study in Infrared and X-ray \citep[MYStIX,][]{Feigelson_13} is shown with a color bar in units of observed stars pc$^{-2}$. The brown contours show the increase in surface density by factors of 1.5. The ellipses mark the core regions of the isothermal ellipsoids. To identify the subclusters and to estimate the model parameters of the young stars a finite mixture model with a maximum likelihood estimation was used. The left panel shows NGC 2264 while the right panel shows the Carina star-forming region. This figure was published as part of Fig. 2 in \citet{Kuhn_14}.}
\label{fig:subcluster}
\end{figure}

Large star-forming regions such as 30 Dor in the LMC \citep{Sabbi_12} or the Carina Nebula Complex \citep{Smith_08,Zeidler_16a} are highly substructured and show a multitude of individual star clusters and associations of various masses and ages (e.g., Trumpler 14, 16, NGC~3324, and the Treasure Chest in the Carina Nebula Complex or R136, NGC 2070, and Hodge 301 in 30 Dor). These regions are dominated by feedback processes from massive stars, supernova explosions, and the formation of new stars in the surrounding GMCs. The individual clusters within these star-forming regions span wide mass and age ranges, i.e., \citet{Grebel_00} derived an age of 10--25\,Myr for Hodge~301, while R136 contains stars as young as 0.5\,Myr \citep{Walborn_97}. Individual, isolated star clusters or star clusters within these larger star-forming regions may themselves show sub clustering and highly complex structures \citep[see Fig.~\ref{fig:subcluster} and e.g.,][]{Kuhn_14}.

The by far best studied young star-forming region is the Orion Nebula (Messier 42 and NGC 1976) and its associated star cluster, the ONC. Although not very massive \citep[$4.6\times10^3\,M_\odot$,][]{Hillenbrand_98}, its close proximity \citep[$\sim440$\,pc,][]{O'Dell_08b} makes it a perfect target to study the stellar and gas content. The ONC revealed the first direct imaging of protoplanetary disks \citep[proplyds,][]{McCaughrean96, O'Dell_08a}. A recent 3D kinematic study by \citet{Zari_19} using Gaia DR2 data \citep{Gaia_16,Gaia_18} showed that the ONC is highly sub-structured, with stellar ensembles of different ages in several kinematic groups, mixed in 3D space, which are overlapping in projection. \citet{Jerabkova_19} suggested that these YSCs may harbor multiple populations. Using OmegaCAM photometry they identified three populations with an age difference of 3\,Myr between the oldest and the youngest sequence. These sequences cannot be described with binary or triplet systems alone leading to the conclusion that they are real, which is in agreement with the above findings by \citet{Zari_19} and suggesting star formation happens sequentially possibly triggered by the luminous OB stars.

Although multiple populations have not been detected in any other YSCs, mainly due to observational limitations, the majority of clusters and star-forming regions is still highly sub-structured showing multiple smaller clumps and are far from a spherical shape. In Wd2, \citet{Zeidler_18} recently discovered that the cluster stellar population shows multiple velocity components using MUSE observation to measure stellar radial velocities (RVs). These components appear to be spatially correlated with its two coeval subclumps \citep{Hur_15,Zeidler_15}, suggesting that they are, given the young age \citep[$\sim1\,\rm Myr$,][]{Zeidler_15}, an imprint of the formation history of the cluster. Other clusters, such as NGC~346 in the SMC are constituted even more complicated and show more than 16 individual sub clusters \citep{Sabbi_07}. Wd1 is with $\sim 5$\,Myr older and dynamically more evolved, is elongated, which is probably a product of the past merging of former subclumps \citep{Crowther_06, Gennaro_11}. Other young massive star clusters in a similar mass and age range, such as NGC~3603 \citep{Stolte_06,Pang_13} or R136 \citep{Hunter_95,Sung_04,Sabbi_12,Crowther_16} do not show present sub clustering. Their spherical shape may be explained through a spherical, burst-like single star formation event or through dynamical evolution suggesting that both hierarchical and in-situ cluster formation may be possible.

\subsection{The stellar mass function}

The IMF is a key parameter that affects almost all the fields of astrophysics from stellar populations up to formation of first galaxies and galaxy evolution in general. Empirical studies in the Milky Way and the Magellanic Clouds have revealed a remarkably constant IMF, regardless of location, age, or metallicity \citep[e.g.,][]{Chabrier_03,Bastian_10,Offner_14}. This lead to the idea that the IMF is constant across the Universe, which implies constant and somehow regulated star formation processes. More recent observational studies in more extreme, extra galactic star-forming regions, low-metallicity star clusters, and likelihood studies have started to challenge this view \citep{vanDokkum_10,Dib_14,Kalari_18}.

\begin{figure}
\center\includegraphics[width = 0.8\linewidth]{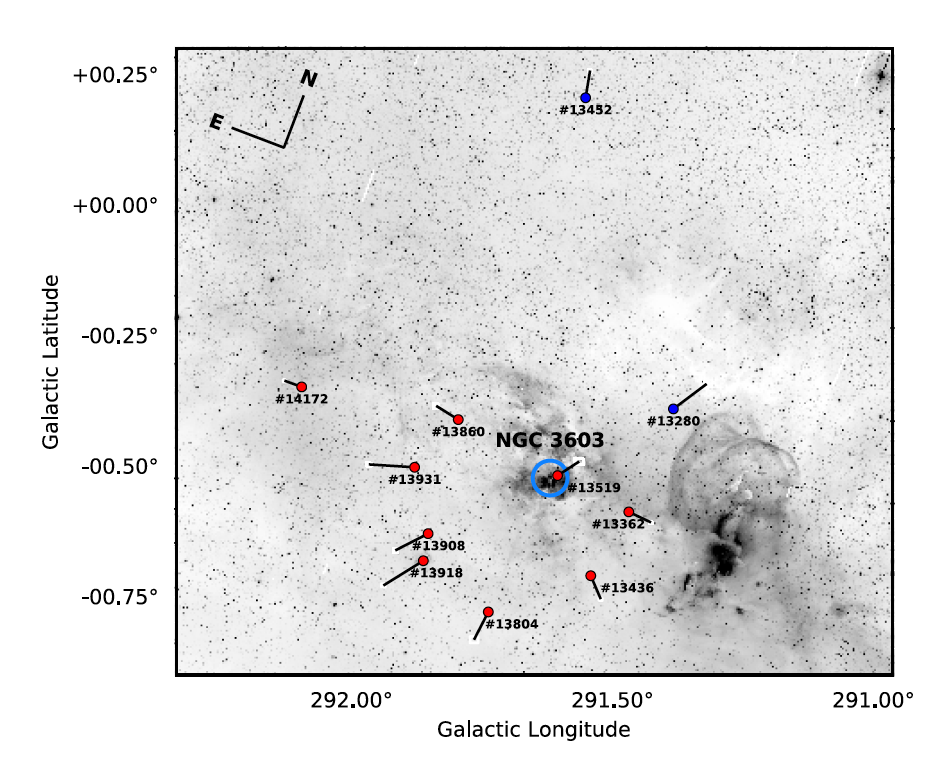}
\caption{This figure shows the larger area around the YSC NGC~3603 showing O-star candidates with a high probabilty to be ejected from the cluster core. Objects coloured in red are those with a high ejection probability while the two stars marked in blue have a trajectory consistent with a reduced probability to originate from the star cluster. The lines mark the direction of movement. This figure was published as Fig.7 in \citet{Drew_19}.}
\label{fig:ejected_stars}
\end{figure}

While the high-mass end of the IMF is relatively well-studied via simple star counts, this is more difficult for the low-mass end of the IMF due to observational limitations. Therefore, the shape of the IMF below a critical mass remains uncertain. Studying the IMF in in the Milky Way and Magellanic Cloud clusters showed that massive stars appear to be over-abundant compared to the expectations from a standard \citet{Salpeter_55} slope resulting in a slightly top-heavy IMF. This holds for star-formingregions in varying environments \citep[e.g.,][]{Zeidler_17,Schneider_18,Kalari_18,Hosek_19}. Most of the studies on the cluster IMF assume that if the cluster is massive and young enough $<2-3$\,Myr the upper main sequence is fully populated. Yet, even if this assumption holds (even the most massive stars have lifetimes of a few million years) most observations have a limited survey area and, therefore, fast runaway stars may have left the immediate vicinity of the cluster and the survey area. Recent studies of several massive clusters showed that a significant fraction of O and B stars may have been ejected from the cluster center within the last million years \citep[see e.g., Fig.~\ref{fig:ejected_stars} and][]{Lennon_18, Drew_18, Drew_19}. Although the reasons for the ejection are not clear these studies show that a significant number of stars can be missed using the traditional method of star counts in the closer cluster region. This argument though would lead to an even more top-heavy IMF.

The majority of YSCs are highly mass segregated. Mass segregation describes the over-abundance of high-mass stars relative to low-mass stars in the cluster center, compared to the outer regions of a star cluster. Mass segregation can have a significant influence on the cluster evolution and survival. The majority of massive stars will go supernova with the first $\sim5$\,Myr. These stars are located deeper in the cluster's gravitational potential well and in the case of remaining gas within the cluster, these supernova explosions and the resulting abrupt mass ejection may disrupt the cluster faster. Additionally, the low mass stars that are on larger orbits around the center can be stripped away more easily while moving through the ISM. Both effects lead to an accelerated cluster dispersal. The origin of mass segregation is assumed to be two-fold: 1) primordial mass segregation, where more massive stars have formed in the cluster center (suggested for e.g., Wd2, \citealt{Zeidler_17}, or NGC~346, \citealt{sabbi2008}) and 2) dynamical mass segregation,  where the high-mass stars migrated inwards due to two-body relaxation driving the cluster towards (but never fully reaching, see e.g., \citealt{trenti13,bianchini16,parker16,spera16}) energy equipartition (e.g., the Arches cluster, \citealt{Habibi_13}, or NGC 3603, \citealt{Sung_04}). The origin of mass segregation for a specific cluster is often difficult to establish due to their dynamical evolution. Sub-clustering and the none-spherical distribution of stars highly influences the determination of the cluster mass, as well as the crossing time \citep[e.g.,][]{Binney_87}, both on which the mass segregation time scale depends. \citet{McMillan_07} argue that merging sub-clusters keep their mass segregation imprint, the individual sub-clusters are less massive, and therefore, have shorter dynamical mass-segregation time scales. Conclusively, the origin of mass segregation of a star cluster that formed through hierarchical merging may be dynamical although the system as a whole suggests otherwise. Yet, the low-number high-mass stars in each individual sub-clump makes difficult to observe this effect.

\subsection{Feedback processes and stellar and gas dynamics}

 Stellar feedback dramatically modifies the appearance of the region where star clusters form. During the first $\sim 10$s of Myr, feedback from massive stars (mass $> 8$ \msun), in the form of photoionisation and mechanical feedback (radiation pressure, stellar winds, SN explosions) ionises the left-over gas in the region and it imparts energy and momentum on the dust and gas out of which the star clusters form. We refer the interested reader to \citet{dale15b} and to \citet{Krause_20}, another review in this series, for theoretical and numerical reviews of the stellar feedback from young star clusters. We summarise here some key observations of local massive star-forming regions in the Milky Way and Magellanic clouds carried out in recent years.

\begin{figure}
  \includegraphics[width = \linewidth]{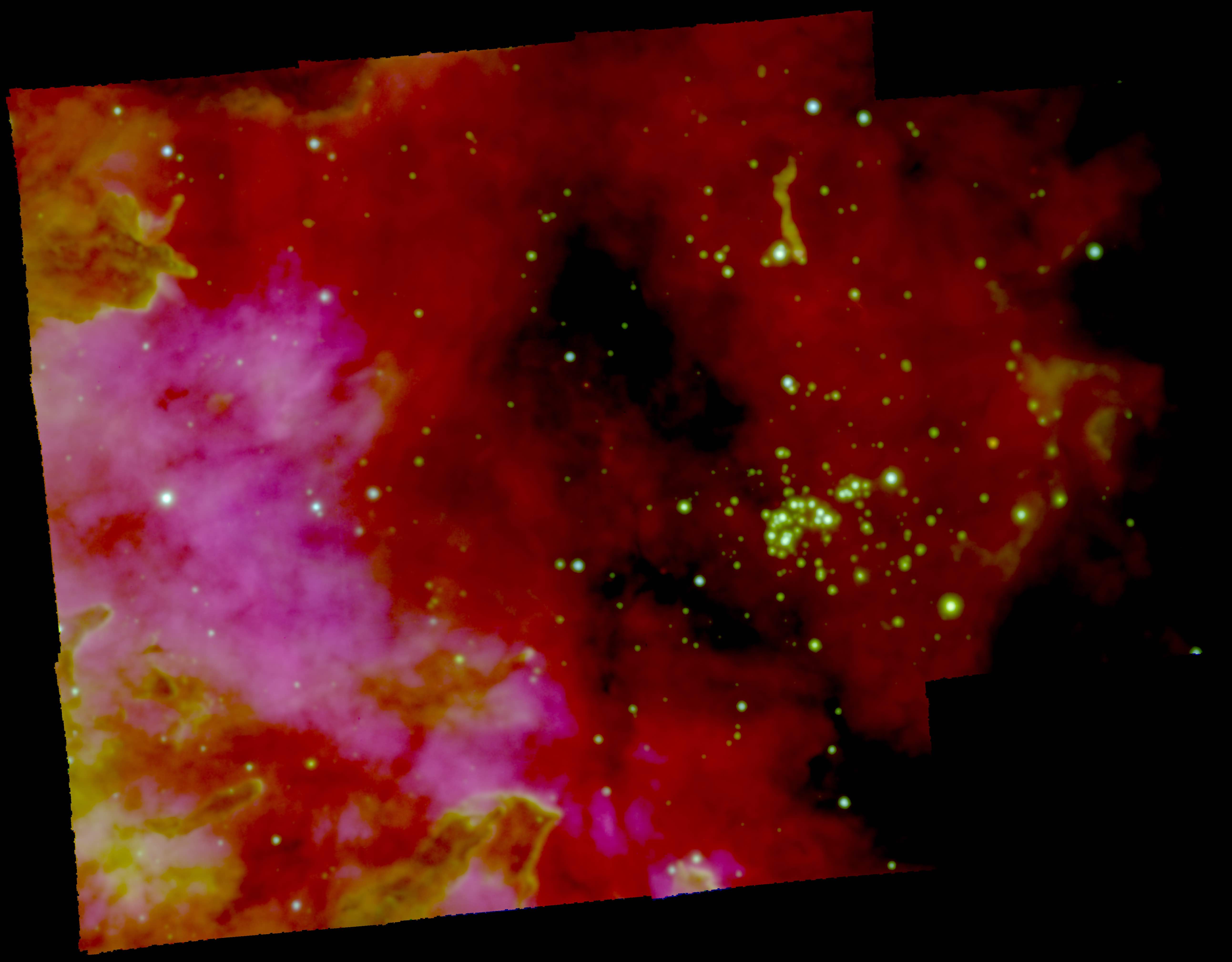}
\caption{The Galactic young massive star cluster Westerlund 2 as it is seen by the integral field spectrograph MUSE. The image represents H$\alpha$, NII$\lambda5583$, and [OIII$\lambda5007$] in red, green, and blue, respectively. This image was similarily published in \citet{Zeidler_18}.}
\label{fig:Wd2MUSE}
\end{figure}

 The combination of ground-based, AO supported telescopes using optical and NIR IFUs and space-based photometry and FUV spectroscopy provide astonishing insights into the feedback processes of these YSCs (e.g., see Fig.~\ref{fig:Wd2MUSE}). FUV, NUV, and optical spectroscopy of massive OB stars allow accurate spectral typing, stellar wind strength measurements, and FUV flux determinations, providing information about the ionizing flux budget of the central star cluster \citep[e.g.][]{smith2016}. The early feedback originating within the star cluster could be the main driver for possible, triggered secondary star formation events \citep{McLeod_18,Crowther_16,Zeidler_18} in the shell of gas and dust still surrounding the cluster, which emphasized the necessity of a detailed analysis of the gas.

Mapping optical strong and forbidden line ratios yield insightful information on the properties of the ionised gas. Typical Hydrogen Balmer line decrements (e.g., H$\alpha/$H$\beta$) provide extinction information of the star-forming regions, leading to the reconstruction of the relative 3D location of the stars within the cluster, the individual PDRs, and gas and dust rims \citep{McLeod_15}. Studies show \citep[e.g.,][]{McLeod_16b} that gas pillars need a minimum density to survive a given, local ionizing radiation level, confirming existing, theoretical mass-loss rate models. Combining multiple optical gas emission lines (i.e., the Balmer lines or [SII] 6717,6731) lead to the detection of embedded objects in the gas, like Herbig- Haro (HH) jets and bipolar outflows. Recently this has become feasible also outside the Milky Way, namely in the Magellanic Clouds \citep{McLeod_16a,McLeod_18}.

Line ratios like [OIII]/[OII], or [OIII]/[SII] are routinely used to map the optical depth of the medium surrounding stars massive enough to produce Ly continuum radiation \citep[e.g.][]{pellegrini2012}. [OIII] line emission is very strong in proximity of the ionising source and quickly decline at the edge of the HII region, where [OII] and [SII] emission increases. Therefore, a decline in the high ionisation species and increase in the low ionisation lines (e.g. decreasing [OIII]/[OII], or [OIII]/[SII] ratios), will correspond to an increase in the optical depth of the HII region and, therefore, smaller chances of for the ionising radiation to escape the region. In the LMC, \cite{pellegrini2012} show the power of this technique, by tracing regions that are ionisation bound (optically thick to their ionising radiation) and/or density bound (optically thin, therefore leaking ionizing radiation). It is quite complex to relate the 2D picture of a star-forming region provided by IFU observations to the real status of the system. Numerical simulations show that high ionisation channels might form in low gas density regions created by turbulence in the gas \citep{Dale_15}. Recent, high-resolution cosmological simulations have reported to find that most leaked ionizing photons are from star-forming regions that usually contain a feedback-driven kpc-scale superbubble surrounded by a dense shell. Young stars in the bubble and near the edge of the shell can fully ionize some low-column-density paths pre-cleared by feedback, allowing a large fraction of their ionizing photons to escape \citep{ma19, ma2020}. 

Another leap forward will be made with JWST, both for detecting and probing accreating protostars and for investigating the effect of stellar feedback in the photo-dissociated (PDRs) and most dense gas regions that remains inaccessible at optical wavelengths.  JWST will have the necessary spatial resolution in the NIR and MIR to detect YSOs in the gas rim, to study their properties in detail, and to determine to which extent the central ionizing cluster drives star formation into the surrounding gas cloud as seen e.g., for NGC~602 \citep{Carlson_07}. It will also give insight into the evolution and distribution of disk-bearing objects throughout the cluster. Observations \citep[e.g.,][]{deMarchi_11,Zeidler_16b} hint that the close proximity to the OB-star population accelerates mass accretion processes in protostellar disks, leading to a faster disk dispersal, and eventually hinders planet formation, confirming various theoretical studies \citep[e.g.,][]{Clarke_07,Anderson_13,winter20}. JWST will also allow us to map the gradual evolution of the gas and dust within the star-forming region as a function of its ionising stellar population. Polycyclic aromatic hydrocarbon (PAH) emissions, combined with photoionisation line emissions in the MIR, will provide an extinction-free view of the earliest phases of interaction between the source of feedback and the surrounding ISM. These studies are currently limited to star-forming regions in the Milky Way and Magellanic Clouds \citep[e.g.,][]{chevance2016,chevance2020C}, but these will be extended beyond the Local Group, enabling a much more complete understanding of the early phases of star formation in a large variety of galactic environments and physical properties.

\subsection{The young star cluster population of the Milky Way; properties of open clusters}
\label{sec:open_clusters}

Although the number of massive YSCs in the Milky Way are limited, our Galaxy hosts numerous open clusters. The detection of those clusters can be challenging due to their potential lower stellar density (surface densities are not much higher than the those of the field stars) and the lack of gas. Extensive all-sky star catalogs, such as the ASCC-2.5 bright star catalog \citep{Kharchenko_01}, the two Micron All Sky Survey \citep[2MASS][]{Skrutskie_06}, or the Gaia catalog \citep{Gaia_16,Gaia_18} are necessary to attempt detecting those star clusters. Various studies, dating back to the 1970s \citep[e.g.,][]{Becker_71,Kharchenko_05a,Kharchenko_05b, Kharchenko_12,Schmeja_14,Scholz_15,Castro-Ginard_20} attempt to collect and classify a complete sample of open star clusters in the Milky Way. 

\begin{figure}[htb]
\includegraphics[width = 0.9\linewidth]{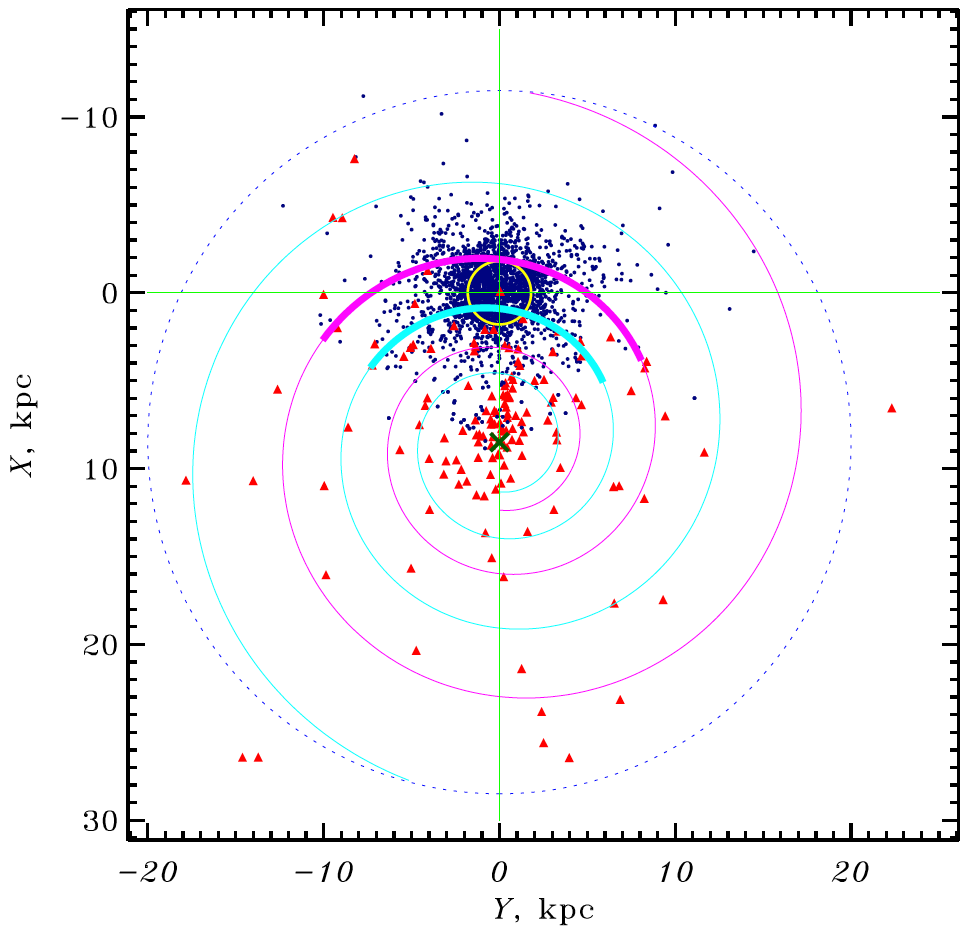}
\caption{The distribution of star clusters onto the Galactic XY-plane. Blue dots represent open clusters and associations while red triangles mark GCs. The border of the Galactic disk is represented by the dotted circle (diameter 20 kpc). The local spiral arms are marked as thick solid cyan and magenta sections and are formally formally extended to the edge of the disk with thin solid curves. The light (yellow) thick circle around the Sun with radius of 1.8\,kpc marks the completeness limit of the survey. The cross at (X, Y) = (8.5, 0) kpc indicates the position of the Galactic center. This figure was published as Fig.2 in \citet{Kharchenko_13}.}
\label{fig:OC_Kharchenko}
\end{figure}

\citet{Kharchenko_13} present a sample of 3006 open clusters (see Fig. \ref{fig:OC_Kharchenko}). This catalog was further extended by another 202 clusters by \citet{Schmeja_14} and \citet{Scholz_15} leading to a total number of 3061 open and 147 GCs. The analyzed open clusters cover a wide range of ages \citep[$6.0 \le \log(t) \le 9.8$,][]{Kharchenko_16} with absolute integrated $K_S$-band magnitudes between -11.7\,mag and 0.6\,mag. \citet{Kharchenko_16} also analyzed the cluster luminosity function (CLF) with respect to cluster age and distance to the Galactic center using 2MASS photometry. The slope of the CLF appears to decrease with increasing age up to an age of $\log(t) \approx 7.2$ slightly increase for $8.3 < \log(t) < 8.8$ and decrease again for older ages. This behaviour may be explained by stellar evolution, changing the relative number of red giant stars in the individual clusters, which dominate the luminosity in the NIR. Additionally, \citet{Kharchenko_16} found that the CLF slope increases from the inner to the outer Galactic disk, which may indicate that massive clusters tend to be located more in the inner disk. One needs to caution a possible bias due to the limited depth of the survey toward the fainter end of the CLF, especially in the direction of the Galactic center, where extinction is higher.

The Gaia DR2 \citep{Gaia_16,Gaia_18}, subsequent data releases, and machine learning techniques will increase the sample of open clusters by the analysis of the 5D phase space ($l,b,\varpi,\mu_\alpha,\mu_\delta$). Many recent studies of the Gaia DR2 reported the findings of unknown open clusters using a variation of detection techniques. For example, \citet{Cantat-Gaudin_19} reported 41 new open clusters using a Gaussian-mixture model, while \citet{Sim_19} applied a visual inspection in proper motion and $l$ -- $b$ space identifying 206 new open clusters within a distance of 1\,kpc of the Sun. \citet{Liu_19} discovered 76 new star clusters within 4\,kpc of the Sun using a friend-of-friend-based cluster finder method. \citet{Castro-Ginard_20} detected 582 unknown open clusters using a deep learning artificial neural network.

Many of these newly detected clusters are located closer than 2\,kpc from the Sun and, although, these studies probably consist of overlapping samples it shows that the statement of having compiled an almost complete sample (out to 1.8\,kpc from the Sun) by \citet{Kharchenko_16} has to be handled with care. It also shows that with the further data releases of the Gaia mission and an introduction of new techniques in machine learning and neural networks using big data may reveal many more clusters in the near future.


\section{Star cluster populations in the local Universe; a statistical view of their formation and evolution}
\label{sec:local_universe_clusters}

As we move away from the Local Group, we lose resolution on the single components of star clusters but gain a viewpoint into whole cluster populations forming into a much larger  spectrum of galactic environments than offered by the Local Group.

YSCs form in the densest regions of GMCs.  Turbulent energy regulates the density structure and distribution of the cold gas. When gravitational fragmentation takes over, the densest regions in a cloud start to collapse. The interplay between turbulence and gravity results in a clustered and hierarchically distributed star formation. However, only gravitationally bound stellar systems, with stellar densities sufficient to overcome the tidal field of the galaxy and the destabilising gravitational pull of the remaining gas \citep{elmegreen10, kruijssen11} will move away from their birth place and survive for a certain time span within their host galaxies.

We will introduce the main properties of GMC populations and, in general, the conditions of dense gas in local galaxies (Section~\ref{gmcpop}). We will then focus on the statistical properties of cluster populations. We will show how YSCs trace the clustering properties of star formation and the largest coherent regions of star formation in different galaxies (Section~\ref{yscclustering}). We will discuss the cluster size-mass relation as determined from measurement of cluster populations in local galaxies and its implications for cluster formation and evolution (Section~\ref{yscsize}). To link GMC to YSC populations it is necessary to take into account that only a fraction of the dense gas forming stars will result in bound stellar systems. To date, contrasting evidence are reported in the literature both in favour and against a variation of the fraction of stars forming in bound clusters as a function of galactic environment. We will combine results available in the literature, with recent multi-band imaging survey of a large spectrum of local galaxies. Namely, we will refer to the analyses based on the HST treasury program Legacy ExtraGalactic UV Survey \citep[LEGUS,][]{Calzetti_15, adamo2017, Sabbi_18, cook2019}. LEGUS sampled typical star-forming galaxies within 4 and 18 Mpc. The other project is Hubble imaging probe of extreme conditions and clusters (HiPEEC, Adamo et al 2020a to be subm.). HiPEEC consists of 6 starburst/merger systems observed and analysed in a similar fashion as the LEGUS targets. The distances of these systems are between 30 and 80 Mpc and SFRs are above 10~\msunyr, i.e. this program extend the cluster analysis from the LEGUS galaxy spectrum to highly efficiently star-forming systems.   We will summarise the current status of the field from the observational, theoretical and numerical point of view (Section~\ref{yscgamma}). After formation, the masses of the newborn YSCs follow mass distributions which have a power-law shape of index close to $-2$. However, the description of the YSC mass function requires the addition of an upper mass truncation (\mc) that we will discuss more in detail in Section~\ref{yscmassfunction}. Finally, to fully describe YSC populations in local galaxies we also need to account for their dissolution rates. We will provide a description of the possible models put forward and how they are reflected in the literature in Section~\ref{yscsurvival}.  

\begin{figure}
  \includegraphics[width = 0.5\linewidth]{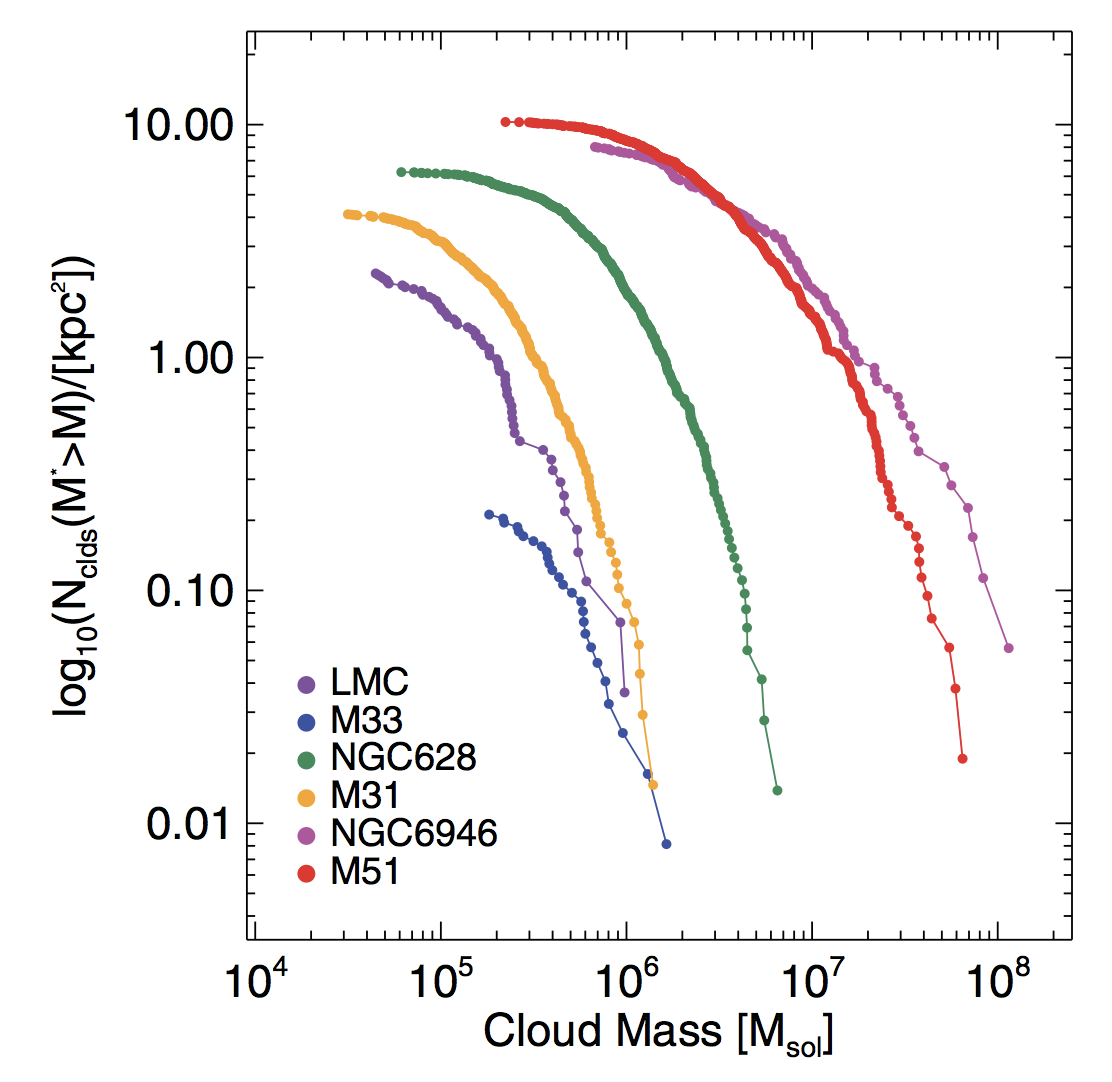}
  \includegraphics[trim=0mm 0mm 8.9mm 0mm, clip=true, width = 0.5\linewidth]{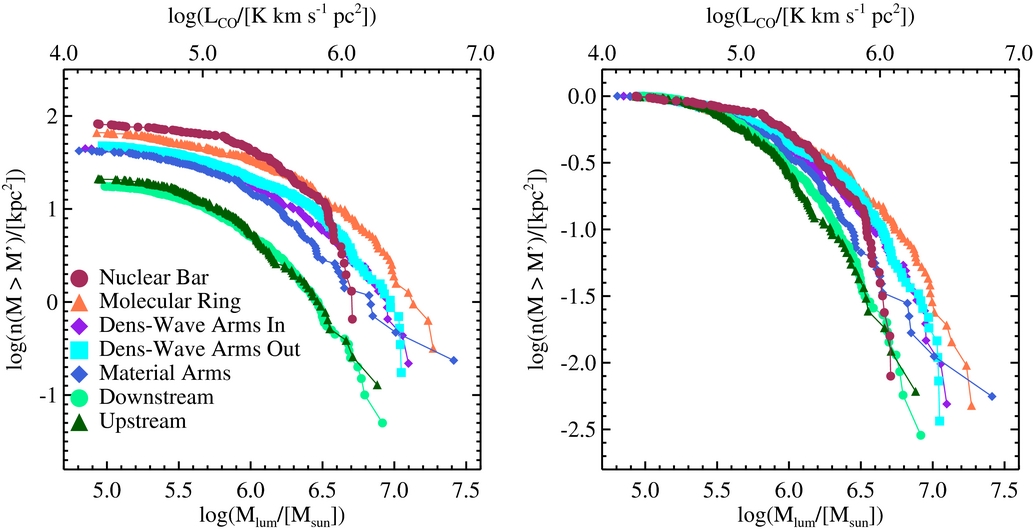}
\caption{Cumulative GMC mass spectra normalised by the observed area for different galaxies (left; from \citealt{hughes2016}) and for different regions of the spiral galaxy M51 (right; from \citealt{Colombo2014}). The differences in slopes and maximum masses in different galaxies and for different kinematic environments of a given galaxy suggest an environmental dependence of the GMC mass distribution.}
\label{fig:mass_spectrum}
\end{figure}

\begin{figure}
\centering
  \includegraphics[width = 0.7\linewidth]{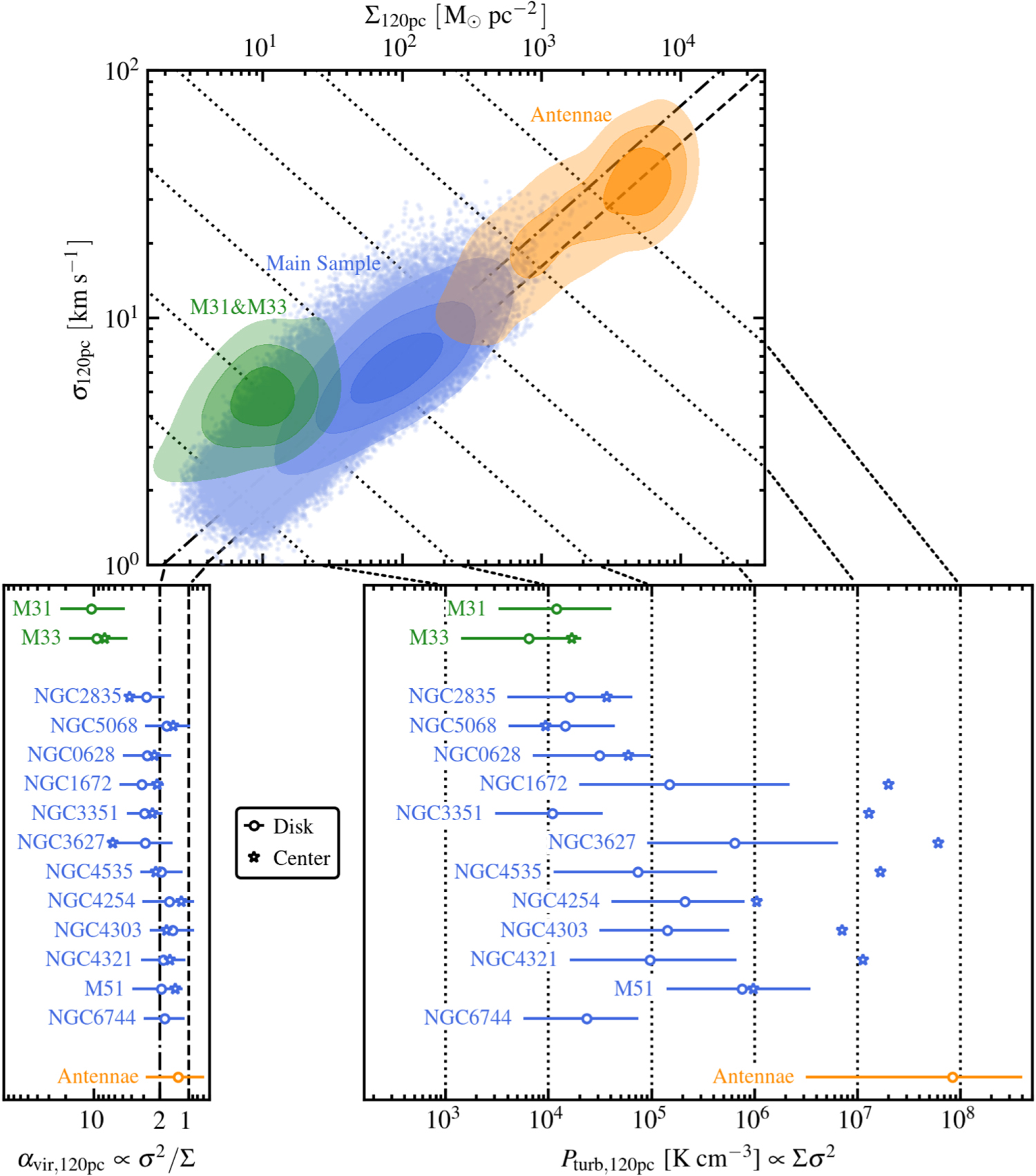}
\caption{Figure from \cite{sun18}. The top panel shows the relation between the CO line width $\sigma$ and the gas surface density $\Sigma$ at a common resolution of 120pc for the discs of a sample of 15 nearby galaxies. The bottom left panel presents the mass-weighted distribution of the virial parameter $\alpha_{\rm vir}$ and the bottom right panel the distribution of turbulent pressure $P_{\rm turb}$ for the disc (circles) and center (star symbols) of all galaxies. The spread in molecular gas dynamical state and internal turbulent pressure is clearly visible within and between galaxies, in particular when comparing normal star-forming disc galaxies with a merger system such as the Antennae, or more quiescent galaxies such as M31.}
\label{fig:sun18}
\end{figure}

\section{Properties and conditions of the molecular gas in local galaxies}
\label{gmcpop}

To understand the conditions under which YSCs form, it is important to understand how the properties and spatial distributions of GMCs depend on the environment (i.e. galaxy structure and dynamics, ISM pressure, etc) and how these are linked to the properties of YSCs. From early Milky Way observations, GMCs seem to have relatively uniform properties and follow the relations described by \cite{larson81}, showing correlations between their sizes, line-widths and luminosities \citep[e.g.,][]{solomon1987, heyer2009}. These relations describe clouds as having constant surface densities, being in virial equilibrium and following a size-line width relation. However, the universality of GMC properties and of Larson's relations has been questioned since. Early theoretical works already predicted an environmental dependence of cloud properties, such as their surface density, velocity dispersion, mass, and size distributions (e.g., \citealt{Elmegreen1989}). However, until recently, it has been challenging to extend GMC observations to other galaxies. To probe cloud properties and their dynamical state requires surveys of the molecular gas in nearby galaxies at high sensitivity and resolution, with a large coverage to provide a statistically significant sample, in a large variety of environments. Recent sub-millimetre facilities such as the Plateau de Bure Interferometer (PdBI), ALMA and the NOrthern Extended Millimeter Array (NOEMA) have now made this possible.

Several observational studies have since then revealed significant variations of the molecular cloud properties in nearby star-forming galaxies \citep[e.g][]{Koda2009, hughes2013, donovan2013, leroy2013, Colombo2014, hughes2016, leroy2016, sun18, schruba2019}, as well as in starburst galaxies and merging systems \citep{downes1998, wilson2003,wei12} and it is now clear that there exists an environmental dependence of cloud properties. For instance, in the nearby spiral galaxy M51, \cite{Colombo2014} highlight the change of GMC properties in different regions of the galaxy (e.g.\ bar, bulge, disk, spiral arms, inter-arm regions). In particular, the GMC mass spectrum is found to vary (in terms of slope, normalisation and maximum mass; see Fig.~\ref{fig:mass_spectrum}) between arms and inter-arm regions: the population of clouds in the inter-arm regions is dominated by lower-mass objects (with a power law slope of the mass spectrum steeper than $-2$), than the population located in the arms (with a slope shallower than $-2$). This suggests that different mechanisms are regulating the growth and destruction of GMCs in different regions. Differences between the mass spectra of GMCs have also been observed between galaxies. In particular, \cite{hughes2016} show that the mass distribution is flatter in denser and more massive galaxies (e.g.\ M51) compared to lower mass galaxies (such as M33; see Fig.~\ref{fig:mass_spectrum}).

The physical state of GMCs in extragalactic environments has also been compared to that seen in the Milky Way. Clouds in extragalactic environments seem in general to follow the Milky Way size-linewidth relation relatively well \citep{bolatto2008, faesi2018}. In addition, molecular clouds in other galaxies are typically observed to be bound or marginally bound structures, with a viral parameter close to unity \citep[e.g.,][and references therein]{krumholz_araa19}, although there are exceptions, especially in low surface density galaxies. This agrees with observations of GMCs in our Galaxy. \cite{sun18} observed relatively universal virial parameters throughout a sample of nearby galaxies ($\alpha_{\rm vir}=1{-}3$, excluding M31 and M33; see Fig.~\ref{fig:sun18}), suggesting that molecular clouds are close to virial energy balance. However, they find a wide range of different turbulent pressures, with ranges of $\sim 1-2$\,dex within galaxies and a variation across the sample of four orders of magnitude. In particular, in the gas-rich, turbulent environment of the Antennae galaxies, which is the nearest major merger, the internal pressure of the gas is considerably elevated by the merging process compared to disc galaxies. However, this does not seem to significantly affect the dynamical state of the gas -- the measured scaling relation between the CO line width $\sigma$, and the gas surface density $\Sigma$ in the Antennae galaxies follows the average relation observed in the discs of star-forming galaxies. This example is particularly interesting, because it possibly forms a bridge to the extreme environmental conditions of high-redshift galaxies. As shown by \citet{DZ2019}, the GMC population detected in a typical star-forming galaxy at $z \sim 1$ has physical properties similar to those detected in local starbursts. \cite{schruba2019} extend this result by showing that, statistically speaking, GMCs are in ambient pressure-balance virial equilibrium. Clouds are near energy equipartition in high-pressure (molecular-dominated) environments ($\alpha_{\rm vir} \sim 1-2$, considering self-gravity only) and pressure confined by the diffuse ambient medium in low-pressure (atomic-dominated) environments, leading to higher viral parameters ($\alpha_{\rm vir} \sim 3-20$).

The environmental dependence of ISM structure and molecular cloud properties also affects the process of star formation and feedback. Recent work by \cite{chevance20} analysing a sample of galaxies from the PHANGS survey (Physics at High Angular resolution in Nearby Galaxies; \citealt{sun18,schinnerer19}; Leroy et al. in prep) shows that the molecular cloud lifetime is not constant between and within galaxies, suggesting that the cloud lifecycle, star formation, and feedback are regulated by different physical processes in different galaxies. Specifically, the lifetimes of CO clouds sitting in environments of high global (kpc-scale) molecular gas surface density ($\geq 8\,$M$_{\odot}$ pc$^{-2}$) are regulated by galactic processes (in particular the gravitational free-fall of the mid-plane ISM and shear, as predicted by \citealt{jeffreson2018} and in agreement with theoretical predictions by \citealt{dobbs2013} and \citealt{reyraposo2017}). By contrast, CO clouds in environments of low global molecular gas surface density ($\leq 8\,$M$_{\odot}$ pc$^{-2}$) decouple from the galactic environment and live for a free-fall time or a crossing time, i.e.\ their lifetime is regulated by internal dynamics. More details about the lifecycle of molecular clouds can be found in \citet{chevance20b}, another review in this series. After the onset of massive star formation, the rate at which YSCs will destroy their parent molecular cloud through feedback is also likely to be environmentally dependent. The duration of this feedback phase has been shown to be relatively short (a few Myr after the formation of the first massive stars; \citealt[e.g.,][]{kawamura2009, whitmore2014, hollyhead2015, Grasha2018, Kruijssen2019d, chevance20}), suggesting a rapid cycling of the gas, with a low integrated efficiency of star formation per formation event \citep{Kruijssen2019d, chevance20}. Future, multi-wavelength, high-resolution observations of the gas during the early phases of star formation and feedback with recent and coming facilities such as MUSE and the JWST in a large variety of environments will help understand how the properties of YSCs are affected by the properties of their natal molecular cloud and by the large-scale galactic environment.

\section{The clustering properties of young star clusters}
\label{yscclustering}

Star formation is clustered, carrying the imprint of the gas from which stars form \citep{Lada_03, krumholz_araa19, ward19}. The gas in galaxies is hierarchically distributed, with  power law mass distributions  measured for molecular clouds \citep[e.g.,][]{elmergeen_falgarone96,romanduval10}, for the gas within both molecular \citep{lombardi15} and pre--molecular clouds \citep{MivilleDeschenes2010}, and for dense cores \citep{Stanke2006,Alves2007} and young stellar objects \citep[e.g.,][]{Schmeja2008}.  In star-forming  regions, the ISM fragments into smaller and smaller substructures, driven by  supersonic turbulence aided by gravity \citep{ElmegreenScalo2004}. At the smallest scales of the hierarchy are the stars, which also form fractal, scale--free structures of increasing density and decreasing scale from large star--forming complexes to bound star clusters \citep{Elmegreen2011}. Observations reveal that young stellar populations, associations, and clusters are in fact clustered \citep[e.g.,][]{Gieles2008, bastian2009, delafuente2009, gouliermis2015, Grasha2015, Grasha2017a, Grasha2017b}. 
YSCs trace the densest peaks of the hierarchy, and can be used to trace the clustering and its relation to the hierarchy of the gas.

Pre-supernova feedback from massive stars, in the form of stellar winds and photoionisation, exposes stellar clusters \citep{hollyhead2015,smith2016} and even disperses molecular clouds within the first $1{-}5$~Myr \citep{kim18a, rahner19, Kruijssen2019d, chevance20}, i.e., well before secular and bulk motions act on star clusters to disperse them out of the parent environment. As a result, emerged star clusters with ages below a few Myr are closely associated with their parent cloud: the median age of the clusters whose location is projected within the area  occupied  by a molecular cloud  is about 4~Myr and 2~Myr in the two galaxies M51 and  NGC 7793, respectively \citep{Grasha2018, Grasha2019}. If the clusters are bound and survive, they tend to migrate away from their parent cloud as they age; in the same two galaxies, clusters that are more than 4 times separated from the closest molecular cloud are about 12 and 3.5 times older, respectively, than those that are coincident with the cloud's footprint. Although the median ages of the star clusters not associated with the molecular clouds are drastically different for the two galaxies,  $\sim$50~Myr in M51 and $\sim$7~Myr in NGC 7793, the differences disappear when the ages are normalized by the median age of the entire young ($<$200~Myr) cluster population. The result is that the amount of time a cluster takes to migrate away from the  parent molecular cloud is a fixed fraction, $\sim$1.1--1.3, of the median age of the cluster population. This result is likely related to the measured sizes of the molecular clouds that host YSCs, as well as to other effects (e.g., the dispersion velocity of the cluster population): the median radius of the clouds is $\sim$10~pc in  NGC 7793 and $\sim$40~pc in M51, which is reflected in the footprints that are used for the association between clouds and clusters. 

The two--point correlation function (TPCF) is a  standard tool for measuring the clustering of a population, by quantifying how much the distribution of pairs deviates from a random distribution as a function of the pairs' separation. According  to this metric, \citet{zhang2001} report that younger star clusters in the Antennae are more clustered and more associated to longer wavelength tracers of star formation. Recently, \citet{Grasha2018, Grasha2019} find that molecular clouds are randomly or almost randomly distributed in M51 and NGC 7793,  but massive clouds are clustered, in agreement with findings  that massive clouds are preferentially  located in the spiral arms and other galactic structures \citep{Koda2009, Colombo2014}. This is  matched by the TPCF of the youngest, $<$10~Myr, star clusters which are as strongly clustered as the massive clouds which they likely originated from.

As the star  clusters age, they disperse or migrate within the galaxy, and this trend is also reflected in their TPCF \citep{Grasha2015, Grasha2017a}. In general, clusters younger than about 40~Myr have a TPCF that is best described as a power law with exponent$\sim -$0.6-- $-$0.8, while star clusters older than $\sim$40--60~Myr are consistent with a random distribution. The age difference between any two pairs of clusters increases for increasing separation, according to  a power law $\Delta (Age)\sim (Sep)^{\alpha}$ with $\alpha\sim$0.3--0.6  \citep{Efremov1998,delafuente2009, Grasha2017b}. For reference, in turbulent--driven star formation $\alpha$=0.5 \citep{Elmegreen1996}.  The power law is truncated at  separations between 200~pc and 1~kpc, depending on the galaxy; this maximum separation marks the maximum  `cell of coherent star formation' present in galaxies \citep[also see][]{Kruijssen2019d}. The size and age difference at the truncation point define a velocity, which is likely related to the average speed at which turbulence moves through the `cell of coherence'. This speed is a constant multiple, about a factor 2--3, of  the velocity difference imparted by shear in each of the galaxies  \citep{Grasha2017b}. Thus, while turbulence is likely responsible for the age--separation relation, the maximum size of the cell of coherent star formation in a galaxy could be determined by its shear. A recent analysis of the nearby flocculent spiral galaxy NGC 300 shows that the cell size closely matches the gas disc scale height, suggesting that in this galaxy the cell size is set by stellar feedback breaking out of the host galaxy disc, rather than shear  \citep{Kruijssen2019d}. It remains an important open question which physical mechanisms set the length scale for the independent building blocks of galaxies as a function of the galactic environment.

\section{The cluster mass-radius relation; insights into the dynamical state of young star clusters}
\label{yscsize}

The radius of a star cluster is usually expressed in the effective radius ($\reff$), defined as  the radius containing either half the cluster light (for unresolved clusters) or half the number of observed stars (for resolved clusters). The mass-radius relation of cluster populations at various evolutionary stages provides insight in cluster formation and evolution. From early {\it HST} observations of young massive clusters in NGC~3256, \citet{1999AJ....118..752Z} reported a surprisingly shallow size-luminosity relation: $\reff\propto L^{0.07}$, i.e. a nearly constant radius. \citet{2004A&A...416..537L} found a similar shallow slope between $\reff$ and cluster mass for young clusters ($\lesssim100~$Myr) in several spiral galaxies, with a typical $\reff\sim3~$pc. A near constant radius was also found for clusters in several other galaxies \citep[e.g.,][]{2007A&A...469..925S,2015MNRAS.452..525R,2017ApJ...841...92R}. The near constant radius implies that massive clusters are denser than low-mass clusters and it is not clear whether this relation is the result of nature or nurture. These findings are surprising, because molecular clouds -- from which clusters form --  have a constant surface density (i.e. a radius increasing as the square-root of the mass). But a word of caution is in place, because for these extra-galactic samples the resolution limit imposes a constant lower limit to the values of  $\reff$  that can be resolved, possibly biasing the mass-radius relation to a constant value. In addition, in most of these samples, clusters with a range of ages are included, making it difficult to separate formation from evolution effects. 

Both the resolution and age effect can be addressed by looking at the youngest Galactic clusters. For Galactic embedded clusters with 10-100 stars, \citet{2006ApJ...641..504A} find a steep mass-radius relation of the form $\reff\propto N_*^{1/2}$, where $N_*$ is the number of stars, i.e. a constant surface density. Because these clusters  still have gas associated  with them, this is likely as close we can get to observing the initial mass-radius relation of star clusters. The selection procedure of these clusters likely puts a lower limit on the observable surface density, possibly biasing the results to this steep relation.  The slightly older Galactic open clusters in the catalogue of \citet{2013A&A...558A..53K}, with masses derived from the tidal radii by \citet{2007A&A...468..151P} show a near constant volume density (i.e. $\reff\propto M^{1/3}$). It is important to realise that the masses and radii are simultaneously determined from \citet{1962AJ.....67..471K} profile fits, possibly introducing a correlation. 
The radii found by Piskunov et al. are also  substantially smaller ($\lesssim1~$pc) then what was recently found with  {\it Gaia} (average radii of $2-3~$pc, \citealt{2018A&A...618A..93C}). The average radii of M31 clusters are also somewhat smaller \citep[$1-2$~pc,][]{2012ApJ...752...95J} than the other extra-galactic samples. Combined with their  slightly lower masses ($10^{2-4}~\msun$), this may point at a slight mass-radius correlation. However, because of the increasing spatial resolution limit with galaxy distance, it is difficult to make definite statements about this relation.

By splitting in age, \citet{portegies10} showed that a sample with with clusters younger than 10~Myr contains clusters with $M\lesssim10^5~\msun$ and $\reff\lesssim1~$pc, which are not found in the older sample $10-100~$Myr. This may point at an expansion, something that was also noticed from the expansion with age of  the core radii of extra-galactic clusters \citep{2008MNRAS.389..223B} and the radii of Galactic clusters and OB associations \citep{2009A&A...498L..37P}. This expansion could be the result of residual gas expulsion \citep{2017A&A...597A..28B} or internal two-body relaxation (\citealt{2010MNRAS.408L..16G}). In Section 6 of \citet{Krause_20}, another review in this series, we discuss in details the dynamics of stars within a cluster (we refer the interested reader to that review for more information). It is here important to point out that two-body relaxation leads to a faster expansion of low-mass clusters, potentially erasing a  mass-radius correlation, or even inverting it to an anti-correlation. 

\section{The fraction of stars forming in candidate bound young star clusters}
\label{yscgamma}

\subsection{Observational constraints}
After more than a decade, it has yet not been possible to reach a final agreement on whether or not the fraction of stars that form in bound stellar clusters will depend on the intensity of the star formation event and on the general physical properties of the galactic environment where clusters form. 

Already from the very beginning, thanks to the HST high-spatial resolution optical/UV imaging, it was observed that  SSCs preferentially formed in galaxies experiencing starburst events, like merger systems \citep[e.g.,][]{meurer1995, whitmore95} or in dwarf galaxies \citep{Billett2002}. 
However, it was soon recognised that galaxies with higher SFR would likely host more massive (luminous) star clusters, simply because a larger number of clusters are formed and, therefore, the likelihood of sampling the cluster mass function at the high-mass end increases \citep{whitmore2000, larsen2002}. These relations simply describe a ``size--of--sample effect''. On the other hand, an increase in the fraction of stars forming in bound clusters implies a change in the clustering nature of star formation and in the efficiency at which bound stellar structures can be formed. We will quantify this process defining the cluster formation  efficiency (hereafter CFE or $\Gamma$) as the fraction of total stellar mass formed in clusters per unit time over a given age interval (cluster formation rate, CFR in units of \msunyr) divided by the SFR of the galaxy or region of the galaxy where clusters have been detected \citep[e.g.,][]{bastian2008}. 

\begin{figure*}
  \includegraphics[width=0.5\textwidth]{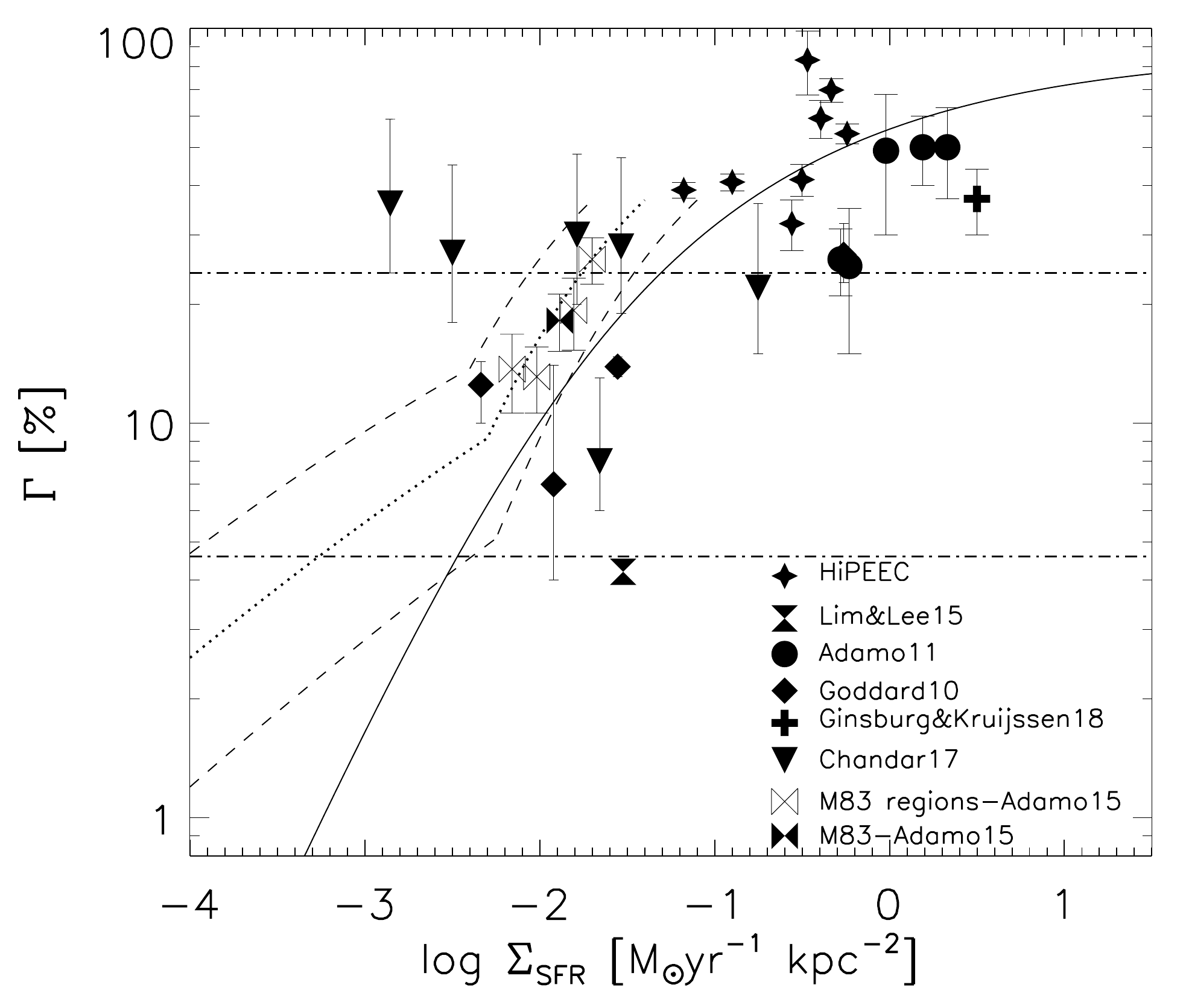}
  \includegraphics[width=0.5\textwidth]{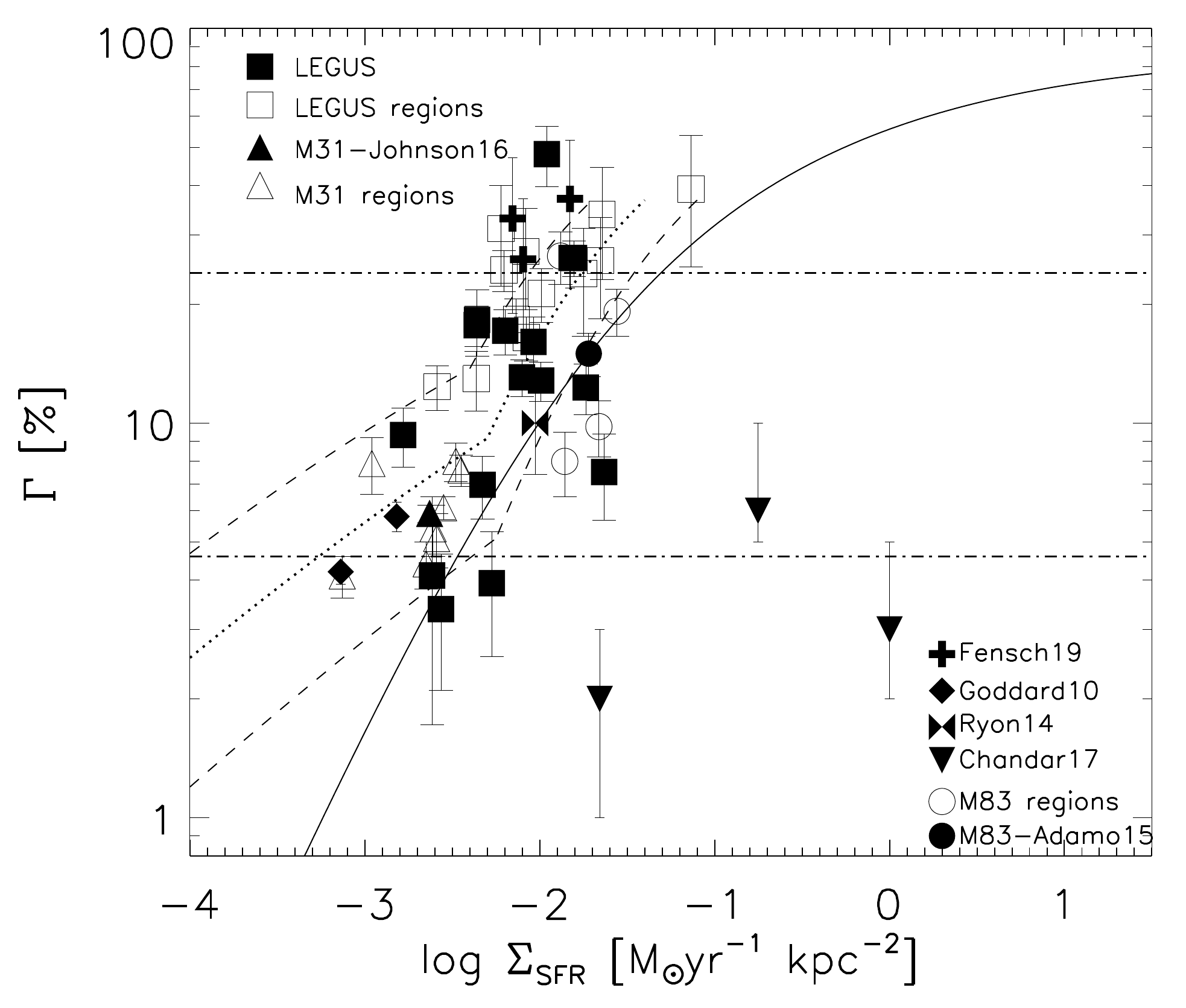}

\caption{{\it Left}: observed cluster formation efficiency,$\Gamma$, vs. star formation rate density, $\Sigma_{\rm SFR}$, relation obtained by combining data available in the literature. This plot contains only $\Gamma$ values determined with YSCs between 1--10 Myr. The data used are published by \citet{goddard10, adamo2011, adamo2015, lim2015, chandar17, GK2018}, HiPEEC, Adamo et al (2020a, to be subm.). The solid line reproduces the fiducial model by \citet{kruijssen2012} using the Kennicutt-Schmidt law, the dotted and the dashed lines reproduce the same model but applying the \citet{bigiel08} conversion from $\Sigma_{\rm gas}$ to $\Sigma_{\rm SFR}$. The suggested constant $\Gamma$ (24 and 4.6 \% estimated for the age range 1--10 and 10--100 Myr) values by \citet{chandar17} are included as horizontal dot--dashed lines. {\it Right}: Same as the left plot but for different age ranges, including clusters with ages $>$10 Myr (see text for details). We only include $\Gamma$ available in the literature that are estimated over a longer age interval. For details on the used cluster catalogues see  \citet{goddard10, ryon2014, adamo2015, chandar17, fensch19, johnson2016} for the M31 data, and the LEGUS sample by Adamo et al (2020b, to be subm.). In both plots, empty symbols are used for plotting $\Gamma$ values detected on sub-regions of galaxies. See main text for discussion.}
\label{fig:gamma}       
\end{figure*}

The pioneering work by \cite{goddard10} suggested that the CFE would steadily increase in galaxies with higher SFR per unit area. Since then, several observational works in the literature have extended this positive correlation both at high and low $\Sigma_{\rm SFR}$ and galactic and sub-galactic scales \citep[e.g.,][among many others, see references in Figure~\ref{fig:gamma}]{adamo2011, cook2012, ryon2014, adamo2015, johnson2016, GK2018}. \cite{kruijssen2012} derived an analytical model that reproduces the positive correlation between the two physical quantities ($\Gamma$ and $\Sigma_{\rm SFR}$). In this theoretical framework, bound star clusters form in the high-density peaks of the hierarchically organised ISM, where the free-fall time is shorter and the star formation efficiency higher. Additionally, it includes a prescription for how tidal perturbations caused by the encounters with dense GMCs set a minimum limit for the formation of bound star clusters. Overall, the model predicts the $\Gamma$ vs. $\Sigma_{\rm SFR}$ relation given three observable galactic properties, the gas surface density $\Sigma_{\rm gas}$, the Toomre parameter $Q$, and the angular velocity $\Omega$, by converting $\Sigma_{\rm gas}$ into $\Sigma_{\rm SFR}$ with a star formation relation (e.g.\ the Schmidt-Kennicutt relation or the \citealt{bigiel08} formulation, see e.g.\ \citealt{kennicuttevans2012}).

However, it is important to note that not all the data reported in the literature support the $\Gamma$ vs. $\Sigma_{\rm SFR}$ relation \citep[e.g.,][]{chandar17, fensch19}. \citet{chandar17} raised one important point regarding the observed $\Gamma$ vs. $\Sigma_{\rm SFR}$ relation. The data at high $\Sigma_{\rm SFR}$ have $\Gamma$ preferentially estimated using short time scales (e.g., 1--10 Myr), while data at low $\Sigma_{\rm SFR}$ are estimated over a longer time range (e.g., up to 100 Myr). In their work they report to find a constant CFE at formation (over an age range of 1--10 Myr) close to 24\%. The CFE constantly and rapidly declines to few percents in the age range 10--100 and 100-400 Myr, because of rapid cluster disruption, equally affecting the overall cluster populations of their sample. Therefore, they conclude that the observed $\Gamma$ vs. $\Sigma_{\rm SFR}$ relation is driven simply by mixing data in the literature that have CFR derived over different time ranges. 

We take now this discussion a step further. These contrasting observational results may be understood in light of limitations and assumptions that go into the estimate of the CFE and $\Sigma_{\rm SFR}$. The estimates of $\Gamma$ relies on:
\begin{enumerate}
    \item 
a significant fraction of cluster candidates used to estimate the CFR being gravitationally bound;
    \item
reliable cluster age and mass determinations and detection limits; 
    \item
a SFR tracer that is sensitive to the same age interval as the cluster population.
\end{enumerate}
Hence, the challenge to estimate $\Gamma$ resides in the difficulty to create homogeneous reliable cluster catalogues combined with a star formation tracer that is sensitive to variations on time scales of tens of million years. In a recent review on star clusters, \cite{krumholz_araa19} carefully discuss the data available in the literature in light of the different assumptions made to estimate the CFE. As pointed out in their review, different works take different steps in constructing their cluster catalogues. For instance, the \cite{chandar17} results are based on catalogues considered ``inclusive'', following the terminology of \cite{krumholz_araa19}. This means that catalogues are constructed automatically. Some steps are taken to remove potential stellar systems and interlopers, but no human visual inspection takes place. The latter task is undertaken by several studies in the literature as a necessary step to clean the cluster catalogues by fake cluster candidates, i.e. systems that may appear to have a light spread function larger than a stellar one, but simply because of interposition chances along the line of sight. This is a clear problem for detection of cluster candidates, as it is very likely to have these spurious detections in regions with higher and clustered stellar densities, i.e. typical star-forming regions in local galaxies that are also the place where bound clusters are formed \citep{adamo2017}. Visual inspection has then become an important step in mitigating the contamination of unbound or spurious systems \citep[see for example][]{bastian2012, ryon2014, johnson2015, adamo2015, adamo2017, cook2019}.  These cluster catalogues are referred to as ``exclusive'' in the terminology of \cite{krumholz_araa19}. Analyses on the boundness criterion \citep[$\Pi$,][and discussed in Section~1.1]{Gieles_11}, confirm that the majority of candidate clusters contained in the exclusive catalogues satisfy the $\Pi>1$ boundness condition \citep[e.g.,][]{bastian2012, ryon2014, ryon2017, johnson2015}. Independent approaches, like the one used by \citet{Sabbi_18}, confirm indeed that the fraction of massive stars forming in clustered star-forming regions are constant with SFR and that the time scales for dissolution of these regions is of $\sim$10~Myr. This implies that gravitationally unbound associations will heavily contaminate the fraction of bound star formation measured in `inclusive' catalogues at early age ranges and more so in galaxies with lower $\Sigma_{\rm SFR}$,  where the overall CFE is only a few percent and star formation is therefore dominated by unbound systems. At high $\Sigma_{\rm SFR}$, the bound clusters instead dominate the SFR, such that the inclusion of unbound systems only has a relatively small effect on $\Gamma$.

In recent years, to overcome the ``human intervention and subjectivity'' on the cluster catalogues and improve upon the reproducibility of these catalogues some attempts have been done to introduce supervised training of neural network algorithms capable of classify sources into cluster candidates or contaminants \citep[e.g.,][]{messa2018a, Grasha2019,wei19}. These first attempts report $\sim$70\% agreement between machine learning and human morphological classifications, similar to the agreement reported among several human classifiers according to \citet{Grasha2019}. In the near future, improvements on the cluster catalogues by better training sets and recognition algorithms will certainly open the way to huge advancements in our understanding of cluster formation and evolution.

Finally, before presenting the observed $\Gamma$ vs. $\Sigma_{\rm SFR}$ relation, by using all data available in the literature to date, it is also important to note that at distances beyond 20~Mpc, clusters become point-like sources even at HST resolution.  It requires different assumptions and approaches to produce cluster candidate catalogues \citep[e.g.,][]{adamo2010, goddard10, linden2017}. We assume that the light spread function of the compact object is dominated by the stellar cluster within the region. However, clusters are never formed in isolation but in a star-forming region with elevated stellar clustering. The reader can think of the 30 Doradus region (described in Section~\ref{sec:cluster_stats}). 30 Doradus hosts a very massive cluster R136 of about $10^5$~\msun only a few Myr old \citep{zinnecker2007} with its light dominated by very massive stars \citep{Crowther2019}. Other, less massive clusters have been found within distances of tens of pc. For example, Hodge 301 is significantly older, containing red super giant stars. At a distance of 80 Mpc the entire region will fit within a few HST pixels. In UV and optical bands the light of the region will be dominated by the O stars within R136. At the NIR, the red super-giant stars in Hodge 301 will dominate the integrated flux of the region. With increasing galactic distance the approximation of having a single cluster within the compact source becomes weaker and weaker \citep{randria2013} and eventually even the approximation of single stellar population fails. At distance beyond 80 Mpc star-forming regions with the size of 30 Doradus become unresolved, we enter the domain of the so-called stellar clumps \citep[e.g., ][]{messa2019} studied up to redshift $z\sim6$ \citep[e.g.,][]{shibuya2016}.

In Fig.~\ref{fig:gamma}, we collect data on the CFE available in the literature. We divide them into two groups accordingly to the age range used to estimate $\Gamma$. Galaxies for which $\Gamma$ is estimated within 1--10 Myr range are showed in the left side plot. On the right side, we plot all the $\Gamma$ estimates obtained using a longer age interval. The interval 10-100 Myr is used in values extracted in the LMC and SMC by \citet{goddard10}; NGC4214, Antennae, NGC 3256 by \citet{chandar17}; M31 by \citet{johnson2016}. \citet{fensch19} uses an age interval of 1--30 Myr. \citet{adamo2015} selects clusters in the range 10--50 Myr. The $\Gamma$ values in the LEGUS sample have been derived within an age interval of 1--100 Myr (Adamo et al 2020b, to be subm.). In both plots we use filled symbols to indicate that values have been estimated over a large fraction (or the whole) galaxy, while empty symbols show $\Gamma$ values estimated in sub-regions of a given galaxy. We do not separate the sample into inclusive vs. exclusive. However, we note that most of the data showed in the left plot for $\Gamma(1-10{\rm Myr})$ are obtained with inclusive cluster catalogues except for the M83 data points \citep{adamo2015} and the Sagittarius B2 complex in our own galaxy \citep{GK2018}. Except for Sagittarius B2, data with $\log(\Sigma_{\rm SFR}[$\dsfr$]) \gtrsim -1$ have all been estimated in galaxies with distances $\gtrsim 20$ and lower than 80 Mpc. On the right side plot, most of datapoints have been derived with exclusive cluster catalogues except for the estimates by \cite{fensch19} (plus symbols) and \cite{chandar17} (lowerside triangles). Both plots include the fiducial model by \cite{kruijssen2012} (solid and dotted lines with dashed line intervals) derived assuming averaged $\Sigma_{\rm gas}$, $Q$, and $\Omega$ values, typical of local galaxies. The same model results in different curves if the Kennicutt-Schmidt law (derived for galactic scales) or the Bigiel formulation (derived using kpc-regions within galaxies) is used to convert  $\Sigma_{\rm gas}$ into  $\Sigma_{\rm SFR}$ \citep{kruijssen2012, johnson2016}. The horizontal dotted-dashed lines show the constant $\Gamma(1-10\,{\rm Myr})$ and $\Gamma(10-100\,{\rm Myr})$ proposed by \citet{chandar17}. 

Overall, we observe that in spite of the large scatter (in part introduced by the different approaches to sample definition), both age intervals are statistically consistent with a positive correlation between $\Gamma$ and $\Sigma_{\rm SFR}$. We do not see a drastic decline in $\Gamma$ in the longer age interval (right plot) confirming the results reported in the literature that $\Gamma$ values estimated at different age range are similar within uncertanties \citep{adamo2015, johnson2016, messa2018b}. This latter result also suggests that in galaxies with $\log(\Sigma_{\rm SFR}[$\dsfr$]) \gtrsim -1$ cluster disruption is not significantly affecting $\Gamma$ estimates within the age range 1--100 Myr. We also notice that $\Gamma$ estimated in sub-regions of galaxies (empty symbols) tend to occupy the region of the Kruijssen model (dashed and dotted line) obtained by using the Bigiel formulation of the Kennicutt--Schidmit relation, derived for kiloparsec--size regions within galaxies.

\subsection{Comparison to numerical simulations}

So far, we have discussed the evidence in support of and limitations in the data that can affect the observed $\Gamma$ vs. $\Sigma_{\rm SFR}$ relation. Another powerful tool to test whether such a relation arises from the physical properties of the star formation process is to use numerical approaches. In recent years, the increasing computational power combined with improved numerical recipes that account for sub-galactic scale physics such as stellar feedback and self-consistently evolving multi-phase ISM, has made possible to follow cluster formation and evolution in combination with galaxy evolution \citep[e.g.,][]{kruijssen11, kruijssen2019a, renaud2015, Li2017, choksi2018, pfeffer2018, lahen2019, li2019}. However, computational resources are not unlimited and the initial generation of works modelling YSC populations focus on isolated or merging galaxies in idealized, non-cosmological simulations \citep[e.g.,][]{kruijssen11, renaud2015, lahen2019}. These setups were designed for developing methodologies and numerical methods, but they lack the cosmological context that determines and drives galaxy assembly history. The dependence of star cluster formation on galactic-scale properties means that modelling the formation of realistic star cluster populations also requires modelling the formation and evolution of galaxies and their environment. We will provide a more detailed description of different approaches to simulate cluster populations in a cosmological context in Sections~\ref{sec:redshift_formation_GC} and~\ref{sec:simulations} and we refer the interested reader to \citet{Forbes2018} for a careful review of the field. 

\begin{figure*}
  \includegraphics[width=0.50\textwidth]{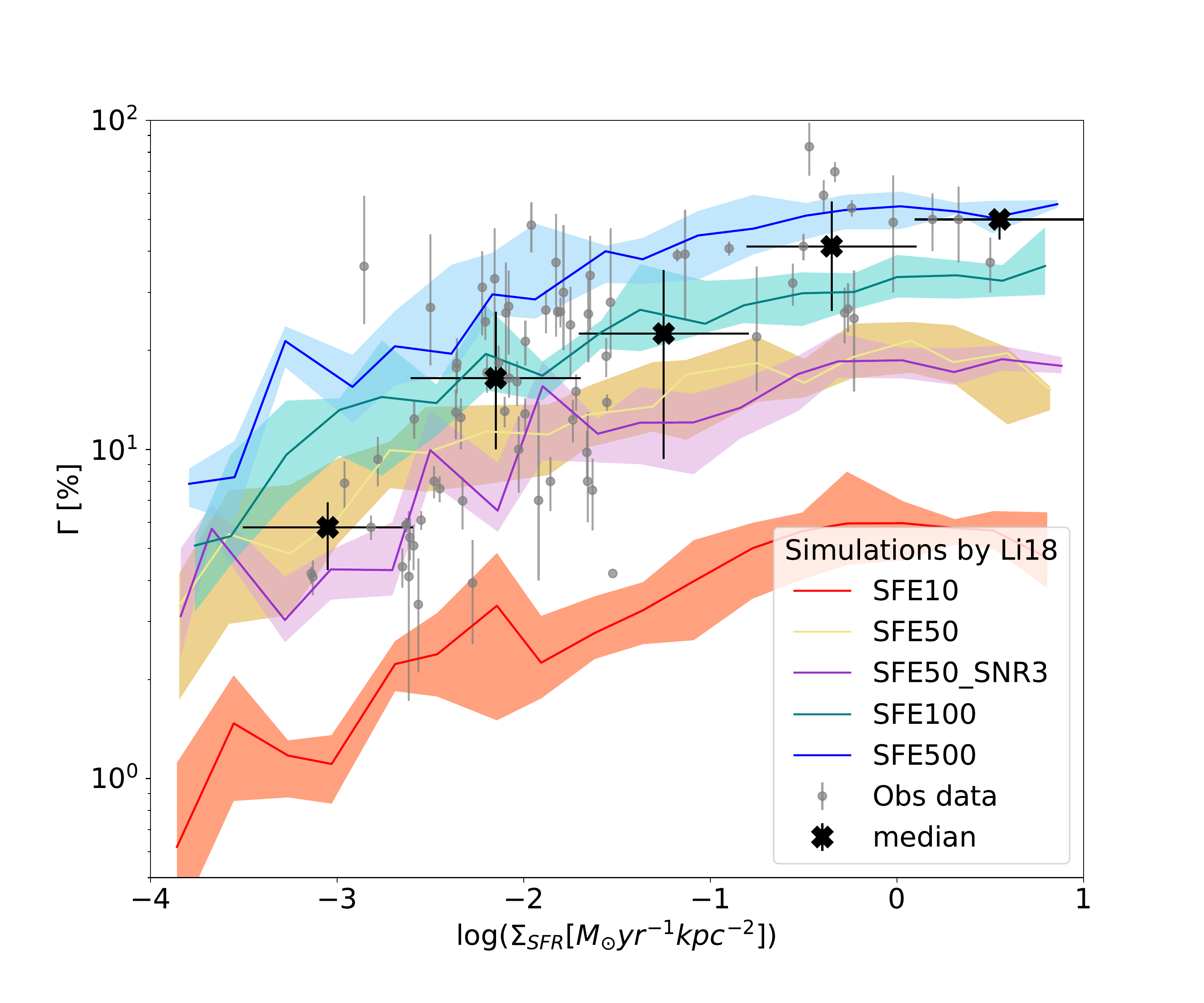}
  \includegraphics[width=0.49\textwidth]{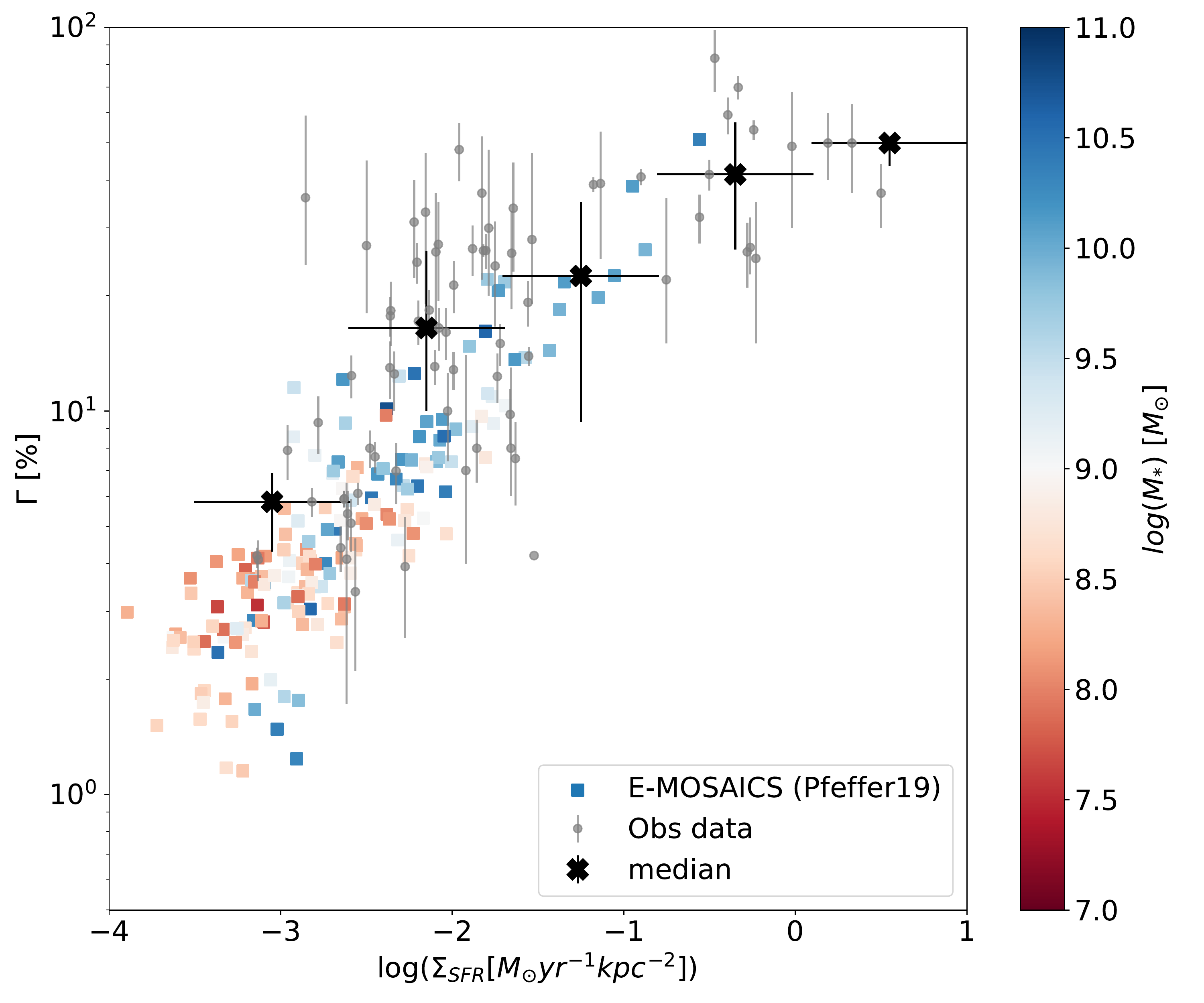}

\caption{$\Gamma$ vs. $\Sigma_{\rm SFR}$ plane. We combine the observed data available in the literature (filled grey symbols), i.e, the data are not separated by age ranges used to estimate $\Gamma$ (see Section~\ref{yscgamma} and Figure~\ref{fig:gamma} for description of the observed data). We also include median and 25 and 75 \% quartiles of the observed data (filled black symbols) estimated in $\Sigma_{\rm SFR}$ bins, of size indicated by the horizontal bars. {\it Left}: We include the resulting $\Gamma$ vs. $\Sigma_{\rm SFR}$ extracted from the simulation sets by \citet{Li2018}. The curves show the median and quartiles of different sets of simulations using different feedback prescriptions. {\it Right}: We include the extracted $\Gamma$ and $\Sigma_{\rm SFR}$ for the E-MOSAICS simulations (filled square symbols) by \citet{pfeffer2019b}. The values have been extracted from galaxies at redshift 0 and are plotted color-coded by the galaxy stellar mass in the same snapshot.}
\label{fig:gammasimul}       
\end{figure*}

In this section, we compare observational data compiled from the literature to two different sets of cosmological simulations, that use radically different approaches and numerical recipes.  In Fig.~\ref{fig:gammasimul} we show the $\Gamma$ vs. $\Sigma_{\rm SFR}$ space.  We combine all the observational data (black dots) included in the two plots of Fig.~\ref{fig:gamma}. \citet{chandar17} reports several $\Gamma$ estimates using different age ranges for each galaxy in their sample. In this figure, we report $\Gamma$ estimated at the youngest age interval for each galaxy to avoid contamination by disruption. We report in Table~\ref{tab:2} the median and quartiles of the observed $\Gamma$ values estimated in  $\Sigma_{\rm SFR}$ intervals. These values are also included in Figure~\ref{fig:gammasimul} to facilitate the comparison with simulations. It is important to notice that in order to estimate the median trends of the observational results, we have combined heterogeneous datasets that does not uniformly sample the $\Sigma_{\rm SFR}$ space and, therefore, these trends should not be over-interpreted.

\begin{table}
\centering
\setlength{\tabcolsep}{3pt}
\caption{Estimated $\Gamma$ median and quartiles as a function of $\Sigma_{\rm SFR}$ of all the values reported in the literature. These values are showed in Figure~\ref{fig:gammasimul}.}
\label{tab:2}       
\begin{tabular}{cccccc}
\hline
 & &&&&\\
$\Sigma_{\rm SFR}$ & [$-3.5$, $-2.6$] & [$-2.6$, $-1.7$] & [$-1.7$, $-0.8$] &[$-0.8$, $0.1$] & [$0.1$, $1.0$] \\

[M$_\odot$yr$^{-1}$kpc$^{-2}$] & &&&&\\
 & &&&&\\
\hline
 & &&&&\\
median $\Gamma$[\%] &  5.8    & 16.5  & 22.5  & 41.4 & 50.0 \\
 & &&&&\\
25, 75\% $\Gamma$[\%] &  4.3, 6.9 &  10.0, 26.2 & 9.3, 35.1 & 26.3, 54.6 &   43.5,  50.0\\
 & &&&&\\
\hline
\end{tabular}
\end{table}

Overplotted on the left side of Fig.~\ref{fig:gammasimul} are the resulting $\Gamma$ vs $\Sigma_{\rm SFR}$ obtained by the galaxy simulations of \citet{Li2018}. On the right side, we overplot the $\Gamma$ and $\Sigma_{\rm SFR}$ reported by \citet{pfeffer2019b} obtained in the E-MOSAICS simulations \citep[MOdelling Star cluster population Assembly In Cosmological Simulations within EAGLE][]{pfeffer2018, kruijssen2019a}.  Despite the fact that both sets of simulations use vastly different numerical approaches, the resulting fraction of stars forming in clusters increases with $\Sigma_{\rm SFR}$ in both works. 

If we look into the details of each simulation we can try to understand how they can inform us on the origin and physical meaning of the $\Gamma$ vs. $\Sigma_{\rm SFR}$ relation. \citet{Li2018} simulate an isolated overdense region within a box of 4 comoving Mpc across, using a Eulerian gas dynamics and N-body adaptive refinement tree (ART) code. They incorporate many state-of-the-art physical processes, such as non-equilibrium chemical networks and radiative transfer at very high spatial resolution (about 5 pc at the redshift range of the run). Due to the computational cost, the simulation runs included in the left plot of Fig.~\ref{fig:gammasimul} stop at redshift $z\sim1.5$. Each run has different assumptions for the star formation efficiency within each grid cell \citep[see][for more details]{Li2018}. Clusters are not analytically implemented in the run but ``form'' in the cells with the highest gas densities. Only a fraction of the formed stars belong to a cluster, and this fraction is determined locally by considering the star formation efficiency and gas condition within each cell. The $\Gamma$ and $\Sigma_{\rm SFR}$ are determined at each snapshot within the region of the simulation that is gravitationally part of the central galaxy. \citet{Li2018} note that different prescriptions for the SFE have no significant impact on the SFH of the central galaxy, which also follow the Schmidt-Kennicutt law, but it significantly affects the normalisation of the positive relation between $\Gamma$ and $\Sigma_{\rm SFR}$. This means that models with higher star formation efficiency have higher fraction of stars belonging to bound clusters. Therefore, as already suggested by the observational data, the change in the gas conditions (reflected by the observational quantity $\Sigma_{\rm SFR}$) will change the clustering efficiency of the star formation.  A direct comparison between the literature data and the simulations by \citet{Li2018} shows that simulations with low SFE (SFE10) cannot explain the currently observed $\Gamma$ and $\Sigma_{\rm SFR}$. Yet, from observations it is not clear how clustered star formation proceeds at the very low $\Sigma_{\rm SFR}$ ranges ($<-3.5$ in log scale). On the other hand, the remaining sets of simulations reproduce quite well the space occupied by the observations. They suggest that the observed scatter in the data might be the result of varying SFE (as predicted by \citealt{kruijssen2012}).

The simulations (square symbols color-coded accordingly to their host galaxy stellar mass at $z=0$) in the right plot of Fig.~\ref{fig:gammasimul} are taken from the E-MOSAICS simulations and were presented in \citet{pfeffer2019b}. The E-MOSAICS project \citep{pfeffer2018, kruijssen2019a} couples the cosmological, hydrodynamical simulations EAGLE \citep[][]{crain2015, schaye2015} to cluster formation and evolution via analytical implementations \citep{kruijssen11,kruijssen2012, reinacampos2017}. They utilize a sub-grid model where a fraction of the stellar particles formed by the simulation are converted into a cluster population. The analytical model that predicts CFE by \citet{kruijssen2012} is used to set the fraction of the stellar mass formed in clusters. The fiducial runs (showed in Fig.~\ref{fig:gammasimul}) use the analytical model of \citet{reinacampos2017} to predict the maximum mass that a cluster population can have and sample the cluster mass distributions according to a \citet{schechter76} initial cluster mass function. Once formed, clusters evolve together with their host systems. The simulations also account for cluster mass loss and disruption  because of interactions with the tidal field of their host galaxy, and internal processes such as stellar evolution and tidal evaporation.

The subgrid cluster formation models used in the E-MOSAICS simulations have an environmental dependence on the local gas conditions (density, pressure, dynamical state) within each galaxy. In particular, the data included in the plot here are published in \citet{pfeffer2019b} and are obtained by a suite of zoom-in re-simulations of 25 galaxies selected at redshift $z=0$ to have Milky Way-mass halos. The CFE is determined locally in each simulation based on the gas conditions and used to determine what fraction of stellar mass ``forms'' in clusters. The values reported in the right plot of Fig.\ref{fig:gammasimul} are estimated in galaxies at $z=0$, using clusters with ages younger than 300 Myr. In general, we observe that the majority of the E-MOSAICS data cluster around a CFE between 1 and 10 \%, quite close to the values reported for spiral galaxies like the Milky Way and the Andromeda galaxy. A fraction of the galaxies at $z=0$ have also more elevated CFE, tracing the median trend reproduced from the observed data. Due to the choice of simulated galaxies, the simulations do not significantly cover the higher $\Sigma_{\rm SFR}$ values. Nonetheless, the resulting $\Gamma$ vs. $\Sigma_{\rm SFR}$ relation from the E-MOSAICS simulations has a similar normalisation and slope as recovered in the observed data, thereby confirming the dependence of the CFE on the physical parameters used in the model, i.e.\ $\Sigma_{\rm gas}$, $Q$, and $\Omega$. The inclusion of a third parameter, i.e., the galaxy stellar mass, $M_*$, can help us to make a few more considerations. Galaxies with low stellar masses ($<10^9$ \msun\, e.g.,dwarf systems with sub-solar metallicties) rarely reach CFE of 10\%. This implies that cluster formation in these systems is highly stochastic, as indeed reported by \citet{cook2012, cook2019}. On the other hand, CFE can change from a few to very high rates in galaxies with larger stellar masses, indicating that the positions of galaxies in the $\Gamma$ vs. $\Sigma_{\rm SFR}$ relation will change as a function of the evolutionary phase that the galaxy experiences. This was indeed recently outlined also in the simulations analysed by \citet{lahen2019}. 

\section{The mass distributions of young star clusters in local galaxies}
\label{yscmassfunction}

The mass distribution is a fundamental observable of a star cluster population because it links the formation of bound stellar systems to the star formation process. As already discussed in Section~\ref{gmcpop}, GMC populations have mass distributions described by a power-law slope close to $\alpha=-2$ \citep[][]{kennicuttevans2012}. This slope is similar to the slope recovered for the luminosity function of HII regions and star-forming clumps in the local Universe \citep[e.g.,][]{thilker2002, bradley2006, elmegreen2006} and mass distributions of stellar clumps at redshift $\sim2$ \citep{DZA2018}.

From the very beginning, YSC luminosity and mass functions appeared to be consistent with a power law function of slope $-2$ \citep[see review by][]{adamobastian18}. The power-law shape that characterises the distributions of masses and luminosities, from largest coherent star-forming regions to the densest and compact structures such as star clusters, down to clumps of proto--stars inside embedded clusters, is very likely the result of fragmentation produced by the balance between gravitational collapse and turbulence compression \citep{Elmegreen2011}.

Typically, the shape of the YSC (and OC in our Milky Way) mass function is thought to closely follow the initial intrinsic mass function at formation. Evolved systems, like GCs, have mass distributions that are better described by a log-normal function, peaked at a luminosity (mass) values that remains almost unchanged across the local Universe \citep{BS2006}. Whether the GC mass function is the result of cluster mass loss and dissolution, remains still under debate. 

\subsection{Observations of the cluster mass function from local galaxies}

\begin{figure*}
  \includegraphics[width=0.5\textwidth]{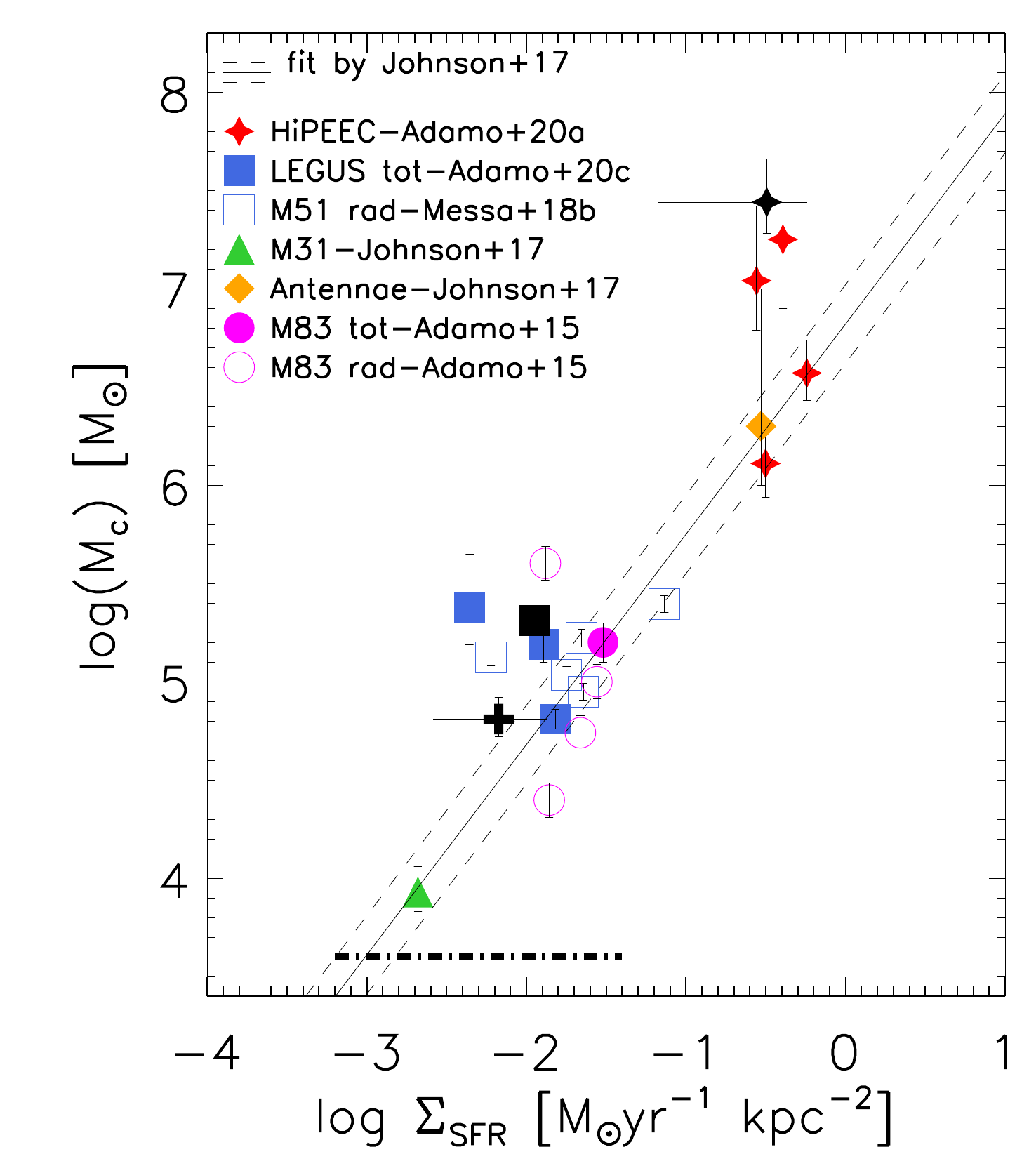}
  \includegraphics[width=0.5\textwidth]{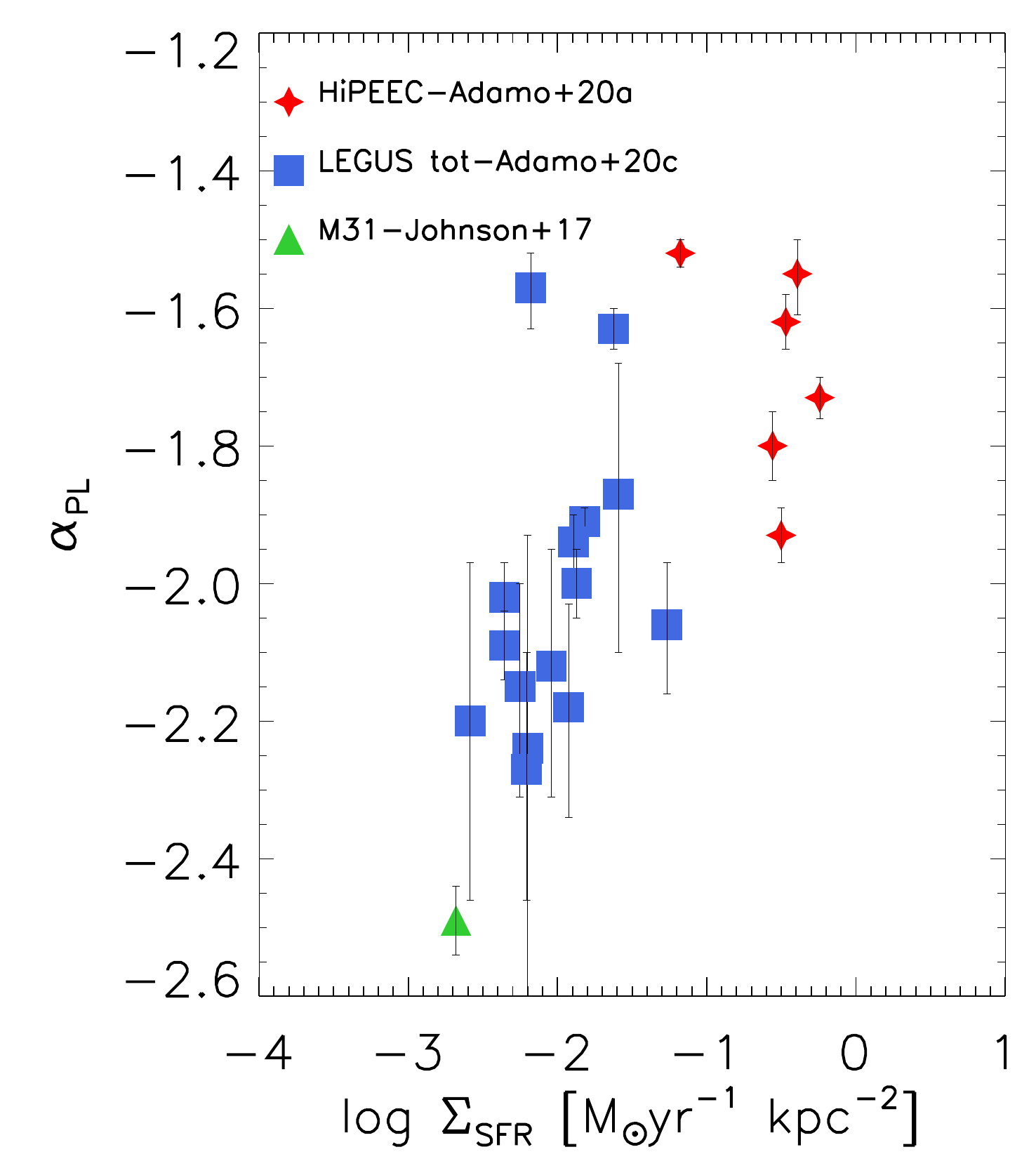}

\caption{ {\it Left}: Truncation mass, \mc, vs. the star formation rate density plane. To the initial data points used by \cite{johnson17}, we have added  literature data \citep{messa2018b} and newly determined \mc\, in single galaxies from the LEGUS (blue filled squares, Adamo et al, 2020c to be subm.) and HiPEEC (red filled diamonds, Adamo et al 2020a to be subm.) sample. Only statistically significant determinations of \mc\, are reported. Black symbols are the recovered \mc\, reported in Table~\ref{tab:1} and obtained by combining the cluster catalogues of 8 LEGUS spiral galaxies with SFR$<0.5$ \msunyr\,({\it lowSFR} spirals, black cross), 6 spiral galaxies with SFR$>0.5$ \msunyr\,({\it highSFR} spirals, black square), 6 merger systems belonging to the HiPEEC sample (SFR$>$ 10 \msunyr, black diamond). The horizontal bar associated with each black symbol shows the $\Sigma_{\rm SFR}$ interval that each sample spans (see main text). The range of $\Sigma_{\rm SFR}$ of the dwarf sample is plotted as a dot-dashed horizontal line. {\it Right}: Power-law slope of the cluster mass function vs. star formation rate surface density. Values plotted in this plot are reported by \citet{johnson17}, and Adamo et al (2020a, c, to be subm.).}
\label{fig:MFobs}       
\end{figure*}

It is widely accepted that the shape of the cluster mass function can be describe to first order by a power law. However what has emerged in the last decade is that a pure power-law distribution is not be sufficient to explain the dearth of very massive clusters both in our own and in local galaxies \citep{larsen2006, larsen2009}. A power law with a truncation at the high-mass end of the cluster mass distribution, in the shape of an exponential cut-off above a characteristic mass, \mc (typically referred to as a Schechter function) might be a more realistic representation of the true cluster mass function. First evidence of a possible truncation at the high-mass end was suggested by \citet{gieles2006} and \citet{gieles2009}. It was also suggested that the \mc\, would change as a function of galactic environment, i.e. higher \mc\, would be found in galaxies experiencing elevated star formation events \citep[like the Antennae, see e.g.,][]{larsen2009,portegies10}. 

In the literature, however, discrepant conclusions are reached by different authors. From the cluster mass distribution analyses in two grand-design spiral galaxies, M51 and M83, \citet{chandar2014, chandar2016} report that there is no clear evidence of the presence of a truncation in the cluster mass function and that a pure power-law shape with slope consistent with $-2$ is statistically the preferred solution. Similar conclusions are reached by \cite{mok2019} on a small sample of local galaxies. On the other hand, \cite{bastian2012, adamo2015, messa2018a}, report to have determined a \mc$\sim 10^5$ \msun for both M83 and M51 cluster populations. Variations in the recovered \mc\, have been found in M83 \citep{adamo2015}, where \mc\, decreases at increasing galactocentric distance. No significant variations have been found across M51 cluster disk population \citep{messa2018b}. \cite{johnson17} reported the lowest yet determined \mc$=8.5\times10^3$\msun in the cluster population of M31.

The exact description of the cluster mass function has paramount implications not only for our understanding of cluster formation and evolution, but also for stellar feedback. Numerical simulations that aim at understanding the the link between the multi-phase ISM and the stellar feedback have shown that the clustering of massive stars and SNe, like in massive star clusters, is a key player in the star-formation cycle of galaxies \citep{krause2013, gentry2017, kim2017}. Because star clusters host a large fraction of the massive stars forming in the host galaxy, especially massive clusters are fundamental units to maintain a multi-phase ISM and regulate the star formation process. The difference between a pure power-law mass function and a Schechter-type function provide very different predictions on the number of massive clusters that will form in a galaxy \citep[e.g., see discussion in][]{johnson17, adamo2017}. 

To understand why contrasting conclusions are reached on the intrinsic shape of the cluster mass function, it is important to make some considerations both on the size of the sample typically analysed and the methods used. As mentioned above, cluster formation is a stochastic process, which implies that galaxies with higher SFR will be able to form more numerous clusters and therefore better sample the mass function at the high-mass end. The implications are twofold.
\begin{enumerate}
    \item 
Very often the number of detected clusters in the field of view barely reach 100 objects \citep[e.g., see][or Adamo et al 2020c to be subm]{cook2019} either because the galaxies intrinsically have small cluster populations or because the HST imaging coverage is limited to a portion of the disk of the system. Small numbers implies a poor sampling of the mass function and thus a degeneracy when fitting for two parameters (\mc\, and $\alpha$) instead of one \citep[$\alpha$, see appendix in][]{messa2018b}.
    \item
Until very recently, the fit is done on binned distributions. However, as pointed out in \cite{adamo2017}, in case of small number statistics a binned distribution always bias the fit against the upper mass-end of the distribution. Equal size bins will always have higher error bars at the high-mass end because of the small number of objects they contained and therefore weight less significantly on the resulting slope. Equal-number of object bins have also a biased impact because they tend to become very large at the high-mass end, washing out the presence of a truncation.
\end{enumerate}
To overcome these limitations, different methods have recently been implemented. For example, by fitting cumulative mass distributions \citep{adamo2015}, using maximum-likelihood methods applied to cumulative distributions \citep{adamo2017, messa2018a}, and a maximum-likelihood analysis combined with a Markov Chain Monte Carlo (MCMC) technique to sample the posterior probability distributions of the Schechter and power-law mass function parameters \citep[e.g., see][for more details]{johnson17, messa2018b}. In the latter case, the analysis does not depend on the binning technique used or the functional distribution applied, therefore, overcomes several limitations discussed above. However, as pointed out in \cite{johnson17} and \cite{messa2018b} the convergence of the fit will still be affected by small number statistics, i.e. if the cluster sample is small, the \mc\, parameter remains unconstrained and single power-law fit is preferred.

In Fig.~\ref{fig:MFobs}, we compile a sample of statistically significant recovered \mc\, and $\alpha$ by fitting cluster catalogues available in the literature with the method introduced by \cite{johnson17}. For the sake of homogeneity and consistency, we include in these plots only data that have been fitted with the same method. The \mc\, best values are sampled out of the marginalized posterior probability distribution function (PDF) for each of the Schechter function parameters,  accompanied by a 1$\sigma$ confidence interval defined by the 16th to 84th percentile range. For $\alpha_{\rm PL}$ we report the median and 1$\sigma$ confidence interval for the single parameter, $\alpha$ \citep[see respective source papers][Adamo et al 2020a,c; for details on the cluster mass function analysis]{johnson17, messa2018b}. In the left plot of Fig.~\ref{fig:MFobs}, we plot \mc\, vs. $\Sigma_{\rm SFR}$. The Antennae (orange diamond) and the M83 data (magenta circles) are taken from the literature and included in the plot for completeness because they were used by \cite{johnson17} to derive an analytical formulation of the positive correlation between \mc\, and $\Sigma_{\rm SFR}$. As reported in Adamo et al (2020a), of the 6 HiPEEC galaxies analysed only 4 have a statistically meaningful constraint on \mc. Those values are included in the plot along with the \mc\, value (black diamond) obtained by fitting the combined cluster catalogues of the 6 merger systems (we report in Tab.~\ref{tab:1} the SFR range of the galaxies, the age range of the combined clusters, the resulting number after a conservative mass limit cut at  M$=5\times10^4$ \msun has been applied to mitigate the effect of incompleteness due to detection limits). Within the LEGUS survey we have analysed in total 31 galaxies. We report here the results obtained by using only systems classified as class 1 and 2 (compact cluster candidates), with masses M$>5000$ \msun and ages younger than 200 Myr. First, the marginalized posterior probability distribution analysis of the Schechter function parameters of the cluster population of each of the 17 dwarf systems published by \cite{cook2019} does not produce a statistically meaningful constraint on the \mc. It is also important to note that the fit to the combined cluster catalogues of the 17 LEGUS dwarf galaxies does not provide a tight constraint on \mc, as reported in Tab.~\ref{tab:1} and in \citet{cook2019}. In total 14 LEGUS spiral galaxies have complete cluster catalogues. Of those only 3 galaxies (M51, NGC 628, NGC 1566) have converging marginalised PDF for the \mc. These values are reported in the plot as blue diamonds. The spiral sample is then divided in two groups based on their SFR (determined within the HST field of view where the cluster population has been analysed). The combination of cluster samples has the advantage of increasing the number of clusters for a given SFR range so that it mitigates the size-of-sample effects mentioned above. Indeed, the convergence of the marginalized posteriors of the Schechter function parameters in both samples confirms that the number statistics is the main problem in determining these parameters in local galaxies as also pointed out by \citet{larsen2009}, \citet{adamo2017}, and \citet{Elmegreen2018}.

\begin{table}
\centering

\setlength{\tabcolsep}{3pt}
\caption{Maximum-likelihood fit outputs of the cluster mass function in diverse galaxy environments (see Section~\ref{yscmassfunction}). Clusters populations of similar galaxy types (see column 1) have been combined to increase the statistical significance of the fit. The cluster catalogues of dwarf and spirals systems have been obtained from the LEGUS survey. The merger/starburst system catalogues from the HiPEEC survey. The columns list, in order, the galaxy SFR range, the cluster age range used, the total number of clusters included in the fit, the determined \mc\, and $\alpha_{Sch}$, the slope of the power-law fit, $\alpha_{\rm PL}$. The last column summarises whether the two parameters (\mc\, and $\alpha_{Sch}$) describing the Schechter function are uniquely determined (therefore, statistically significant) or cluster mass distribution is better described by a power-law function (Sch or PL, respectively). $a$. Data published by \cite{johnson17}; $b$. \cite{cook2019}; $c$. data published in Adamo et al (2020c to be subm.); $d$ catalogues published in Adamo et al (2020a subm.).}
\label{tab:1}       
\begin{tabular}{lccccccc}
\hline\noalign{\smallskip}
\multicolumn{1}{c}{Galaxy} & \multicolumn{1}{c}{SFR range} & \multicolumn{1}{c}{age range} & \multicolumn{1}{c}{N$_{cl}$} & \multicolumn{1}{c}{$\log({\rm M}_{\rm c})$} & \multicolumn{1}{c}{$\alpha_{\rm Sch}$}  & \multicolumn{1}{c}{$\alpha_{\rm PL}$} & \multicolumn{1}{c}{sig} \\
\multicolumn{1}{c}{sample} & \multicolumn{1}{c}{($M_\odot/$yr)} & \multicolumn{1}{c}{(Myr)} & & \multicolumn{1}{c}{($M_\odot$)} & & & \\
\noalign{\smallskip}\hline\noalign{\smallskip}
M31$^a$ & 0.28 & 10--300 & 840 & 3.93$^{+0.13}_{-0.10}$ & $-1.99^{+0.12}_{-0.12}$ & $-2.49^{+0.05}_{-0.05}$ & Sch \\

17 dwarfs$^b$ & 0.003--0.758 & 1--200 & 375 & 5.94$^{+1.25}_{-0.38}$  & $-1.92^{+0.09}_{-0.08}$ & $-2.00^{+0.05}_{-0.05}$ & PL \\

8 lowSFR spirals$^c$ & 0.088--0.342& 1--200 & 497 & 4.81$^{+0.11}_{-0.09}$&     $-1.4^{+0.11}_{-0.11}$ & $-1.97^{+0.04}_{-0.04}$ & Sch\\

6 highSFR spirals$^c$ & 0.59--2.76 & 1--200 & 2478 &  5.31$^{+0.07}_{-0.06}$ & $-1.7^{+0.05}_{-0.05}$& $-2.07^{+0.02}_{-0.02}$ & Sch \\ 

 6 mergers$^d$ & 5.8--47.4 & 1--100 & 1237 & 7.44$^{+0.22}_{-0.16}$ & $-1.7^{+0.03}_{-0.03}$  & $-1.77^{+0.02}_{-0.02}$ & Sch\\
\noalign{\smallskip}\hline
\end{tabular}
\end{table}

The \mc\ retrieved for both sub-samples of spirals is included in the left plot of Fig.~\ref{fig:MFobs} as black square (high SFR spiral sample) and cross (low SFR spiral sample). The addition of new datapoints confirm the overall positive correlation between \mc\, and the $\Sigma_{\rm SFR}$ of the host galaxy as proposed by \cite{johnson17}. If we consider the $\Sigma_{\rm SFR}$ as a tracer of gas surface density (and pressure), the relation suggests that galaxies experiencing elevated star formation episodes, high gas density (pressure) have higher probability to form massive star clusters. From Tab.~\ref{tab:1}, we can see that dwarf galaxies have modest SFR values, which explains the small number of clusters that are available for the fit. However, dwarf galaxies have also very compact  star-forming regions, resulting in $\Sigma_{\rm SFR}$ overlapping with populous spiral galaxy systems. Extrapolating the \mc\, vs. $\Sigma_{\rm SFR}$ relations for the starburst dwarfs, like NGC 4449, NGC 5253, NGC 4656, with the highest $\Sigma_{\rm SFR}$, we expect them to have \mc$\sim$ a few times 10$^5$ $M_{\odot}$. However, their combined population counts $\sim 200$ clusters (as opposed to the almost 2500 clusters for the high SFR spirals), i.e. too small to provide a good constraint on the \mc. So cluster formation in dwarf galaxies is either fundamentally different than in spiral and merger systems, or simply more affected by stochastic effects and low number statistics which prevent to make definitive conclusions. On the right side of Fig.~\ref{fig:MFobs} we collect the recovered $\alpha_{\rm PL}$ obtained by marginalising the posterior distributions assuming a power-law function of 17 LEGUS galaxies and 6 HiPEEC systems. We remind the reader that this is in many galaxies not the best description of their cluster mass distributions. We assume a power-law function because for a Schechter function the \mc\, and $\alpha$ are degenerate parameters, therefore, we want to prevent mixing slopes derived by assuming different functions. A similar plot was presented by \citet{krumholz_araa19} using data published in the literature. There, they noticed that the slope of the cluster mass function appears to fattens in environments with increasing $\Sigma_{\rm SFR}$. The advantage, here, is that we adopt the same method across a large galaxy spectrum and control the lower mass limit of each sample against completeness issues. As one can see from the values listed in Tab.~\ref{tab:1}, and showed by numerical exercises \citep{adamo2015, johnson17, messa2018b}, $\alpha_{\rm PL}$ is systematically steeper than $\alpha_{Sch}$ obtained for the same sample. Therefore, the trend observed in the right plot of Fig.~\ref{fig:MFobs}, would not disappear even if we could ideally determine both \mc\, and $\alpha$ for the entire sample. Our homogeneous sample confirms the trend outlined by \citet{krumholz_araa19}. The change in the star formation conditions (higher $\Sigma_{\rm SFR}$ corresponds to increasingly elevated $\Sigma_{\rm gas}$ and therefore pressures) are reflected in shallower slopes and a larger number of massive clusters. 

 It is interesting to compare the properties of the cluster mass function to those of the GMC mass spectrum reported in Section~\ref{gmcpop}. There is not a one--to--one correspondence between a GMC and the formation of a star cluster. The fragmentation process combined with the star formation efficiency within each collapsing core could result in one bound cluster, several, or none.  Irrespectively, some fraction of these dense gas regions will be able to form star clusters. The mass distributions and other properties of cluster populations (like the presence of a truncation mass) correlate in some way with the variations observed in the GMC mass function, as a function of galaxy scale dynamics, gas content, and $\Sigma_{\rm SFR}$. For completeness, we report that different conclusions have been recently reached by \citet{mok2020}, who suggest that GMC and YSC populations are unrelated to the global properties of the galaxies where they form. Clearly, the field has not yet converged on a single interpretation and definitive answers may require larger samples subjected to a homogeneous analysis.

Finally, the presence of a truncation \mc\, has also been observed in GC populations of galaxies belonging to the Virgo Galaxy cluster \citep[][]{jordan2007}. The GC \mc\, appears to positively scale with galaxy stellar mass. It is not yet clear whether the GC \mc\,is a property of the initial cluster mass function of the progenitor clusters or the results of evolution. Observationally both \citet[][]{johnson17} and Adamo et al. (2020c, to be subm.) report no significant evolution within the uncertanties on the recovered \mc\, as a function of age bins. Numerical works agree that significant fractions of massive clusters that could be considered GC progenitors are formed during the  merger events that determine the assembly history of the hosts \citep[e.g.,][]{forbes10, reinacampos2019}. However, numerical simulations also agree that in spite of their large masses these systems have very slim chances to survive if they remain trapped in the dense gas environments where they form \citep[e.g.,][]{kruijssen11, kruijssen12c, renaud2015}. Based on this evidence \citet{lamers2017} argue that the recovered \mc\, of GC populations may be the result of nurture (i.e. cluster disruption).

\subsection{The origin of the \mc; from observations to theory}

To understand what regulates the formation of the most massive clusters and the origin of the \mc\, different theoretical models have been put forward.  

In the model initially proposed by \citet{kruijssen14} the maximum cloud mass and the stellar \mc\,  might have a common origin and it may correspond to the Toomre mass. The latter depends on the maximum size of a region that will overcome the shear of the disk and the kinetic pressure of the gas and start collapsing \citep{toomre1964}. In this model for a given gas surface density, this length-scale directly provides the Toomre mass. This shear-dependent model was subsequently refined by \citet{reinacampos2017}. In addition to the shear--limited maximum mass model they take into account that feedback from young stars might disrupt the cloud before the global collapse of the shear-limited area is completed. If the feedback time (i.e.\ the time it takes for the stellar feedback to destroy the cloud) is smaller than the free--fall time of the shear--limited region the resulting collapsed mass is smaller than the shear-limited Toomre mass. The model is able to reproduce the observed trend of maximum GMC mass and maximum cluster mass as a function of increasing galactocentric distance in M31 (maximum GMC and cluster mass peak at the star-forming ring), M83 \citep[declining maximum GMC and \mc\, as a function of distance from the centre][]{adamo2015, freeman2017}, and M51 \citep[only small variations as a function of distance from the centre][]{messa2018b}. The \citet{reinacampos2017} model has analytically been implemented in the E-MOSAICS simulations \citep{pfeffer2018}. In \citet{pfeffer2019b} the authors recover similar \mc\, and  $\alpha$ vs. $\Sigma_{\rm SFR}$ relations as showed here for observations in Fig.~\ref{fig:MFobs}. They suggest that not only gas surface density and pressure (via the dependence from $\Sigma_{\rm SFR}$) play a role in determining the maximum GMC mass and maximum cluster mass, but also the angular velocity of the host galaxy from which the Toorme size depends. 

Recently, \citet{trujillogomez19} have extended the \citet{reinacampos2017} model to investigate the effect of feedback on the formation of low mass clusters, hence the intrinsic shape of the cluster mass function at the low mass end. The model evaluates which parts of the star-forming region remain bound given the time--scales for gravitational collapse, star formation, and stellar feedback that also determine the upper mass distribution. In this model, galaxies like the dwarf Fornax in the Local Group, with high specific frequency of GCs, might have never formed the lower mass counterpart expected for a power-law cluster mass distribution.

Based on observational evidence, \citet{Elmegreen2018} derives an analytical model that is based on the minimum SFR and $\Sigma_{\rm SFR}$ necessary to form a GC progenitor, i.e.\ a star cluster of $\sim10^6$ $M_{\odot}$. A SFR of 1 \msunyr\, is necessary in a region to sample the cluster mass function to the high--mass end. A minimum $\Sigma_{\rm SFR}$ of  1 \dsfr\, will ensure that the gas density and therefore pressure is high enough to form bound stellar systems with stellar densities similar to GCs. It is therefore the gas pressure in the ambient medium that determine \mc. If the SFR is elevated but spread over a large area (low $\Sigma_{\rm SFR}$) then the pressure-limit mass, that correspond to the \mc, is never reached and the most massive cluster is simply determined by size--of--sample effects. If both conditions are satisfied, galaxies will be able to sample their cluster mass function up to the \mc, which should then be traced in the cluster mass function analysis. The predicted trend of the \citet{Elmegreen2018} models agree with the observed distribution of observed datapoints in the \mc\, vs. $\Sigma_{\rm SFR}$ plane of Fig.~\ref{fig:MFobs} reinforcing the idea that pressure could be a driving parameter for the change of \mc.

Finally it is interesting to note that, at difference of the E-MOSAICS simulations where the cluster mass function is analytically implemented, the cosmological simulations of \citet{Li2018} find steeper cluster mass functions than the canonical $-2$ slope, which they interpret as a possible presence of a truncation, though quantitatively it remains unconstrained.

\section{The survival rates of young star clusters forming in local galaxies}
\label{yscsurvival}

Observationally, cluster disruption has traditionally been traced using the demographics of the cluster population, such as their combined age and mass distributions or cluster frequencies per age dex \citep[e.g.,][]{lamers05,lamers05b,whitmore07,bastian2012,fouesneau14,chandar2014, chandar17, messa2018b, cook2019}. As summarised in recent reviews \citep[e.g.,][]{portegies10, adamobastian18, krumholz_araa19}, there is no observational consensus on the way clusters disrupt. Observationally, different teams reach different conclusions by counting numbers of clusters in age and mass bins. The debate has focused on whether clusters simply disrupt because gas expulsion has destabilised their potential  \citep[e.g.,][]{Fall2005, Chandra2010} and therefore disrupt independently of their mass and environment, or mainly because they lose mass via interaction with their galactic environment \citep[e.g.,][among many others]{baumgardt03,lamers05, lamers05b, gieles2006,gieles07}, resulting in a gradual mass loss, otherwise known as `mass dependent cluster disruption'. While observationally there is no consensus yet, \citet{adamobastian18} noted, based on a compilation of results available in the literature, that the cluster disruption rate may change as a function of environment, becoming progressively higher in galaxies with higher gas surface densities (and thus higher external pressures), as suggested by \citet{elmegreen10} and \citet{kruijssen11}. Indeed, \citet[][]{messa2018b} analysing the number of clusters per age bin and as a function of position in the disc of M51, find a positive correlation between the increasing number rate and the gas surface density of the region considered \citep[also see][]{miholics17}. We are therefore moving away from the initial dichotomy (mass-dependent as apposed to mass-independent disruption) into an environmental dependence of the disruption rate of clusters, that goes from considerably longer dissolution timescales in low-pressure galaxies to very fast disruption rates in highly star-forming galaxies like the Antennae.

From a theoretical perspective, after their formation clusters lose mass due to stellar evolution, two-body relaxation-driven evaporation in the host galaxy's tidal field, and tidal perturbations (`shocks') resulting from GMC passages, spiral arm passages, pericentre passages, and disc crossings \citep[e.g.,]{spitzer58,gnedin97,baumgardt03,gieles2006,gieles07,kruijssen11,Krause_20}. Recent analyses of {\it Gaia}'s second data release reveal tidal tails emerging from clusters in the Galactic disc \citep[e.g.,][]{fuernkranz19,meingast19,roeser19}. After carrying out a completeness correction (which is non-trivial), the detailed phase space information of these tails can potentially be used to directly measure dynamical mass loss rates. A rich variety of theoretical work has investigated dynamical cluster mass loss. Next to predictions for cluster mass loss rates from analytical theory \citep[e.g.,][]{gnedin97,lamers05b,Kruijssen_15,gieles16}, a wide variety of numerical approaches has been used, ranging from direct $N$-body simulations \citep[e.g.,][]{baumgardt03,gieles08,renaud2013,rieder13,miholics16,webb19} to hydrodynamical simulations that link cluster mass loss to the formation and evolution of the host galaxy \citep[e.g.,][]{kruijssen11,kruijssen2019a,miholics17,pfeffer2018,reinacampos18,reinacampos19b,li2019}.

Modeling cluster disruption is critical for reconstructing the properties of cluster populations at the time of their formation from the observed cluster demographics. This holds for young cluster populations, which can experience significant attraction due to tidal shock-driven disruption on timescales as short as 50--100~Myr \citep{gieles2006,miholics17,pfeffer2019b}, but even more so for old GC populations, which have undergone $\sim10$~Gyr of evolution before attaining their present-day properties. The physical mechanisms driving cluster disruption are discussed in more detail in \citet{Krause_20}, another review in this series. As stated above, these are a combination of stellar evolution, tidal evaporation, and tidal shocks. Numerical models need to include all of these processes, including their environmental dependence. This means that a complete model for cluster formation and destruction requires self-consistent simulations of galaxy formation and evolution, including descriptions for star formation, feedback, chemical enrichment, cluster formation, cluster disruption, and stellar evolution. Recent years have seen major steps in this direction \citep[e.g.,][]{pfeffer2018,kruijssen2019a,li2019}, and we will discuss these efforts in more detail in Sections~\ref{sec:redshift_formation_GC} and~\ref{sec:simulations}.


\section{Current and future observations and simulations of massive cluster formation across cosmic time}
\label{sec:redshift_formation_GC}

Direct detections of proto--GC candidates are reported in the literature by \citet{Vanzella_etal17, Bouwens_etal17arXiv, JohnsonT2017, Vanzella_etal19}, and \citet{vanzella2020}. All these detections rely on the aid of gravitational lensing, which can magnify several times the fluxes of these very young systems in their UV and blue optical rest-frames. Indeed because of their compact sizes these massive and compact star-forming regions have larger chances to be magnified and detected and might become the only signposts of their otherwise faint host galaxies \citep{zick2018}. The intrinsic sizes of these systems are very often uncertain because they rely on lensing models and are limited to upper limits of a few tens of parsec, i.e. the sizes of entire star-forming regions. Spectroscopic studies of a handful of these lensed proto-GCs at redshift $\sim 2$--3, show that their FUV light is dominated by spectral signatures of very massive stars (with lifetimes shorter than 4--5 Myr), WR HeII broad wind emission, and direct evidence of Lyman continuum escape radiation \citep{Rivera2019, vanzella2020}. These initial results put proto-GCs as potential contributors to the reionisation era, or towards the end of it ($\lesssim 7$). Indirect evidence of proto-GCs formation at high redshift is produced by the physical properties of stellar clumps. Stellar clumps dominate the UV rest-frame of their host systems \cite[e.g.,][]{elmegreen2013, shibuya2016, messa2019} and have very high SRF surface densities, making them a natural sight for very massive cluster formation \citep{Elmegreen2018}. Indeed, assuming that proto--GCs form in stellar clumps following a Schechter mass function distribution and accounting for cluster disruption,  it produces a first order calculation of the surving GC pupulations at redshift $z=0$ that are comparable to the number and mass functions of the GCs detected around galaxies like the Milky Way and M31 \citep[e.g.,][]{shapiro10, adamo2013}.

At least in part motivated by the impending launch of JWST (and facilitated by technical improvements in hardware and modelling), a wide variety of recent studies has made predictions of the properties of young GC populations at high redshift. These range from back-of-the-envelope estimates, assuming simple scaling relations \citep[e.g.,][]{katz13,Renzini2017,boylankolchin18}, to expectations based on observations of nearby GC populations \citep[e.g.,][]{zick2018} and high-redshift star formation \citep[e.g.,][]{Vanzella_etal17}, and predictions from numerical simulations of galaxy formation, either by `painting on' GCs using an ad-hoc model \citep[e.g.,][]{renaud17,halbesma19,madau19,phipps19} or by including a physical description for GC formation and evolution \citep[e.g.,][]{Li2018,pfeffer2019a,reinacampos2019}. The resulting predictions for when GCs formed and what their corresponding detectability is with future observations with JWST or 30-m class telescopes vary greatly between these different approaches. The reason is two-fold.
\begin{enumerate}
    \item
    Different models assume different formation scenarios for GCs. Recently, numerous works suggest that GCs are the natural byproduct of `normal' star formation under the high-pressure conditions in gas-rich high-redshift galaxies \citep[e.g.][]{shapiro10,Kruijssen_15,Kruijssen2019e,Li2017,pfeffer2018,kruijssen2019a,elbadry19,ma19,keller20}. However, additional formation scenarios that have been considered are mergers of galaxies or dark matter substructure \citep[e.g.][]{kim18,li2019,madau19}, or special conditions during reionisation \citep[e.g.][]{katz14,trenti15,kimm16,ricotti16,creasey19}. All of these scenarios predict increased GC formation towards high redshift, because of an increase in gas pressure, merger rate, or specific early-Universe conditions, but the specific redshift range in which GCs are predicted to have formed still varies, from slightly preceding (but largely tracing) the cosmic star formation history \citep[e.g.][]{reinacampos2019} to extending deeply into the epoch of reionisation \citep[e.g.][]{katz14}. In the latter case, it is possible that GCs may have played an important role in reionisation \citep[e.g.][]{ricotti02,katz13,katz14,he2020}. 
    \item
    Even within families of models adopting a similar formation scenario for GCs, there exist considerable differences in terms of GC formation redshifts and luminosity functions. For instance, numerical simulations of galaxy formation that describe GCs as the natural outcome of regular star formation roughly fall into two categories.\footnote{For readers interested in the details of the various galaxy formation simulations predicting GC demographics, we refer to \citet{Forbes2018}.} The first category of models takes a `normal' galaxy formation simulation and uses some set of conditions (e.g.\ cuts in age, metallicity, or halo mass) to associate GCs to star particles \citep[e.g.,][]{renaud17,halbesma19,madau19,phipps19}. This `particle tagging' technique has had difficulties to reproduce the observed demographics of GCs, such as their total number per unit galaxy mass and their age or metallicity distribution. If an age cut is made, this obviously directly defines the redshift range where GCs are expected to be observed. The second category of models employs a sub-grid model for stellar cluster formation and disruption within galaxy formation simulations \citep[e.g.,][]{Li2017,pfeffer2018,kruijssen2019a}. In general, these models predict a lower formation redshift of GCs \citep[e.g.,][]{reinacampos2019}, as well as lower initial numbers and masses of proto-GCs than in other models \citep[e.g.,][]{pfeffer2019a}.
\end{enumerate}

Unfortunately, age measurements of old GCs in the local Universe have uncertainties of $\sim1$~Gyr, which limits their diagnostic power to distinguish between these ideas. However, the observation that GC formation extends well down to $z\sim1$ \citep{marinfranch09,forbes10,leaman13} and still takes place in the present-day Universe \citep[e.g.,][]{ashman92,elmegreen97,kruijssen14} suggests that regular star formation under high-pressure conditions can indeed lead to GC formation. While it is currently under discussion as to whether or not the requirement of a high gas pressure still makes GCs analogues to young massive clusters in the local Universe \citep{renaud19}, a consensus is emerging that there is no evidence for two `modes' of cluster formation, but GCs rather seem to form at the extreme end of a continuum. Occam's Razor thus suggests that the models describing GCs as the outcome of regular star formation provide the most accurate description of GC formation, without a major contribution by merger-induced cluster formation \citep{keller20}. While some of the more exotic formation mechanisms may have contributed a handful of GCs, it is unlikely that they dominated the formation of the GC population in galaxies \citep{keller20}. Fortunately, the great variety of predictions for the initial demographics of the GC population, and in particular their UV luminosities (compare e.g.\ \citealt{Renzini2017} and \citealt{pfeffer2019a}) and the redshift range of their formation (see the above discussion), means that future observations of young GCs in the early Universe with JWST, the ELT, the TMT, and the GMT will be able to distinguish between the currently considered formation scenarios.


\section{Cluster populations as tracers of galaxy assembly}
\label{sec:simulations}

The discussion in this review so far has emphasised the prominent environmental dependence of both cluster formation and their dynamical disruption. This environmental dependence is critical in the context of galaxy formation. Firstly, it implies that cluster formation and evolution cannot be described in isolation, but require the multi-scale coupling to the galaxy formation context. Secondly, it implies that (especially old, i.e.\ globular) cluster populations can be used as tracers of galaxy formation and evolution, provided that a comprehensive model for GC formation and evolution can be constructed.

The ultimate goal of comprehensive models of GC formation and evolution in the context of galaxy formation is to simultaneously reproduce the demographics of the evolved GC populations observed in the local Universe and those of the initial GC populations that will be observed with JWST and 30-m class telescopes (see Section~\ref{sec:redshift_formation_GC}). Thanks to considerable efforts aimed at integrating GC formation and evolution in cosmological simulations, this goal is now starting to come within reach, and at a very opportune moment, given the impending arrival of the next generation of major facilities. However, despite significant progress, current numerical simulations still face a number of shortcomings.

As stated, GC formation and evolution is one of the greatest multi-scale problems in astrophysics, spanning the scales of black hole binaries ($<0.1$~pc) to galaxy formation ($\sim1$~Mpc). This large dynamic range implies a great computational cost when trying to resolve the dynamics of individual stars within GCs, and doing so self-consistently in relation to galaxy formation and evolution for the entire GC population of a galaxy throughout its history. This cost is so prohibitive, that numerical studies need to choose to either model the formation and evolution of individual clusters at high resolution \citep[e.g.,][]{kim18,ma19}, or to employ a sub-grid prescription for the formation and evolution of the entire cluster population \citep[e.g.,][]{pfeffer2018,kruijssen2019a,li2019}. Both approaches are entirely complementary, but they suffer from different problems.
\begin{enumerate}
    \item 
    Simulations that focus on maximally resolving the relevant physics obviously provide the most fundamental descriptions of GC formation, but they cover a limited range in terms of the number of clusters and the redshift interval, prohibiting predictions for the demographics of the GC population at any redshift, such that the $z=0$ GC population is completely out of reach. In addition, these studies often need to focus on a single formation environment, which means that the dependence of GC formation and evolution on the formation and assembly history of the host galaxy is not accounted for. Therefore, the results may not be statistically representative.
    \item
    Simulations that instead employ a sub-grid model for cluster formation and evolution can predict the demographics of the cluster population at all redshifts \citep{Li2017,Li2018,li2019,pfeffer2018}. If a sufficiently large number of galaxies is modelled, they can even do so as a function of the galaxy formation and assembly history \citep{kruijssen2019a}. However, these simulations are fundamentally limited in terms of the (unresolved) physical process that are described by the sub-grid physics. Most critically, GC formation and evolution relies on the interaction with the cold ISM, either because it sets the gas pressure from which clusters form \citep[e.g.,][]{elmegreen97,kruijssen2012,johnson2016,johnson17}, or because it sets the disruption rate of the resulting clusters \citep[e.g.,][]{gieles2006,Kruijssen_15}. To date, simultaneously modelling the physics of the cold ISM and GC formation down to $z=0$ has been too expensive to enable multiple galaxies to be simulated \citep[e.g.,][]{li2019}. Even the most sophisticated suites of cosmological simulations modelling large numbers of galaxies and their GC populations currently lack a cold ISM \citep{pfeffer2018}, which most prominently limits their ability to accurately describe cluster disruption, especially at high metallicities (see \citealt{kruijssen2019a} for a discussion).  The next generation of galaxy formation simulations will overcome this problem (Reina-Campos et al.\ in prep.).
\end{enumerate}

Despite the need for further development, the current generation of galaxy formation simulations describing the formation and evolution of the GC population that include a large number of galaxies (developed as part of the E-MOSAICS project, \citealt{pfeffer2018,kruijssen2019a}) has already revealed a number of quantitative connections between the present-day properties of the GC population and the assembly history of the host galaxy. In many ways, this is the most explicit realisation to date of the promise that GCs can be used as tracers of galaxy formation and assembly. Many of these predicted correlations rely on the diagnostic power of the age-metallicity distribution of GCs \citep[e.g.,][also see \citealt{choksi2018} for an earlier discussion on the importance of the GC age-metallicity distribution in this context]{kruijssen2019a}. Specific examples of strong correlations are:
\begin{enumerate}
    \item 
    The total number of mergers experienced by the host galaxy is traced by the slope of the GC age-metallicity distribution and the total number of GCs.
    \item
    The fraction of these mergers taking place before $z=2$ can be estimated by measuring the median age of the GC population in a galaxy.
    \item
    The median age of the GC population also probes the redshift by which certain fractions of the host galaxy mass have been assembled, thus tracing its assembly rate.
\end{enumerate}
These examples are non-exhaustive, and upcoming works will extend the range of correlations beyond the GC age-metallicity distribution, also including GC kinematic and spatial distributions \citep[][Reina-Campos et al.\ in prep.]{trujillo20}. Most importantly, these results are not affected by the current omission of a cold ISM in these simulations, because the global GC demographics trace the assembly history of the host galaxy even if GC disruption is not sufficiently efficient.

\citet{kruijssen2019b} applied the insights from E-MOSAICS to the GC population of the Milky Way, with the goal of constraining our galaxy's assembly history and reconstructing its merger tree. The analysis showed that the Milky Way experienced a total of $\sim15$ mergers throughout its history, with the last major merger having taken place at $z>4$, thus sharpening previous limits of $z>2$ \citep[e.g.,][]{wyse01,hammer07,stewart08}. Improved determinations of the ages of extragalactic GCs may enable the application of these correlations to galaxies beyong the Milky Way in the near future \citep{usher19}. In addition, the analysis presented by \citet{kruijssen2019b} predicted the existence of the satellite Gaia-Enceladus/Sausage (GES), which was discovered only weeks later in the data from {\it Gaia}'s second data release \citep[e.g.,][]{myeong18,helmi18}, as well as the enigmatic galaxy {\it Kraken}, which together with the GES accretion event forms the two most massive galaxies ever accreted by the Milky Way. Recent analyses of the phase space distribution of GCs in the {\it Gaia} data by \citet{massari19} suggest that the relics of {\it Kraken} have been found \citep[][]{kruijssen20, pfeffer20}.

An extremely promising avenue of research is the association of GCs in the Galactic halo with fossil stellar streams from disrupted satellite dwarf galaxies that were accreted by the Milky Way. {\it Gaia} now enables the identification of these systems, and galaxy formation simulations including a model for the GC population provide quantitative predictions for the properties and demographics of such streams and their GCs \citep[e.g.,][]{hughes19}. In the long run, extremely sensitive, wide-field photometry with LSST will aid the efforts to link the GC population to accretion events, not just in the Milky Way, but also in other galaxies.

In summary, the falsification and validation of GC formation models in the context of galaxy formation and evolution are now becoming possible. The combination of revolutionary observational data from {\it Gaia}, JWST, LSST, and 30-m class telescopes will provide a comprehensive picture of GC populations in the local Universe and their progenitors at high redshift, for the first time enabling models to be tested simultaneously at the time of GC formation and after nearly a Hubble time of evolution. In addition, observations spanning a wide redshift range will provide direct tests of the connection between GC demographics and the evolution of the host galaxy. It is to be expected that such observations will cement the use of GCs as quantitative tracers of galaxy formation and assembly.

\section{Open questions, outlook, and future steps}
\label{sec:outlook}

The development of new observational techniques such as large IFUs and the increased low-cost computational capacities have made a large impact on the studies of resolved star-forming regions. Over the past decade is has become feasible to study the multi dimensional phase space using 3D positions and velocities of stars to measure velocity dispersions to an accuracy of a few km\,s$^{-1}$ and refining stellar distances and cluster membership probabilities. Also it allows us to study the star-gas interactions and feedback processes in unprecedented details. Yet, many aspects of the earliest stages of star cluster formation and evolution  still remain hidden due to observational limitations leaving unanswered questions: How is the low-mass and brown dwarf mass function constituted? How do massive stars, binary, and higher-order systems influence planet formation? To what extend does star formation progress into the HII regions around a massive star clusters with a high ionizing potential?; to name a few.

To address these unsolved questions, larger and more sensitive telescopes are necessary. Over the next few years some revolutionizing missions will go online: JWST (spring 2021) a new generation 6.5\,m class space telescope will observe from the Lagrange point 2. This joint mission between NASA, ESA, and CSA will allow us to observe (photometry and spectroscopy) in the near and mid infrared at a spatial resolution rivaling that of HST. The E-ELT is scheduled to have first light in 2025 observing in the optical and NIR. This 39\,m ground-based telescope will be equipped with state-of-the-art adaptive optics systems for high-resolution photometry and spectroscopy. The Wide-Field Infrared Space Telescope (WFIRST) will be a HST-class ultra wide field surveyor designed to settle essential questions in the areas of dark energy, exoplanets, and infrared astrophysics. These new telescopes in combination with the existing ones will allow us to study the lowest-mass star and planet formation processes and track the earliest stages of star formation (Class 0 - III) inside their parental molecular clouds. This will be even pushed further with telescopes planned for the late 2020s and 2030s, such as the TMT and future flagship space missions like the Large UV/Optical/IR Surveyor (LUVOIR), a 9-16\,m class space telescope combining the capabilities of HST and JWST at an unprecedented precision, the Origins Space Telescope (OST), the Habitable Exoplanet Observatory (HabEx), or the Lynx X-ray observatory.

Our current understanding of cluster formation and evolution remains still fragmented even if we limit ourself to the local Universe. Thanks to the multi-wavelength and high-spatial resolution coverage offered by the HST we begin to witness what unique carrier of information cluster populations are. YSCs are potentially the link between small scale events like the formation of stars and the multiphase-ISM necessary to regulate the star formation process throughout the galaxy assembly history. During their formation and early evolution they can be considered fundamental units of stellar feedback, they are the nurse of very massive stars and, thus, carriers of radiative and mechanical energy. While aging they will keep records of the gas conditions where they formed provided we understand how they evolve. GC populations have encoded in them the history of their host galaxies, yet we are not able to fully map their genetic sequence. 

The question is \emph{how to move forward}. While HST reaches its endpoint, we are gaining a large FUV and optical archive that will allow us to reinforce our studies of cluster populations across all the star-forming galactic spectra available in the local Universe. We will loose for more than a decade our FUV window to the local Universe. We will, however, gain, with the JWST and E-ELT, access to the same window at redshift 2 to 10, where we expect to be able to detect proto-GCs.

For the local Universe, we gain a new wavelength regime at spatial resolutions never achieved before on such large field of views. The NIR and MIR spectral wavelength carry nearly dust-free information into the onset of cluster formation and provide direct statistical constraints on the time scales a cluster requires to emerge from its natal cloud. These time scales can be obtained across a vast sample of galaxies (we are no more limited by the \textit{Spitzer} resolution to our Local Group),  with improved cluster ages (by the use of NIR and MIR indicators) and improved cluster masses (by breaking the extinction-age-metallicity degeneracy plaguing cluster analysis). Multi-object spectroscopy and IFU capabilities of JWST and E-ELT will allow to get spectra of a significant fraction of the cluster population of a galaxy on a time scale comparable to what imaging requires nowadays. Such information is still highly distance dependent and remains limited to a handful of bright YSCs in other galaxies. As we prepare for the next decades of new observatories and instruments, we also need to take major steps in 1. improving stellar population models to account for binary fractions and stochastic sampling of the IMF; 2. introduce cognitive algorithms at all phases of the cluster analysis, from detection to constraining of their physical parameters. 

Observations of cluster populations in local and high redshift galaxies are vital to inform theoretical and numerical modeling of galaxy formation and evolution that are challenged to link parsec scale physics to inter-galactic scale processes.

From the theoretical perspective, the past years have seen a major effort in developing numerical simulations of galaxy formation and evolution that simultaneously describe the formation and evolution of GCs at high redshift and of YSCs in the local Universe. The impending observational revolution with the arrival of JWST, LSST, and 30-m class telescopes, combined with the recent observations of molecular gas with ALMA and galactic archaeological relics with {\it Gaia}, will for the first time allow the demographics of the cluster population predicted by these state-of-the-art numerical simulations to be tested {\it as a function of cosmic time}. This comprehensive range of tests is unprecedented and will lead to the most transformational change in our understanding of cluster formation across cosmic history that the field has ever experienced. In order to realise this major step, a number of key developments are necessary, both on the theoretical and observational fronts. 
\begin{enumerate}
    \item 
    Numerical simulations of galaxy formation must include an accurate description of the cold ISM. The descriptions of cluster formation and evolution must account for the expected environmental dependences of these processes, and include a model for cluster disruption by tidal shocks. Suites of such simulations must cover a sufficiently large number of galaxies to probe how these processes and the resulting cluster demographics depend on the formation and assembly history of the host galaxy. Finally, in order to fully develop our understanding of tidal-shock driven cluster disruption, systematic suites of direct $N$-body simulations resolving the collisional stellar dynamics are needed. These should fully map a comprehensive parameter space of shock durations, strengths, and successions.
    \item
    Observational programmes targeted at mapping the properties of proto-GC populations at high redshift with the next generation of telescopes should be allocated sufficient observing time to obtain a comprehensive census of the demographics of the cluster population, measuring the luminosity function, peak masses, and redshift distribution of the young GC population. Simultaneously, deep observations with 30-m class telescopes will be able to detect the brightest GCs at intermediate redshifts, revealing how the correlations between GC population properties (e.g.\ their numbers and the luminosities of the brightest GCs) and the host galaxy properties (e.g.\ mass, radius, metallicity) evolve with cosmic time. Finally, galactic archaeological surveys in the Milky Way and other galaxies must develop ways of statistically associating GCs with the enormous richness of fossil stellar streams that will be discovered, and thereby reconstructing the properties (e.g.\ masses and chemical compositions) of the disrupted galaxies that brought GCs into the halo of the central galaxy.
\end{enumerate}
By connecting the above developments, the field will achieve a step-change in our understanding of cluster formation and evolution throughout cosmic history. In particular:
\begin{enumerate}
    \item 
    We will have a complete picture of the physics driving the cluster demographics as a function of cosmic time and environment.
    \item
    We will be considerably closer to uncovering the intricate connection between stellar cluster populations and their host galaxies.
    \item 
    We will be able to use cluster populations as the fossils revealing the formation and assembly histories of galaxies.
\end{enumerate}
HST brought about a true revolution in studies of stellar cluster populations, both young and old. While a much wider range of telescopes is necessary to make the next major step, and this will require a much closer connection to theoretical models than has been common practice in the past, this holistic approach will hopefully make the impact on our understanding of cluster formation and evolution even greater.


\begin{acknowledgements}
 We thank the anonymous referee for a constructive and careful revision of this work. We thank the staff at ISSI for their generous hospitality, and the editors to invite us to contribute to this topic collection. We are grateful to Hui Li and Joel Pfeffer for sharing their data and Nate Bastian for providing comments on the manuscript. A.A. thanks Matteo Messa, Katie Hollyhead and the LEGUS and HiPEEC teams for support with the analyses showed in this review. A.A. acknowledges the support of the Swedish Research Council, Vetenskapsr{\aa}det, and the Swedish National Space Agency (SNSA). P.Z. acknowledges support by the Forschungsstipendium (grant number ZE 1159/1-1) of the German Research Foundation (DFG). J.M.D.K. and M.C. gratefully acknowledge funding from the German Research Foundation (DFG) in the form of an Emmy Noether Research Group (grant number KR4801/1-1) and a DFG Sachbeihilfe Grant (grant number KR4801/2-1). J.M.D.K. gratefully acknowledges funding from the European Research Council (ERC) under the European Union's Horizon 2020 research and innovation programme via the ERC Starting Grant MUSTANG (grant agreement number 714907), and from Sonderforschungsbereich SFB 881 ``The Milky Way System'' (subproject B2) of the DFG. C.C.acknowledges funding from the Swiss National Foundation (SNF grants 200021\_169125 and 200020\_192039).
\end{acknowledgements}

%
%


\bibliographystyle{spbasic}      
\small
\bibliography{bibliography}   

\begin{thebibliography}{399}
\providecommand{\natexlab}[1]{#1}
\providecommand{\url}[1]{{#1}}
\providecommand{\urlprefix}{URL }
\expandafter\ifx\csname urlstyle\endcsname\relax
  \providecommand{\doi}[1]{DOI~\discretionary{}{}{}#1}\else
  \providecommand{\doi}{DOI~\discretionary{}{}{}\begingroup
  \urlstyle{rm}\Url}\fi
\providecommand{\eprint}[2][]{\url{#2}}

\bibitem[{{Adamo} and {Bastian}(2018)}]{adamobastian18}
{Adamo} A and {Bastian} N (2018) {The Lifecycle of Clusters in Galaxies},
  Astrophysics and Space Science Library, vol 424, p~91

\bibitem[{{Adamo} et~al.(2010){Adamo}, {{\"O}stlin}, {Zackrisson}, {Hayes},
  {Cumming}, and {Micheva}}]{adamo2010}
{Adamo} A, {{\"O}stlin} G, {Zackrisson} E, et~al. (2010) {Super star clusters
  in Haro 11: properties of a very young starburst and evidence for a
  near-infrared flux excess}. \mnras 407(2):870--890

\bibitem[{{Adamo} et~al.(2011){Adamo}, {{\"O}stlin}, and
  {Zackrisson}}]{adamo2011}
{Adamo} A, {{\"O}stlin} G, and {Zackrisson} E (2011) {Probing cluster formation
  under extreme conditions: massive star clusters in blue compact galaxies}.
  \mnras 417(3):1904--1912

\bibitem[{{Adamo} et~al.(2013){Adamo}, {{\"O}stlin}, {Bastian}, {Zackrisson},
  {Livermore}, and {Guaita}}]{adamo2013}
{Adamo} A, {{\"O}stlin} G, {Bastian} N, et~al. (2013) {High-resolution Study of
  the Cluster Complexes in a Lensed Spiral at Redshift 1.5: Constraints on the
  Bulge Formation and Disk Evolution}. \apj 766(2):105

\bibitem[{{Adamo} et~al.(2015){Adamo}, {Kruijssen}, {Bastian}, {Silva-Villa},
  and {Ryon}}]{adamo2015}
{Adamo} A, {Kruijssen} JMD, {Bastian} N, et~al. (2015) {Probing the role of the
  galactic environment in the formation of stellar clusters, using M83 as a
  test bench}. \mnras 452(1):246--260

\bibitem[{{Adamo} et~al.(2017){Adamo}, {Ryon}, {Messa}, {Kim}, {Grasha},
  {Cook}, {Calzetti}, {Lee}, {Whitmore}, {Elmegreen}, {Ubeda}, {Smith},
  {Bright}, {Runnholm}, {Andrews}, {Fumagalli}, {Gouliermis}, {Kahre}, {Nair},
  {Thilker}, {Walterbos}, {Wofford}, {Aloisi}, {Ashworth}, {Brown}, {Chandar},
  {Christian}, {Cignoni}, {Clayton}, {Dale}, {de Mink}, {Dobbs}, {Elmegreen},
  {Evans}, {Gallagher}, {Grebel}, {Herrero}, {Hunter}, {Johnson}, {Kennicutt},
  {Krumholz}, {Lennon}, {Levay}, {Martin}, {Nota}, {{\"O}stlin}, {Pellerin},
  {Prieto}, {Regan}, {Sabbi}, {Sacchi}, {Schaerer}, {Schiminovich}, {Shabani},
  {Tosi}, {Van Dyk}, and {Zackrisson}}]{adamo2017}
{Adamo} A, {Ryon} JE, {Messa} M, et~al. (2017) {Legacy ExtraGalactic UV Survey
  with The Hubble Space Telescope: Stellar Cluster Catalogs and First Insights
  Into Cluster Formation and Evolution in NGC 628}. \apj 841(2):131

\bibitem[{{Adams} et~al.(2006){Adams}, {Proszkow}, {Fatuzzo}, and
  {Myers}}]{2006ApJ...641..504A}
{Adams} FC, {Proszkow} EM, {Fatuzzo} M, et~al. (2006) {Early Evolution of
  Stellar Groups and Clusters: Environmental Effects on Forming Planetary
  Systems}. \apj 641(1):504--525

\bibitem[{{ALMA Partnership} et~al.(2015){ALMA Partnership}, {Brogan},
  {P{\'e}rez}, {Hunter}, {Dent}, {Hales}, {Hills}, {Corder}, {Fomalont},
  {Vlahakis}, {Asaki}, {Barkats}, {Hirota}, {Hodge}, {Impellizzeri}, {Kneissl},
  {Liuzzo}, {Lucas}, {Marcelino}, {Matsushita}, {Nakanishi}, {Phillips},
  {Richards}, {Toledo}, {Aladro}, {Broguiere}, {Cortes}, {Cortes}, {Espada},
  {Galarza}, {Garcia-Appadoo}, {Guzman-Ramirez}, {Humphreys}, {Jung}, {Kameno},
  {Laing}, {Leon}, {Marconi}, {Mignano}, {Nikolic}, {Nyman}, {Radiszcz},
  {Remijan}, {Rod{\'o}n}, {Sawada}, {Takahashi}, {Tilanus}, {Vila Vilaro},
  {Watson}, {Wiklind}, {Akiyama}, {Chapillon}, {de Gregorio-Monsalvo}, {Di
  Francesco}, {Gueth}, {Kawamura}, {Lee}, {Nguyen Luong}, {Mangum}, {Pietu},
  {Sanhueza}, {Saigo}, {Takakuwa}, {Ubach}, {van Kempen}, {Wootten},
  {Castro-Carrizo}, {Francke}, {Gallardo}, {Garcia}, {Gonzalez}, {Hill},
  {Kaminski}, {Kurono}, {Liu}, {Lopez}, {Morales}, {Plarre}, {Schieven},
  {Testi}, {Videla}, {Villard}, {Andreani}, {Hibbard}, and
  {Tatematsu}}]{ALMA_15}
{ALMA Partnership}, {Brogan} CL, {P{\'e}rez} LM, et~al. (2015) {The 2014 ALMA
  Long Baseline Campaign: First Results from High Angular Resolution
  Observations toward the HL Tau Region}. \apjl 808(1):L3

\bibitem[{{Alves} et~al.(2007){Alves}, {Lombardi}, and {Lada}}]{Alves2007}
{Alves} J, {Lombardi} M, and {Lada} CJ (2007) {The mass function of dense
  molecular cores and the origin of the IMF}. \aap 462(1):L17--L21

\bibitem[{{Anderson}(2002)}]{Anderson_02}
{Anderson} J (2002) {Main-Sequence Observations with HST}, Astronomical Society
  of the Pacific Conference Series, vol 265, p~87

\bibitem[{{Anderson} et~al.(2013){Anderson}, {Adams}, and
  {Calvet}}]{Anderson_13}
{Anderson} KR, {Adams} FC, and {Calvet} N (2013) {Viscous Evolution and
  Photoevaporation of Circumstellar Disks Due to External Far Ultraviolet
  Radiation Fields}. \apj 774:9

\bibitem[{{Ashman} and {Zepf}(1992)}]{ashman92}
{Ashman} KM and {Zepf} SE (1992) {The formation of globular clusters in merging
  and interacting galaxies}. \apj 384:50--61

\bibitem[{{Bacon} et~al.(2010){Bacon}, {Accardo}, {Adjali}, {Anwand}, {Bauer},
  {Biswas}, {Blaizot}, {Boudon}, {Brau-Nogue}, {Brinchmann}, {Caillier},
  {Capoani}, {Carollo}, {Contini}, {Couderc}, {Daguis{\'e}}, {Deiries},
  {Delabre}, {Dreizler}, {Dubois}, {Dupieux}, {Dupuy}, {Emsellem}, {Fechner},
  {Fleischmann}, {Fran{\c c}ois}, {Gallou}, {Gharsa}, {Glindemann}, {Gojak},
  {Guiderdoni}, {Hansali}, {Hahn}, {Jarno}, {Kelz}, {Koehler}, {Kosmalski},
  {Laurent}, {Le Floch}, {Lilly}, {Lizon}, {Loupias}, {Manescau}, {Monstein},
  {Nicklas}, {Olaya}, {Pares}, {Pasquini}, {P{\'e}contal-Rousset}, {Pell{\'o}},
  {Petit}, {Popow}, {Reiss}, {Remillieux}, {Renault}, {Roth}, {Rupprecht},
  {Serre}, {Schaye}, {Soucail}, {Steinmetz}, {Streicher}, {Stuik}, {Valentin},
  {Vernet}, {Weilbacher}, {Wisotzki}, and {Yerle}}]{Bacon_10}
{Bacon} R, {Accardo} M, {Adjali} L, et~al. (2010) {The MUSE second-generation
  VLT instrument}. In: Ground-based and Airborne Instrumentation for Astronomy
  III, \procspie, vol 7735, p 773508

\bibitem[{{Banerjee} and {Kroupa}(2015)}]{Banerjee_15}
{Banerjee} S and {Kroupa} P (2015) {The formation of NGC 3603 young starburst
  cluster: `prompt' hierarchical assembly or monolithic starburst?} \mnras
  447:728--746

\bibitem[{{Banerjee} and {Kroupa}(2017)}]{2017A&A...597A..28B}
{Banerjee} S and {Kroupa} P (2017) {How can young massive clusters reach their
  present-day sizes?} \aap 597:A28

\bibitem[{{Barnes} et~al.(2019){Barnes}, {Longmore}, {Avison}, {Contreras},
  {Ginsburg}, {Henshaw}, {Rathborne}, {Walker}, {Alves}, {Bally}, {Battersby},
  {Beltr{\'a}n}, {Beuther}, {Garay}, {Gomez}, {Jackson}, {Kainulainen},
  {Kruijssen}, {Lu}, {Mills}, {Ott}, and {Peters}}]{barnes19}
{Barnes} AT, {Longmore} SN, {Avison} A, et~al. (2019) {Young massive star
  cluster formation in the Galactic Centre is driven by global gravitational
  collapse of high-mass molecular clouds}. \mnras 486(1):283--303

\bibitem[{{Bastian}(2008)}]{bastian2008}
{Bastian} N (2008) {On the star formation rate - brightest cluster relation:
  estimating the peak star formation rate in post-merger galaxies}. \mnras
  390(2):759--768

\bibitem[{{Bastian} and {Goodwin}(2006)}]{Bastian_06}
{Bastian} N and {Goodwin} SP (2006) {Evidence for the strong effect of gas
  removal on the internal dynamics of young stellar clusters}. \mnras
  369:L9--L13

\bibitem[{{Bastian} and {Lardo}(2018)}]{bastianlardo2018}
{Bastian} N and {Lardo} C (2018) {Multiple Stellar Populations in Globular
  Clusters}. \araa 56:83--136

\bibitem[{{Bastian} et~al.(2008){Bastian}, {Gieles}, {Goodwin}, {Trancho},
  {Smith}, {Konstantopoulos}, and {Efremov}}]{2008MNRAS.389..223B}
{Bastian} N, {Gieles} M, {Goodwin} SP, et~al. (2008) {The early expansion of
  cluster cores}. \mnras 389:223--230

\bibitem[{{Bastian} et~al.(2009){Bastian}, {Gieles}, {Ercolano}, and
  {Gutermuth}}]{bastian2009}
{Bastian} N, {Gieles} M, {Ercolano} B, et~al. (2009) {The spatial evolution of
  stellar structures in the Large Magellanic Cloud}. \mnras 392(2):868--878

\bibitem[{{Bastian} et~al.(2010){Bastian}, {Covey}, and {Meyer}}]{Bastian_10}
{Bastian} N, {Covey} KR, and {Meyer} MR (2010) {A Universal Stellar Initial
  Mass Function? A Critical Look at Variations}. \araa 48:339--389

\bibitem[{{Bastian} et~al.(2012){Bastian}, {Adamo}, {Gieles}, {Silva-Villa},
  {Lamers}, {Larsen}, {Smith}, {Konstantopoulos}, and
  {Zackrisson}}]{bastian2012}
{Bastian} N, {Adamo} A, {Gieles} M, et~al. (2012) {Stellar clusters in M83:
  formation, evolution, disruption and the influence of the environment}.
  \mnras 419(3):2606--2622

\bibitem[{{Bastian} et~al.(2013){Bastian}, {Lamers}, {de Mink}, {Longmore},
  {Goodwin}, and {Gieles}}]{Bastian_etal13}
{Bastian} N, {Lamers} HJGLM, {de Mink} SE, et~al. (2013) {Early disc accretion
  as the origin of abundance anomalies in globular clusters}. \mnras
  436(3):2398--2411

\bibitem[{{Baumgardt} and {Makino}(2003)}]{baumgardt03}
{Baumgardt} H and {Makino} J (2003) {Dynamical evolution of star clusters in
  tidal fields}. \mnras 340:227--246

\bibitem[{{Beccari} et~al.(2015){Beccari}, {De Marchi}, {Panagia}, {Valenti},
  {Carraro}, {Romaniello}, {Zoccali}, and {Weidner}}]{Beccari_15}
{Beccari} G, {De Marchi} G, {Panagia} N, et~al. (2015) {Mass accretion rates
  from multiband photometry in the Carina Nebula: the case of Trumpler 14}.
  \aap 574:A44

\bibitem[{{Becker} and {Fenkart}(1971)}]{Becker_71}
{Becker} W and {Fenkart} R (1971) {A catalogue of galactic star clusters
  observed in three colours}. \aaps 4:241

\bibitem[{{Bedin} et~al.(2004){Bedin}, {Piotto}, {Anderson}, {Cassisi}, {King},
  {Momany}, and {Carraro}}]{Bedin_etal04}
{Bedin} LR, {Piotto} G, {Anderson} J, et~al. (2004) {{\ensuremath{\omega}}
  Centauri: The Population Puzzle Goes Deeper}. \apjl 605(2):L125--L128

\bibitem[{{Bianchini} et~al.(2016){Bianchini}, {van de Ven}, {Norris},
  {Schinnerer}, and {Varri}}]{bianchini16}
{Bianchini} P, {van de Ven} G, {Norris} MA, et~al. (2016) {A novel look at
  energy equipartition in globular clusters}. \mnras 458(4):3644--3654

\bibitem[{{Bigiel} et~al.(2008){Bigiel}, {Leroy}, {Walter}, {Brinks}, {de
  Blok}, {Madore}, and {Thornley}}]{bigiel08}
{Bigiel} F, {Leroy} A, {Walter} F, et~al. (2008) {The Star Formation Law in
  Nearby Galaxies on Sub-Kpc Scales}. \aj 136:2846--2871

\bibitem[{{Billett} et~al.(2002){Billett}, {Hunter}, and
  {Elmegreen}}]{Billett2002}
{Billett} OH, {Hunter} DA, and {Elmegreen} BG (2002) {Compact Star Clusters in
  Nearby Dwarf Irregular Galaxies}. \aj 123(3):1454--1475

\bibitem[{{Binney} and {Tremaine}(1987)}]{Binney_87}
{Binney} J and {Tremaine} S (1987) {Galactic Dynamics}. Princton University
  Press

\bibitem[{{Bolatto} et~al.(2008){Bolatto}, {Leroy}, {Rosolowsky}, {Walter}, and
  {Blitz}}]{bolatto2008}
{Bolatto} AD, {Leroy} AK, {Rosolowsky} E, et~al. (2008) {The Resolved
  Properties of Extragalactic Giant Molecular Clouds}. \apj 686:948--965

\bibitem[{{Bouwens} et~al.(2017){Bouwens}, {Illingworth}, {Oesch}, {Maseda},
  {Ribeiro}, {Stefanon}, and {Lam}}]{Bouwens_etal17arXiv}
{Bouwens} RJ, {Illingworth} GD, {Oesch} PA, et~al. (2017) {Very low-luminosity
  galaxies in the early universe have observed sizes similar to single star
  cluster complexes}. arXiv e-prints arXiv:1711.02090

\bibitem[{{Boylan-Kolchin}(2018)}]{boylankolchin18}
{Boylan-Kolchin} M (2018) {The Little Engines That Could? Globular clusters
  contribute significantly to reionization-era star formation}. \mnras
  479(1):332--340

\bibitem[{{Bradley} et~al.(2006){Bradley}, {Knapen}, {Beckman}, and
  {Folkes}}]{bradley2006}
{Bradley} TR, {Knapen} JH, {Beckman} JE, et~al. (2006) {A composite H ii region
  luminosity function in H{\ensuremath{\alpha}} of unprecedented statistical
  weight}. \aap 459(1):L13--L16

\bibitem[{{Brodie} and {Strader}(2006)}]{BS2006}
{Brodie} JP and {Strader} J (2006) {Extragalactic Globular Clusters and Galaxy
  Formation}. \araa 44(1):193--267

\bibitem[{{Calzetti} et~al.(2015){Calzetti}, {Lee}, {Sabbi}, {Adamo}, {Smith},
  {Andrews}, {Ubeda}, {Bright}, {Thilker}, {Aloisi}, {Brown}, {Chandar},
  {Christian}, {Cignoni}, {Clayton}, {da Silva}, {de Mink}, {Dobbs},
  {Elmegreen}, {Elmegreen}, {Evans}, {Fumagalli}, {Gallagher}, {Gouliermis},
  {Grebel}, {Herrero}, {Hunter}, {Johnson}, {Kennicutt}, {Kim}, {Krumholz},
  {Lennon}, {Levay}, {Martin}, {Nair}, {Nota}, {{\"O}stlin}, {Pellerin},
  {Prieto}, {Regan}, {Ryon}, {Schaerer}, {Schiminovich}, {Tosi}, {Van Dyk},
  {Walterbos}, {Whitmore}, and {Wofford}}]{Calzetti_15}
{Calzetti} D, {Lee} JC, {Sabbi} E, et~al. (2015) {Legacy Extragalactic UV
  Survey (LEGUS) With the Hubble Space Telescope. I. Survey Description}. \aj
  149(2):51

\bibitem[{{Cantat-Gaudin} et~al.(2018){Cantat-Gaudin}, {Jordi}, {Vallenari},
  {Bragaglia}, {Balaguer-N{\'u}{\~n}ez}, {Soubiran}, {Bossini}, {Moitinho},
  {Castro-Ginard}, {Krone-Martins}, {Casamiquela}, {Sordo}, and
  {Carrera}}]{2018A&A...618A..93C}
{Cantat-Gaudin} T, {Jordi} C, {Vallenari} A, et~al. (2018) {A Gaia DR2 view of
  the open cluster population in the Milky Way}. \aap 618:A93

\bibitem[{{Cantat-Gaudin} et~al.(2019){Cantat-Gaudin}, {Krone-Martins},
  {Sedaghat}, {Farahi}, {de Souza}, {Skalidis}, {Malz}, {Mac{\^e}do}, {Moews},
  {Jordi}, {Moitinho}, {Castro-Ginard}, {Ishida}, {Heneka}, {Boucaud}, and
  {Trindade}}]{Cantat-Gaudin_19}
{Cantat-Gaudin} T, {Krone-Martins} A, {Sedaghat} N, et~al. (2019) {Gaia DR2
  unravels incompleteness of nearby cluster population: new open clusters in
  the direction of Perseus}. \aap 624:A126

\bibitem[{{Carlson} et~al.(2007){Carlson}, {Sabbi}, {Sirianni}, {Hora}, {Nota},
  {Meixner}, {Gallagher}, {Oey}, {Pasquali}, {Smith}, {Tosi}, and
  {Walterbos}}]{Carlson_07}
{Carlson} LR, {Sabbi} E, {Sirianni} M, et~al. (2007) {Progressive Star
  Formation in the Young SMC Cluster NGC 602}. \apjl 665:L109--L114

\bibitem[{{Carretta} et~al.(2009{\natexlab{a}}){Carretta}, {Bragaglia},
  {Gratton}, and {Lucatello}}]{Carretta_09}
{Carretta} E, {Bragaglia} A, {Gratton} R, et~al. (2009{\natexlab{a}}) {Na-O
  anticorrelation and HB. VIII. Proton-capture elements and metallicities in 17
  globular clusters from UVES spectra}. \aap 505:139--155

\bibitem[{{Carretta} et~al.(2009{\natexlab{b}}){Carretta}, {Bragaglia},
  {Gratton}, {Lucatello}, {Catanzaro}, {Leone}, {Bellazzini}, {Claudi},
  {D'Orazi}, {Momany}, {Ortolani}, {Pancino}, {Piotto}, {Recio-Blanco}, and
  {Sabbi}}]{2009A&A...505..117C}
{Carretta} E, {Bragaglia} A, {Gratton} RG, et~al. (2009{\natexlab{b}}) {Na-O
  anticorrelation and HB. VII. The chemical composition of first and
  second-generation stars in 15 globular clusters from GIRAFFE spectra}. \aap
  505:117--138

\bibitem[{{Carretta} et~al.(2010){Carretta}, {Bragaglia}, {Gratton},
  {Recio-Blanco}, {Lucatello}, {D'Orazi}, and {Cassisi}}]{Carretta_10}
{Carretta} E, {Bragaglia} A, {Gratton} RG, et~al. (2010) {Properties of stellar
  generations in globular clusters and relations with global parameters}. \aap
  516:A55

\bibitem[{{Castro-Ginard} et~al.(2020){Castro-Ginard}, {Jordi}, {Luri},
  {{\'A}lvarez Cid-Fuentes}, {Casamiquela}, {Anders}, {Cantat-Gaudin},
  {Mongui{\'o}}, {Balaguer-N{\'u}{\~n}ez}, {Sol{\`a}}, and
  {Badia}}]{Castro-Ginard_20}
{Castro-Ginard} A, {Jordi} C, {Luri} X, et~al. (2020) {Hunting for open
  clusters in \textit\{Gaia\} DR2: $582$ new OCs in the Galactic disc}. arXiv
  e-prints arXiv:2001.07122

\bibitem[{{Chabrier}(2003)}]{Chabrier_03}
{Chabrier} G (2003) {Galactic Stellar and Substellar Initial Mass Function}.
  \pasp 115:763--795

\bibitem[{{Chandar} et~al.(2010){Chandar}, {Fall}, and
  {Whitmore}}]{Chandra2010}
{Chandar} R, {Fall} SM, and {Whitmore} BC (2010) {New Tests for Disruption
  Mechanisms of Star Clusters: The Large and Small Magellanic Clouds}. \apj
  711(2):1263--1279

\bibitem[{{Chandar} et~al.(2014){Chandar}, {Whitmore}, {Calzetti}, and
  {O'Connell}}]{chandar2014}
{Chandar} R, {Whitmore} BC, {Calzetti} D, et~al. (2014) {Star-cluster Mass and
  Age Distributions of Two Fields in M83 Based on HST/WFC3 Observations}. \apj
  787(1):17

\bibitem[{{Chandar} et~al.(2016){Chandar}, {Whitmore}, {Dinino}, {Kennicutt},
  {Chien}, {Schinnerer}, and {Meidt}}]{chandar2016}
{Chandar} R, {Whitmore} BC, {Dinino} D, et~al. (2016) {The Age, Mass, and Size
  Distributions of Star Clusters in M51}. \apj 824(2):71

\bibitem[{{Chandar} et~al.(2017){Chandar}, {Fall}, {Whitmore}, and
  {Mulia}}]{chandar17}
{Chandar} R, {Fall} SM, {Whitmore} BC, et~al. (2017) {The Fraction of Stars
  That Form in Clusters in Different Galaxies}. \apj 849(2):128

\bibitem[{{Charbonnel}(2016)}]{2016EAS....80..177C}
{Charbonnel} C (2016) {Multiple Stellar Populations and Their Evolution in
  Globular Clusters: A Nucleosynthesis Perspective}. In: {Moraux} E, {Lebreton}
  Y, and {Charbonnel} C (eds) EAS Publications Series, EAS Publications Series,
  vol~80, pp 177--226

\bibitem[{{Chevance} et~al.(2016){Chevance}, {Madden}, {Lebouteiller},
  {Godard}, {Cormier}, {Galliano}, {Hony}, {Indebetouw}, {Le Bourlot}, {Lee},
  {Le Petit}, {Pellegrini}, {Roueff}, and {Wu}}]{chevance2016}
{Chevance} M, {Madden} SC, {Lebouteiller} V, et~al. (2016) {A milestone toward
  understanding PDR properties in the extreme environment of LMC-30 Doradus}.
  \aap 590:A36

\bibitem[{{Chevance} et~al.(2020{\natexlab{a}}){Chevance}, {Kruijssen},
  {Hygate}, {Schruba}, {Longmore}, {Groves}, {Henshaw}, {Herrera}, {Hughes},
  {Jeffreson}, {Lang}, {Leroy}, {Meidt}, {Pety}, {Razza}, {Rosolowsky},
  {Schinnerer}, {Bigiel}, {Blanc}, {Emsellem}, {Faesi}, {Glover}, {Haydon},
  {Ho}, {Kreckel}, {Lee}, {Liu}, {Querejeta}, {Saito}, {Sun}, {Usero}, and
  {Utomo}}]{chevance20}
{Chevance} M, {Kruijssen} JMD, {Hygate} APS, et~al. (2020{\natexlab{a}}) {The
  lifecycle of molecular clouds in nearby star-forming disc galaxies}. \mnras
  493(2):2872--2909

\bibitem[{{Chevance} et~al.(2020{\natexlab{b}}){Chevance}, {Kruijssen},
  {Vazquez-Semadeni}, {Nakamura}, {Klessen}, {Ballesteros-Paredes}, {Inutsuka},
  {Adamo}, and {Hennebelle}}]{chevance20b}
{Chevance} M, {Kruijssen} JMD, {Vazquez-Semadeni} E, et~al.
  (2020{\natexlab{b}}) {The molecular cloud lifecycle}. SSRv in press
  arXiv:2004.06113

\bibitem[{{Chevance} et~al.(2020{\natexlab{c}}){Chevance}, {Madden}, {Fischer},
  {Vacca}, {Lebouteiller}, {Fadda}, {Galliano}, {Indebetouw}, {Kruijssen},
  {Lee}, {Poglitsch}, {Polles}, {Cormier}, {Hony}, {Iserlohe}, {Krabbe},
  {Meixner}, {Sabbi}, and {Zinnecker}}]{chevance2020C}
{Chevance} M, {Madden} SC, {Fischer} C, et~al. (2020{\natexlab{c}}) {The
  CO-dark molecular gas mass in 30 Doradus}. \mnras

\bibitem[{{Choksi} et~al.(2018){Choksi}, {Gnedin}, and {Li}}]{choksi2018}
{Choksi} N, {Gnedin} OY, and {Li} H (2018) {Formation of globular cluster
  systems: from dwarf galaxies to giants}. \mnras 480(2):2343--2356

\bibitem[{{Clarke}(2007)}]{Clarke_07}
{Clarke} CJ (2007) {The photoevaporation of discs around young stars in massive
  clusters}. \mnras 376:1350--1356

\bibitem[{{Colombo} et~al.(2014){Colombo}, {Hughes}, {Schinnerer}, {Meidt},
  {Leroy}, {Pety}, {Dobbs}, {Garc{\'\i}a-Burillo}, {Dumas}, {Thompson},
  {Schuster}, and {Kramer}}]{Colombo2014}
{Colombo} D, {Hughes} A, {Schinnerer} E, et~al. (2014) {The PdBI Arcsecond
  Whirlpool Survey (PAWS): Environmental Dependence of Giant Molecular Cloud
  Properties in M51}. \apj 784(1):3

\bibitem[{{Cook} et~al.(2012){Cook}, {Seth}, {Dale}, {Johnson}, {Weisz},
  {Fouesneau}, {Olsen}, {Engelbracht}, and {Dalcanton}}]{cook2012}
{Cook} DO, {Seth} AC, {Dale} DA, et~al. (2012) {The ACS Nearby Galaxy Survey
  Treasury. X. Quantifying the Star Cluster Formation Efficiency of nearby
  Dwarf Galaxies}. \apj 751(2):100

\bibitem[{{Cook} et~al.(2019){Cook}, {Lee}, {Adamo}, {Kim}, {Chand ar},
  {Whitmore}, {Mok}, {Ryon}, {Dale}, {Calzetti}, {Andrews}, {Aloisi},
  {Ashworth}, {Bright}, {Brown}, {Christian}, {Cignoni}, {Clayton}, {da Silva},
  {de Mink}, {Dobbs}, {Elmegreen}, {Elmegreen}, {Evans}, {Fumagalli},
  {Gallagher}, {Gouliermis}, {Grasha}, {Grebel}, {Herrero}, {Hunter}, {Jensen},
  {Johnson}, {Kahre}, {Kennicutt}, {Krumholz}, {Lee}, {Lennon}, {Linden},
  {Martin}, {Messa}, {Nair}, {Nota}, {{\"O}stlin}, {Parziale}, {Pellerin},
  {Regan}, {Sabbi}, {Sacchi}, {Schaerer}, {Schiminovich}, {Shabani}, {Slane},
  {Small}, {Smith}, {Smith}, {Taibi}, {Thilker}, {de la Torre}, {Tosi},
  {Turner}, {Ubeda}, {Van Dyk}, {Walterbos}, and {Wofford}}]{cook2019}
{Cook} DO, {Lee} JC, {Adamo} A, et~al. (2019) {Star cluster catalogues for the
  LEGUS dwarf galaxies}. \mnras 484(4):4897--4919

\bibitem[{{Crain} et~al.(2015){Crain}, {Schaye}, {Bower}, {Furlong},
  {Schaller}, {Theuns}, {Dalla Vecchia}, {Frenk}, {McCarthy}, {Helly},
  {Jenkins}, {Rosas-Guevara}, {White}, and {Trayford}}]{crain2015}
{Crain} RA, {Schaye} J, {Bower} RG, et~al. (2015) {The EAGLE simulations of
  galaxy formation: calibration of subgrid physics and model variations}.
  \mnras 450(2):1937--1961

\bibitem[{{Creasey} et~al.(2019){Creasey}, {Sales}, {Peng}, and
  {Sameie}}]{creasey19}
{Creasey} P, {Sales} LV, {Peng} EW, et~al. (2019) {Globular clusters formed
  within dark haloes I: present-day abundance, distribution, and kinematics}.
  \mnras 482(1):219--230

\bibitem[{{Crowther}(2019)}]{Crowther2019}
{Crowther} P (2019) {Massive stars in the Tarantula Nebula: A Rosetta Stone for
  Extragalactic Supergiant HII Regions}. arXiv e-prints arXiv:1911.02047

\bibitem[{{Crowther} et~al.(2006){Crowther}, {Hadfield}, {Clark}, {Negueruela},
  and {Vacca}}]{Crowther_06}
{Crowther} PA, {Hadfield} LJ, {Clark} JS, et~al. (2006) {A census of the
  Wolf-Rayet content in Westerlund 1 from near-infrared imaging and
  spectroscopy}. \mnras 372:1407--1424

\bibitem[{{Crowther} et~al.(2016){Crowther}, {Caballero-Nieves}, {Bostroem},
  {Ma{\'{\i}}z Apell{\'a}niz}, {Schneider}, {Walborn}, {Angus}, {Brott},
  {Bonanos}, {de Koter}, {de Mink}, {Evans}, {Gr{\"a}fener}, {Herrero},
  {Howarth}, {Langer}, {Lennon}, {Puls}, {Sana}, and {Vink}}]{Crowther_16}
{Crowther} PA, {Caballero-Nieves} SM, {Bostroem} KA, et~al. (2016) {The R136
  star cluster dissected with Hubble Space Telescope/STIS. I. Far-ultraviolet
  spectroscopic census and the origin of He II {$\lambda$}1640 in young star
  clusters}. \mnras 458:624--659

\bibitem[{{Dale}(2015)}]{dale15b}
{Dale} JE (2015) {The modelling of feedback in star formation simulations}.
  \nar 68:1--33

\bibitem[{{Dale} et~al.(2015){Dale}, {Ercolano}, and {Bonnell}}]{Dale_15}
{Dale} JE, {Ercolano} B, and {Bonnell} IA (2015) {Early evolution of embedded
  clusters}. \mnras 451:987--1003

\bibitem[{{D'Antona} et~al.(2005){D'Antona}, {Bellazzini}, {Caloi}, {Pecci},
  {Galleti}, and {Rood}}]{2005ApJ...631..868D}
{D'Antona} F, {Bellazzini} M, {Caloi} V, et~al. (2005) {A Helium Spread among
  the Main-Sequence Stars in NGC 2808}. \apj 631:868--878

\bibitem[{{de la Fuente Marcos} and {de la Fuente
  Marcos}(2009)}]{delafuente2009}
{de la Fuente Marcos} R and {de la Fuente Marcos} C (2009) {Hierarchical Star
  Formation in the Milky Way Disk}. \apj 700(1):436--446

\bibitem[{{De Marchi} et~al.(2011){De Marchi}, {Panagia}, {Romaniello},
  {Sabbi}, {Sirianni}, {Prada Moroni}, and {Degl'Innocenti}}]{deMarchi_11}
{De Marchi} G, {Panagia} N, {Romaniello} M, et~al. (2011) {Photometric
  Determination of the Mass Accretion Rates of Pre-main-sequence Stars. II. NGC
  346 in the Small Magellanic Cloud}. \apj 740:11

\bibitem[{{de Mink} et~al.(2009){de Mink}, {Pols}, {Langer}, and
  {Izzard}}]{2009A&A...507L...1D}
{de Mink} SE, {Pols} OR, {Langer} N, et~al. (2009) {Massive binaries as the
  source of abundance anomalies in globular clusters}. \aap 507:L1--L4

\bibitem[{{De Silva} et~al.(2007{\natexlab{a}}){De Silva}, {Freeman},
  {Asplund}, {Bland -Hawthorn}, {Bessell}, and {Collet}}]{2007AJ....133.1161D}
{De Silva} GM, {Freeman} KC, {Asplund} M, et~al. (2007{\natexlab{a}}) {Chemical
  Homogeneity in Collinder 261 and Implications for Chemical Tagging}. \aj
  133(3):1161--1175

\bibitem[{{De Silva} et~al.(2007{\natexlab{b}}){De Silva}, {Freeman},
  {Bland-Hawthorn}, {Asplund}, and {Bessell}}]{2007AJ....133..694D}
{De Silva} GM, {Freeman} KC, {Bland-Hawthorn} J, et~al. (2007{\natexlab{b}})
  {Chemically Tagging the HR 1614 Moving Group}. \aj 133(2):694--704

\bibitem[{{Decressin} et~al.(2007{\natexlab{a}}){Decressin}, {Charbonnel}, and
  {Meynet}}]{2007A&A...475..859D}
{Decressin} T, {Charbonnel} C, and {Meynet} G (2007{\natexlab{a}}) {Origin of
  the abundance patterns in Galactic globular clusters: constraints on
  dynamical and chemical properties of globular clusters}. \aap 475(3):859--873

\bibitem[{{Decressin} et~al.(2007{\natexlab{b}}){Decressin}, {Meynet},
  {Charbonnel}, {Prantzos}, and {Ekstr{\"o}m}}]{2007A&A...464.1029D}
{Decressin} T, {Meynet} G, {Charbonnel} C, et~al. (2007{\natexlab{b}}) {Fast
  rotating massive stars and the origin of the abundance patterns in galactic
  globular clusters}. \aap 464:1029--1044

\bibitem[{{Denisenkov} and {Denisenkova}(1990)}]{1990SvAL...16..275D}
{Denisenkov} PA and {Denisenkova} SN (1990) {Correlation Between the Abundances
  of NA and the CNO Elements in Red Giants in Omega-Centauri}. Soviet Astronomy
  Letters 16:275

\bibitem[{{Denissenkov} and {Hartwick}(2014)}]{2014MNRAS.437L..21D}
{Denissenkov} PA and {Hartwick} FDA (2014) {Supermassive stars as a source of
  abundance anomalies of proton-capture elements in globular clusters}. \mnras
  437:L21--L25

\bibitem[{{Dessauges-Zavadsky} and {Adamo}(2018)}]{DZA2018}
{Dessauges-Zavadsky} M and {Adamo} A (2018) {First constraints on the stellar
  mass function of star-forming clumps at the peak of cosmic star formation}.
  \mnras 479(1):L118--L122

\bibitem[{{Dessauges-Zavadsky} et~al.(2019){Dessauges-Zavadsky}, {Richard},
  {Combes}, {Schaerer}, {Rujopakarn}, {Mayer}, {Cava}, {Boone}, {Egami},
  {Kneib}, {P{\'e}rez-Gonz{\'a}lez}, {Pfenniger}, {Rawle}, {Teyssier}, and {van
  der Werf}}]{DZ2019}
{Dessauges-Zavadsky} M, {Richard} J, {Combes} F, et~al. (2019) {Molecular
  clouds in the Cosmic Snake normal star-forming galaxy 8 billion years ago}.
  Nature Astronomy 3:1115--1121

\bibitem[{{Dib}(2014)}]{Dib_14}
{Dib} S (2014) {Testing the universality of the IMF with Bayesian statistics:
  young clusters}. \mnras 444(2):1957--1981

\bibitem[{{Dobbs} and {Pringle}(2013)}]{dobbs2013}
{Dobbs} CL and {Pringle} JE (2013) {The exciting lives of giant molecular
  clouds}. \mnras 432(1):653--667

\bibitem[{{Donovan Meyer} et~al.(2013){Donovan Meyer}, {Koda}, {Momose},
  {Mooney}, {Egusa}, {Carty}, {Kennicutt}, {Kuno}, {Rebolledo}, {Sawada},
  {Scoville}, and {Wong}}]{donovan2013}
{Donovan Meyer} J, {Koda} J, {Momose} R, et~al. (2013) {Resolved Giant
  Molecular Clouds in Nearby Spiral Galaxies: Insights from the CANON CO (1-0)
  Survey}. \apj 772(2):107

\bibitem[{{Downes} and {Solomon}(1998)}]{downes1998}
{Downes} D and {Solomon} PM (1998) {Rotating Nuclear Rings and Extreme
  Starbursts in Ultraluminous Galaxies}. \apj 507(2):615--654

\bibitem[{{Drew} et~al.(2018){Drew}, {Herrero}, {Mohr-Smith}, {Mongui{\'o}},
  {Wright}, {Kupfer}, and {Napiwotzki}}]{Drew_18}
{Drew} JE, {Herrero} A, {Mohr-Smith} M, et~al. (2018) {Massive stars in the
  hinterland of the young cluster, Westerlund 2}. \mnras 480(2):2109--2124

\bibitem[{{Drew} et~al.(2019){Drew}, {Mongui{\'o}}, and {Wright}}]{Drew_19}
{Drew} JE, {Mongui{\'o}} M, and {Wright} NJ (2019) {The O star hinterland of
  the Galactic starburst, NGC 3603}. \mnras 486(1):1034--1044

\bibitem[{{Efremov} and {Elmegreen}(1998)}]{Efremov1998}
{Efremov} YN and {Elmegreen} BG (1998) {Hierarchical star formation from the
  time-space distribution of star clusters in the Large Magellanic Cloud}.
  \mnras 299(2):588--594

\bibitem[{{El-Badry} et~al.(2019){El-Badry}, {Quataert}, {Weisz}, {Choksi}, and
  {Boylan-Kolchin}}]{elbadry19}
{El-Badry} K, {Quataert} E, {Weisz} DR, et~al. (2019) {The formation and
  hierarchical assembly of globular cluster populations}. \mnras
  482(4):4528--4552

\bibitem[{{Elmegreen}(2011)}]{Elmegreen2011}
{Elmegreen} BG (2011) {Star Formation Patterns and Hierarchies}. In:
  {Charbonnel} C and {Montmerle} T (eds) EAS Publications Series, EAS
  Publications Series, vol~51, pp 31--44

\bibitem[{{Elmegreen}(2018)}]{Elmegreen2018}
{Elmegreen} BG (2018) {Two Thresholds for Globular Cluster Formation and the
  Common Occurrence of Massive Clusters in the Early Universe}. \apj 869(2):119

\bibitem[{{Elmegreen} and {Efremov}(1996)}]{Elmegreen1996}
{Elmegreen} BG and {Efremov} YN (1996) {An Extension of Hierarchical Star
  Formation to Galactic Scales}. \apj 466:802

\bibitem[{{Elmegreen} and {Efremov}(1997)}]{elmegreen97}
{Elmegreen} BG and {Efremov} YN (1997) {A Universal Formation Mechanism for
  Open and Globular Clusters in Turbulent Gas}. \apj 480:235--+

\bibitem[{{Elmegreen} and {Falgarone}(1996)}]{elmergeen_falgarone96}
{Elmegreen} BG and {Falgarone} E (1996) {A Fractal Origin for the Mass Spectrum
  of Interstellar Clouds}. \apj 471:816

\bibitem[{{Elmegreen} and {Hunter}(2010)}]{elmegreen10}
{Elmegreen} BG and {Hunter} DA (2010) {On the Disruption of Star Clusters in a
  Hierarchical Interstellar Medium}. \apj 712(1):604--623

\bibitem[{{Elmegreen} and {Scalo}(2004)}]{ElmegreenScalo2004}
{Elmegreen} BG and {Scalo} J (2004) {Interstellar Turbulence I: Observations
  and Processes}. \araa 42(1):211--273

\bibitem[{{Elmegreen} et~al.(1989){Elmegreen}, {Elmegreen}, and
  {Seiden}}]{Elmegreen1989}
{Elmegreen} BG, {Elmegreen} DM, and {Seiden} PE (1989) {Spiral arm amplitude
  variations and pattern speeds in the grand design galaxies M51, M81, and
  M100}. \apj 343:602--607

\bibitem[{{Elmegreen} et~al.(2006){Elmegreen}, {Elmegreen}, {Chand ar},
  {Whitmore}, and {Regan}}]{elmegreen2006}
{Elmegreen} BG, {Elmegreen} DM, {Chand ar} R, et~al. (2006) {Hierarchical Star
  Formation in the Spiral Galaxy NGC 628}. \apj 644(2):879--889

\bibitem[{{Elmegreen} et~al.(2013){Elmegreen}, {Elmegreen}, {S{\'a}nchez
  Almeida}, {Mu{\~n}oz-Tu{\~n}{\'o}n}, {Dewberry}, {Putko}, {Teich}, and
  {Popinchalk}}]{elmegreen2013}
{Elmegreen} BG, {Elmegreen} DM, {S{\'a}nchez Almeida} J, et~al. (2013) {Massive
  Clumps in Local Galaxies: Comparisons with High-redshift Clumps}. \apj
  774(1):86

\bibitem[{{Faesi} et~al.(2018){Faesi}, {Lada}, and {Forbrich}}]{faesi2018}
{Faesi} CM, {Lada} CJ, and {Forbrich} J (2018) {The ALMA View of GMCs in NGC
  300: Physical Properties and Scaling Relations at 10 pc Resolution}. \apj
  857:19

\bibitem[{{Fall} et~al.(2005){Fall}, {Chandar}, and {Whitmore}}]{Fall2005}
{Fall} SM, {Chandar} R, and {Whitmore} BC (2005) {The Age Distribution of
  Massive Star Clusters in the Antennae Galaxies}. \apjl 631(2):L133--L136

\bibitem[{{Feigelson} et~al.(2013){Feigelson}, {Townsley}, {Broos}, {Busk},
  {Getman}, {King}, {Kuhn}, {Naylor}, {Povich}, {Baddeley}, {Bate},
  {Indebetouw}, {Luhman}, {McCaughrean}, {Pittard}, {Pudritz}, {Sills}, {Song},
  and {Wadsley}}]{Feigelson_13}
{Feigelson} ED, {Townsley} LK, {Broos} PS, et~al. (2013) {Overview of the
  Massive Young Star-Forming Complex Study in Infrared and X-Ray (MYStIX)
  Project}. \apjs 209(2):26

\bibitem[{{Fensch} et~al.(2019){Fensch}, {Duc}, {Boquien}, {Elmegreen},
  {Elmegreen}, {Bournaud}, {Brinks}, {de Grijs}, {Lelli}, {Renaud}, and
  {Weilbacher}}]{fensch19}
{Fensch} J, {Duc} PA, {Boquien} M, et~al. (2019) {Massive star cluster
  formation and evolution in tidal dwarf galaxies}. \aap 628:A60

\bibitem[{{Figer} et~al.(1999){Figer}, {McLean}, and {Morris}}]{Figer_99}
{Figer} DF, {McLean} IS, and {Morris} M (1999) {Massive Stars in the Quintuplet
  Cluster}. \apj 514:202--220

\bibitem[{{Forbes} and {Bridges}(2010)}]{forbes10}
{Forbes} DA and {Bridges} T (2010) {Accreted versus in situ Milky Way globular
  clusters}. \mnras 404:1203--1214

\bibitem[{{Forbes} et~al.(2018){Forbes}, {Bastian}, {Gieles}, {Crain},
  {Kruijssen}, {Larsen}, {Ploeckinger}, {Agertz}, {Trenti}, {Ferguson},
  {Pfeffer}, and {Gnedin}}]{Forbes2018}
{Forbes} DA, {Bastian} N, {Gieles} M, et~al. (2018) {Globular cluster formation
  and evolution in the context of cosmological galaxy assembly: open
  questions}. Proceedings of the Royal Society of London Series A
  474(2210):20170616

\bibitem[{{Fouesneau} et~al.(2014){Fouesneau}, {Johnson}, {Weisz}, {Dalcanton},
  {Bell}, {Bianchi}, {Caldwell}, {Gouliermis}, {Guhathakurta}, {Kalirai},
  {Larsen}, {Rix}, {Seth}, {Skillman}, and {Williams}}]{fouesneau14}
{Fouesneau} M, {Johnson} LC, {Weisz} DR, et~al. (2014) {The Panchromatic Hubble
  Andromeda Treasury. V. Ages and Masses of the Year 1 Stellar Clusters}. \apj
  786:117

\bibitem[{{Freeman} et~al.(2017){Freeman}, {Rosolowsky}, {Kruijssen},
  {Bastian}, and {Adamo}}]{freeman2017}
{Freeman} P, {Rosolowsky} E, {Kruijssen} JMD, et~al. (2017) {The varying mass
  distribution of molecular clouds across M83}. \mnras 468(2):1769--1781

\bibitem[{{F{\"u}rnkranz} et~al.(2019){F{\"u}rnkranz}, {Meingast}, and
  {Alves}}]{fuernkranz19}
{F{\"u}rnkranz} V, {Meingast} S, and {Alves} J (2019) {Extended stellar systems
  in the solar neighborhood. III. Like ships in the night: the Coma Berenices
  neighbor moving group}. \aap 624:L11

\bibitem[{{Furukawa} et~al.(2009){Furukawa}, {Dawson}, {Ohama}, {Kawamura},
  {Mizuno}, {Onishi}, and {Fukui}}]{Furukawa_09}
{Furukawa} N, {Dawson} JR, {Ohama} A, et~al. (2009) {Molecular Clouds Toward
  RCW49 and Westerlund 2: Evidence for Cluster Formation Triggered by
  Cloud-Cloud Collision}. \apjl 696:L115--L119

\bibitem[{{Gaczkowski} et~al.(2013){Gaczkowski}, {Preibisch}, {Ratzka},
  {Roccatagliata}, {Ohlendorf}, and {Zinnecker}}]{Gaczkowski_13}
{Gaczkowski} B, {Preibisch} T, {Ratzka} T, et~al. (2013) {Herschel far-infrared
  observations of the Carina Nebula complex. II. The embedded young stellar and
  protostellar population}. \aap 549:A67

\bibitem[{{Gaia Collaboration} et~al.(2016){Gaia Collaboration}, {Prusti}, {de
  Bruijne}, {Brown}, {Vallenari}, {Babusiaux}, {Bailer-Jones}, {Bastian},
  {Biermann}, {Evans}, and et~al.}]{Gaia_16}
{Gaia Collaboration}, {Prusti} T, {de Bruijne} JHJ, et~al. (2016) {The Gaia
  mission}. \aap 595:A1

\bibitem[{{Gaia Collaboration} et~al.(2018){Gaia Collaboration}, {Brown},
  {Vallenari}, {Prusti}, {de Bruijne}, {Babusiaux}, and
  {Bailer-Jones}}]{Gaia_18}
{Gaia Collaboration}, {Brown} AGA, {Vallenari} A, et~al. (2018) {Gaia Data
  Release 2. Summary of the contents and survey properties}. ArXiv e-prints

\bibitem[{{Gascoigne} and {Kron}(1952)}]{Gascoigne_52}
{Gascoigne} SCB and {Kron} GE (1952) {Colors and Magnitudes of Some Star
  Clusters in the Magellanic Clouds}. \pasp 64(379):196

\bibitem[{{Gennaro} et~al.(2011){Gennaro}, {Brandner}, {Stolte}, and
  {Henning}}]{Gennaro_11}
{Gennaro} M, {Brandner} W, {Stolte} A, et~al. (2011) {Mass segregation and
  elongation of the starburst cluster Westerlund 1}. \mnras 412:2469--2488

\bibitem[{{Gennaro} et~al.(2017){Gennaro}, {Goodwin}, {Parker}, {Allison}, and
  {Brandner}}]{Gennaro_17}
{Gennaro} M, {Goodwin} SP, {Parker} RJ, et~al. (2017) {Hierarchical formation
  of Westerlund 1: a collapsing cluster with no primordial mass segregation?}
  \mnras 472(2):1760--1769

\bibitem[{{Gentry} et~al.(2017){Gentry}, {Krumholz}, {Dekel}, and
  {Madau}}]{gentry2017}
{Gentry} ES, {Krumholz} MR, {Dekel} A, et~al. (2017) {Enhanced momentum
  feedback from clustered supernovae}. \mnras 465(2):2471--2488

\bibitem[{{Getman} et~al.(2019){Getman}, {Feigelson}, {Kuhn}, and
  {Garmire}}]{Getman_19}
{Getman} KV, {Feigelson} ED, {Kuhn} MA, et~al. (2019) {Gaia stellar kinematics
  in the head of the Orion A cloud: runaway stellar groups and gravitational
  infall}. \mnras 487(3):2977--3000

\bibitem[{{Gieles}(2009)}]{gieles2009}
{Gieles} M (2009) {The early evolution of the star cluster mass function}.
  \mnras 394(4):2113--2126

\bibitem[{{Gieles} and {Baumgardt}(2008)}]{gieles08}
{Gieles} M and {Baumgardt} H (2008) {Lifetimes of tidally limited star clusters
  with different radii}. \mnras 389:L28--L32

\bibitem[{{Gieles} and {Portegies Zwart}(2011)}]{Gieles_11}
{Gieles} M and {Portegies Zwart} SF (2011) {The distinction between star
  clusters and associations}. \mnras 410:L6--L7

\bibitem[{{Gieles} and {Renaud}(2016)}]{gieles16}
{Gieles} M and {Renaud} F (2016) {If it does not kill them, it makes them
  stronger: collisional evolution of star clusters with tidal shocks}. \mnras
  463:L103--L107

\bibitem[{{Gieles} et~al.(2006){Gieles}, {Larsen}, {Scheepmaker}, {Bastian},
  {Haas}, and {Lamers}}]{gieles2006}
{Gieles} M, {Larsen} SS, {Scheepmaker} RA, et~al. (2006) {Observational
  evidence for a truncation of the star cluster initial mass function at the
  high mass end}. \aap 446(2):L9--L12

\bibitem[{{Gieles} et~al.(2007){Gieles}, {Athanassoula}, and {Portegies
  Zwart}}]{gieles07}
{Gieles} M, {Athanassoula} E, and {Portegies Zwart} SF (2007) {The effect of
  spiral arm passages on the evolution of stellar clusters}. \mnras
  376:809--819

\bibitem[{{Gieles} et~al.(2008){Gieles}, {Bastian}, and
  {Ercolano}}]{Gieles2008}
{Gieles} M, {Bastian} N, and {Ercolano} B (2008) {Evolution of stellar
  structure in the Small Magellanic Cloud}. \mnras 391(1):L93--L97

\bibitem[{{Gieles} et~al.(2010){Gieles}, {Baumgardt}, {Heggie}, and
  {Lamers}}]{2010MNRAS.408L..16G}
{Gieles} M, {Baumgardt} H, {Heggie} DC, et~al. (2010) {On the mass-radius
  relation of hot stellar systems}. \mnras 408:L16--L20

\bibitem[{{Gieles} et~al.(2018){Gieles}, {Charbonnel}, {Krause},
  {H{\'e}nault-Brunet}, {Agertz}, {Lamers}, {Bastian}, {Gualandris}, {Zocchi},
  and {Petts}}]{2018MNRAS.478.2461G}
{Gieles} M, {Charbonnel} C, {Krause} MGH, et~al. (2018) {Concurrent formation
  of supermassive stars and globular clusters: implications for early
  self-enrichment}. \mnras 478:2461--2479

\bibitem[{{Ginsburg} and {Kruijssen}(2018)}]{GK2018}
{Ginsburg} A and {Kruijssen} JMD (2018) {A High Cluster Formation Efficiency in
  the Sagittarius B2 Complex}. \apjl 864(1):L17

\bibitem[{{Glatt} et~al.(2010){Glatt}, {Grebel}, and {Koch}}]{Glatt2010}
{Glatt} K, {Grebel} EK, and {Koch} A (2010) {Ages and luminosities of young
  SMC/LMC star clusters and the recent star formation history of the Clouds}.
  \aap 517:A50

\bibitem[{{Gnedin} and {Ostriker}(1997)}]{gnedin97}
{Gnedin} OY and {Ostriker} JP (1997) {Destruction of the Galactic Globular
  Cluster System}. \apj 474:223--+

\bibitem[{{Goddard} et~al.(2010){Goddard}, {Bastian}, and
  {Kennicutt}}]{goddard10}
{Goddard} QE, {Bastian} N, and {Kennicutt} RC (2010) {On the fraction of star
  clusters surviving the embedded phase}. \mnras 405(2):857--869

\bibitem[{{Gouliermis} et~al.(2015){Gouliermis}, {Thilker}, {Elmegreen},
  {Elmegreen}, {Calzetti}, {Lee}, {Adamo}, {Aloisi}, {Cignoni}, {Cook}, {Dale},
  {Gallagher}, {Grasha}, {Grebel}, {Dav{\'o}}, {Hunter}, {Johnson}, {Kim},
  {Nair}, {Nota}, {Pellerin}, {Ryon}, {Sabbi}, {Sacchi}, {Smith}, {Tosi},
  {Ubeda}, and {Whitmore}}]{gouliermis2015}
{Gouliermis} DA, {Thilker} D, {Elmegreen} BG, et~al. (2015) {Hierarchical star
  formation across the ring galaxy NGC 6503}. \mnras 452(4):3508--3528

\bibitem[{{Grasha} et~al.(2015){Grasha}, {Calzetti}, {Adamo}, {Kim},
  {Elmegreen}, {Gouliermis}, {Aloisi}, {Bright}, {Christian}, {Cignoni},
  {Dale}, {Dobbs}, {Elmegreen}, {Fumagalli}, {Gallagher}, {Grebel}, {Johnson},
  {Lee}, {Messa}, {Smith}, {Ryon}, {Thilker}, {Ubeda}, and
  {Wofford}}]{Grasha2015}
{Grasha} K, {Calzetti} D, {Adamo} A, et~al. (2015) {The Spatial Distribution of
  the Young Stellar Clusters in the Star-forming Galaxy NGC 628}. \apj
  815(2):93

\bibitem[{{Grasha} et~al.(2017{\natexlab{a}}){Grasha}, {Calzetti}, {Adamo},
  {Kim}, {Elmegreen}, {Gouliermis}, {Dale}, {Fumagalli}, {Grebel}, {Johnson},
  {Kahre}, {Kennicutt}, {Messa}, {Pellerin}, {Ryon}, {Smith}, {Shabani},
  {Thilker}, and {Ubeda}}]{Grasha2017a}
{Grasha} K, {Calzetti} D, {Adamo} A, et~al. (2017{\natexlab{a}}) {The
  Hierarchical Distribution of the Young Stellar Clusters in Six Local
  Star-forming Galaxies}. \apj 840(2):113

\bibitem[{{Grasha} et~al.(2017{\natexlab{b}}){Grasha}, {Elmegreen}, {Calzetti},
  {Adamo}, {Aloisi}, {Bright}, {Cook}, {Dale}, {Fumagalli}, {Gallagher},
  {Gouliermis}, {Grebel}, {Kahre}, {Kim}, {Krumholz}, {Lee}, {Messa}, {Ryon},
  and {Ubeda}}]{Grasha2017b}
{Grasha} K, {Elmegreen} BG, {Calzetti} D, et~al. (2017{\natexlab{b}})
  {Hierarchical Star Formation in Turbulent Media: Evidence from Young Star
  Clusters}. \apj 842(1):25

\bibitem[{{Grasha} et~al.(2018){Grasha}, {Calzetti}, {Bittle}, {Johnson},
  {Donovan Meyer}, {Kennicutt}, {Elmegreen}, {Adamo}, {Krumholz}, {Fumagalli},
  {Grebel}, {Gouliermis}, {Cook}, {Gallagher}, {Aloisi}, {Dale}, {Linden},
  {Sacchi}, {Thilker}, {Walterbos}, {Messa}, {Wofford}, and
  {Smith}}]{Grasha2018}
{Grasha} K, {Calzetti} D, {Bittle} L, et~al. (2018) {Connecting young star
  clusters to CO molecular gas in NGC 7793 with ALMA-LEGUS}. \mnras
  481(1):1016--1027

\bibitem[{{Grasha} et~al.(2019){Grasha}, {Calzetti}, {Adamo}, {Kennicutt},
  {Elmegreen}, {Messa}, {Dale}, {Fedorenko}, {Mahadevan}, {Grebel},
  {Fumagalli}, {Kim}, {Dobbs}, {Gouliermis}, {Ashworth}, {Gallagher}, {Smith},
  {Tosi}, {Whitmore}, {Schinnerer}, {Colombo}, {Hughes}, {Leroy}, and
  {Meidt}}]{Grasha2019}
{Grasha} K, {Calzetti} D, {Adamo} A, et~al. (2019) {The spatial relation
  between young star clusters and molecular clouds in M51 with LEGUS}. \mnras
  483(4):4707--4723

\bibitem[{{Gratton} et~al.(2019){Gratton}, {Bragaglia}, {Carretta}, {D'Orazi},
  {Lucatello}, and {Sollima}}]{2019A&ARv..27....8G}
{Gratton} R, {Bragaglia} A, {Carretta} E, et~al. (2019) {What is a globular
  cluster? An observational perspective}. \aapr 27(1):8

\bibitem[{{Grebel} and {Chu}(2000)}]{Grebel_00}
{Grebel} EK and {Chu} YH (2000) {Hubble Space Telescope Photometry of Hodge
  301: An ``Old'' Star Cluster in 30 Doradus}. \aj 119(2):787--799

\bibitem[{{Habibi} et~al.(2013){Habibi}, {Stolte}, {Brandner}, {Hu{\ss}mann},
  and {Motohara}}]{Habibi_13}
{Habibi} M, {Stolte} A, {Brandner} W, et~al. (2013) {The Arches cluster out to
  its tidal radius: dynamical mass segregation and the effect of the extinction
  law on the stellar mass function}. \aap 556:A26

\bibitem[{{Halbesma} et~al.(2019){Halbesma}, {Grand}, {G{\'o}mez}, {Marinacci},
  {Pakmor}, {Trick}, {Busch}, and {White}}]{halbesma19}
{Halbesma} TLR, {Grand} RJJ, {G{\'o}mez} FA, et~al. (2019) {The globular
  cluster system of the Auriga simulations}. \mnras\ submitted arXiv:1909.02630

\bibitem[{{Hammer} et~al.(2007){Hammer}, {Puech}, {Chemin}, {Flores}, and
  {Lehnert}}]{hammer07}
{Hammer} F, {Puech} M, {Chemin} L, et~al. (2007) {The Milky Way, an
  Exceptionally Quiet Galaxy: Implications for the Formation of Spiral
  Galaxies}. \apj 662:322--334

\bibitem[{{Han} et~al.(2009){Han}, {Lee}, {Joo}, {Sohn}, {Yoon}, {Kim}, and
  {Lee}}]{Han_etal2009}
{Han} SI, {Lee} YW, {Joo} SJ, et~al. (2009) {The Presence of Two Distinct Red
  Giant Branches in the Globular Cluster NGC 1851}. \apjl 707(2):L190--L194

\bibitem[{{He} et~al.(2020){He}, {Ricotti}, and {Geen}}]{he2020}
{He} CC, {Ricotti} M, and {Geen} S (2020) {Simulating Star Clusters Across
  Cosmic Time: II. Escape Fraction of Ionizing Photons from Molecular Clouds}.
  \mnras p 162

\bibitem[{{Helmi} et~al.(2018){Helmi}, {Babusiaux}, {Koppelman}, {Massari},
  {Veljanoski}, and {Brown}}]{helmi18}
{Helmi} A, {Babusiaux} C, {Koppelman} HH, et~al. (2018) {The merger that led to
  the formation of the Milky Way's inner stellar halo and thick disk}. \nat
  563(7729):85--88

\bibitem[{{Henshaw} et~al.(2016){Henshaw}, {Longmore}, {Kruijssen}, {Davies},
  {Bally}, {Barnes}, {Battersby}, {Burton}, {Cunningham}, {Dale}, {Ginsburg},
  {Immer}, {Jones}, {Kendrew}, {Mills}, {Molinari}, {Moore}, {Ott}, {Pillai},
  {Rathborne}, {Schilke}, {Schmiedeke}, {Testi}, {Walker}, {Walsh}, and
  {Zhang}}]{henshaw16}
{Henshaw} JD, {Longmore} SN, {Kruijssen} JMD, et~al. (2016) {Molecular gas
  kinematics within the central 250 pc of the Milky Way}. \mnras 457:2675--2702

\bibitem[{{Heyer} et~al.(2009){Heyer}, {Krawczyk}, {Duval}, and
  {Jackson}}]{heyer09}
{Heyer} M, {Krawczyk} C, {Duval} J, et~al. (2009) {Re-Examining Larson's
  Scaling Relationships in Galactic Molecular Clouds}. \apj 699:1092--1103

\bibitem[{Heyer et~al.(2009)Heyer, Krawczyk, Duval, and Jackson}]{heyer2009}
Heyer M, Krawczyk C, Duval J, et~al. (2009) {Re-Examining Larson'S Scaling
  Relationships in Galactic Molecular Clouds}. \apj 699(1985):1092--1103

\bibitem[{{Hilditch} et~al.(2005){Hilditch}, {Howarth}, and
  {Harries}}]{Hilditch_05}
{Hilditch} RW, {Howarth} ID, and {Harries} TJ (2005) {Forty eclipsing binaries
  in the Small Magellanic Cloud: fundamental parameters and Cloud distance}.
  \mnras 357(1):304--324

\bibitem[{{Hillenbrand} and {Hartmann}(1998)}]{Hillenbrand_98}
{Hillenbrand} LA and {Hartmann} LW (1998) {A Preliminary Study of the Orion
  Nebula Cluster Structure and Dynamics}. \apj 492:540--553

\bibitem[{{Hodge}(1961{\natexlab{a}})}]{Hodge_61b}
{Hodge} PW (1961{\natexlab{a}}) {The distribution of stars in the Sculptor
  dwarf galaxy}. \aj 66:384

\bibitem[{{Hodge}(1961{\natexlab{b}})}]{Hodge_61a}
{Hodge} PW (1961{\natexlab{b}}) {The Fornax dwarf galaxy. II. The distribution
  of stars}. \aj 66:249

\bibitem[{{Hollyhead} et~al.(2015){Hollyhead}, {Bastian}, {Adamo},
  {Silva-Villa}, {Dale}, {Ryon}, and {Gazak}}]{hollyhead2015}
{Hollyhead} K, {Bastian} N, {Adamo} A, et~al. (2015) {Studying the YMC
  population of M83: how long clusters remain embedded, their interaction with
  the ISM and implications for GC formation theories}. \mnras 449(1):1106--1117

\bibitem[{{Hollyhead} et~al.(2017){Hollyhead}, {Kacharov}, {Lardo}, {Bastian},
  {Hilker}, {Rejkuba}, {Koch}, {Grebel}, and {Georgiev}}]{2017MNRAS.465L..39H}
{Hollyhead} K, {Kacharov} N, {Lardo} C, et~al. (2017) {Evidence for multiple
  populations in the intermediate-age cluster Lindsay 1 in the SMC}. \mnras
  465:L39--L43

\bibitem[{{Hosek} et~al.(2019){Hosek}, {Lu}, {Anderson}, {Najarro}, {Ghez},
  {Morris}, {Clarkson}, and {Albers}}]{Hosek_19}
{Hosek} J Matthew~W, {Lu} JR, {Anderson} J, et~al. (2019) {The Unusual Initial
  Mass Function of the Arches Cluster}. \apj 870(1):44

\bibitem[{{Hughes} et~al.(2013){Hughes}, {Meidt}, {Colombo}, {Schinnerer},
  {Pety}, {Leroy}, {Dobbs}, {Garc{\'{\i}}a-Burillo}, {Thompson}, {Dumas},
  {Schuster}, and {Kramer}}]{hughes2013}
{Hughes} A, {Meidt} SE, {Colombo} D, et~al. (2013) {A Comparative Study of
  Giant Molecular Clouds in M51, M33, and the Large Magellanic Cloud}. \apj
  779:46

\bibitem[{{Hughes} et~al.(2016){Hughes}, {Meidt}, {Colombo}, {Schruba},
  {Schinnerer}, {Leroy}, and {Wong}}]{hughes2016}
{Hughes} A, {Meidt} S, {Colombo} D, et~al. (2016) {Giant Molecular Cloud
  Populations in Nearby Galaxies}. In: {Jablonka} P, {Andr{\'e}} P, and {van
  der Tak} F (eds) From Interstellar Clouds to Star-Forming Galaxies: Universal
  Processes?, IAU Symposium, vol 315, pp 30--37

\bibitem[{{Hughes} et~al.(2019){Hughes}, {Pfeffer}, {Martig}, {Bastian},
  {Crain}, {Kruijssen}, and {Reina-Campos}}]{hughes19}
{Hughes} ME, {Pfeffer} J, {Martig} M, et~al. (2019) {Fossil stellar streams and
  their globular cluster populations in the E-MOSAICS simulations}. \mnras
  482:2795--2806

\bibitem[{{Hunter}(2001)}]{hunter2001}
{Hunter} DA (2001) {The Stellar Population and Star Clusters in the Unusual
  Local Group Galaxy IC 10}. \apj 559(1):225--242

\bibitem[{{Hunter} et~al.(1995){Hunter}, {Shaya}, {Holtzman}, {Light},
  {O'Neil}, and {Lynds}}]{Hunter_95}
{Hunter} DA, {Shaya} EJ, {Holtzman} JA, et~al. (1995) {The Intermediate Stellar
  Mass Population in R136 Determined from Hubble Space Telescope Planetary
  Camera 2 Images}. \apj 448:179

\bibitem[{{Hur} et~al.(2015){Hur}, {Park}, {Sung}, {Bessell}, {Lim}, {Chun},
  and {Sohn}}]{Hur_15}
{Hur} H, {Park} BG, {Sung} H, et~al. (2015) {Reddening, distance, and stellar
  content of the young open cluster Westerlund 2}. \mnras 446:3797--3819

\bibitem[{{Imara} and {Blitz}(2007)}]{Imara_07}
{Imara} N and {Blitz} L (2007) {Extinction in the Large Magellanic Cloud}. \apj
  662(2):969--979

\bibitem[{{Jeffreson} and {Kruijssen}(2018)}]{jeffreson2018}
{Jeffreson} SMR and {Kruijssen} JMD (2018) {A general theory for the lifetimes
  of giant molecular clouds under the influence of galactic dynamics}. \mnras
  476:3688--3715

\bibitem[{{Jerabkova} et~al.(2019){Jerabkova}, {Beccari}, {Boffin},
  {Petr-Gotzens}, {Manara}, {Prada Moroni}, {Tognelli}, and
  {Degl'Innocenti}}]{Jerabkova_19}
{Jerabkova} T, {Beccari} G, {Boffin} HMJ, et~al. (2019) {When the tale comes
  true: multiple populations and wide binaries in the Orion Nebula Cluster}.
  \aap 627:A57

\bibitem[{{Johnson} et~al.(2012){Johnson}, {Seth}, {Dalcanton}, {Caldwell},
  {Fouesneau}, {Gouliermis}, {Hodge}, {Larsen}, {Olsen}, {San Roman},
  {Sarajedini}, {Weisz}, {Williams}, {Beerman}, {Bianchi}, {Dolphin},
  {Girardi}, {Guhathakurta}, {Kalirai}, {Lang}, {Monachesi}, {Nanda}, {Rix},
  and {Skillman}}]{2012ApJ...752...95J}
{Johnson} LC, {Seth} AC, {Dalcanton} JJ, et~al. (2012) {PHAT Stellar Cluster
  Survey. I. Year 1 Catalog and Integrated Photometry}. \apj 752(2):95

\bibitem[{{Johnson} et~al.(2015){Johnson}, {Seth}, {Dalcanton}, {Wallace},
  {Simpson}, {Lintott}, {Kapadia}, {Skillman}, {Caldwell}, {Fouesneau},
  {Weisz}, {Williams}, {Beerman}, {Gouliermis}, and {Sarajedini}}]{johnson2015}
{Johnson} LC, {Seth} AC, {Dalcanton} JJ, et~al. (2015) {PHAT Stellar Cluster
  Survey. II. Andromeda Project Cluster Catalog}. \apj 802(2):127

\bibitem[{{Johnson} et~al.(2016){Johnson}, {Seth}, {Dalcanton}, {Beerman},
  {Fouesneau}, {Lewis}, {Weisz}, {Williams}, {Bell}, {Dolphin}, {Larsen},
  {Sandstrom}, and {Skillman}}]{johnson2016}
{Johnson} LC, {Seth} AC, {Dalcanton} JJ, et~al. (2016) {Panchromatic Hubble
  Andromeda Treasury. XVI. Star Cluster Formation Efficiency and the Clustered
  Fraction of Young Stars}. \apj 827(1):33

\bibitem[{{Johnson} et~al.(2017){Johnson}, {Seth}, {Dalcanton}, {Beerman},
  {Fouesneau}, {Weisz}, {Bell}, {Dolphin}, {Sandstrom}, and
  {Williams}}]{johnson17}
{Johnson} LC, {Seth} AC, {Dalcanton} JJ, et~al. (2017) {Panchromatic Hubble
  Andromeda Treasury. XVIII. The High-mass Truncation of the Star Cluster Mass
  Function}. \apj 839(2):78

\bibitem[{{Johnson} et~al.(2017b){Johnson}, {Sharon}, {Gladders}, {Rigby},
  {Bayliss}, {Wuyts}, {Whitaker}, {Florian}, and {Murray}}]{JohnsonT2017}
{Johnson} TL, {Sharon} K, {Gladders} MD, et~al. (2017b) {Star Formation at z =
  2.481 in the Lensed Galaxy SDSS J1110 = 6459. I. Lens Modeling and Source
  Reconstruction}. \apj 843(2):78

\bibitem[{{Jord{\'a}n} et~al.(2007){Jord{\'a}n}, {McLaughlin}, {C{\^o}t{\'e}},
  {Ferrarese}, {Peng}, {Mei}, {Villegas}, {Merritt}, {Tonry}, and
  {West}}]{jordan2007}
{Jord{\'a}n} A, {McLaughlin} DE, {C{\^o}t{\'e}} P, et~al. (2007) {The ACS Virgo
  Cluster Survey. XII. The Luminosity Function of Globular Clusters in
  Early-Type Galaxies}. \apjs 171(1):101--145

\bibitem[{{Kalari}(2019)}]{Kalari_19}
{Kalari} VM (2019) {Classical T-Tauri stars with VPHAS+: II: NGC 6383 in Sh
  2-012}. \mnras 484(4):5102--5112

\bibitem[{{Kalari} et~al.(2018){Kalari}, {Carraro}, {Evans}, and
  {Rubio}}]{Kalari_18}
{Kalari} VM, {Carraro} G, {Evans} CJ, et~al. (2018) {The Magellanic Bridge
  Cluster NGC 796: Deep Optical AO Imaging Reveals the Stellar Content and
  Initial Mass Function of a Massive Open Cluster}. \apj 857(2):132

\bibitem[{{Kamann} et~al.(2013){Kamann}, {Wisotzki}, and {Roth}}]{Kamann_13}
{Kamann} S, {Wisotzki} L, and {Roth} MM (2013) {Resolving stellar populations
  with crowded field 3D spectroscopy}. \aap 549:A71

\bibitem[{{Katz} and {Ricotti}(2013)}]{katz13}
{Katz} H and {Ricotti} M (2013) {Two epochs of globular cluster formation from
  deep field luminosity functions: implications for reionization and the Milky
  Way satellites}. \mnras 432(4):3250--3261

\bibitem[{{Katz} and {Ricotti}(2014)}]{katz14}
{Katz} H and {Ricotti} M (2014) {Clues on the missing sources of reionization
  from self-consistent modelling of Milky Way and dwarf galaxy globular
  clusters}. \mnras 444(3):2377--2395

\bibitem[{Kawamura et~al.(2009)Kawamura, Mizuno, Minamidani, {D.
  Fillipovi{\'{c}}}, Staveley-Smith, Kim, Mizuno, Onishi, Mizuno, and
  Fukui}]{kawamura2009}
Kawamura A, Mizuno Y, Minamidani T, et~al. (2009) {the Second Survey of the
  Molecular Clouds in the Large Magellanic Cloud By Nanten. Ii. Star
  Formation}. \apjs 184(1):1--17,
  \urlprefix\url{http://stacks.iop.org/0067-0049/184/i=1/a=1?key=crossref.f468cfe0c9f96aee0201d8a9598a8e47}

\bibitem[{{Keller} et~al.(2020){Keller}, {Kruijssen}, {Pfeffer},
  {Reina-Campos}, {Bastian}, {Trujillo-Gomez}, {Hughes}, and
  {Crain}}]{keller20}
{Keller} BW, {Kruijssen} JMD, {Pfeffer} J, et~al. (2020) {Where did the
  globular clusters of the Milky Way form? Insights from the E-MOSAICS
  simulations}. \mnras\ submitted

\bibitem[{{Kennicutt} and {Evans}(2012)}]{kennicuttevans2012}
{Kennicutt} RC and {Evans} NJ (2012) {Star Formation in the Milky Way and
  Nearby Galaxies}. \araa 50:531--608

\bibitem[{{Kharchenko}(2001)}]{Kharchenko_01}
{Kharchenko} NV (2001) {All-sky compiled catalogue of 2.5 million stars}.
  Kinematika i Fizika Nebesnykh Tel 17(5):409--423

\bibitem[{{Kharchenko} et~al.(2005{\natexlab{a}}){Kharchenko}, {Piskunov},
  {R{\"o}ser}, {Schilbach}, and {Scholz}}]{Kharchenko_05b}
{Kharchenko} NV, {Piskunov} AE, {R{\"o}ser} S, et~al. (2005{\natexlab{a}}) {109
  new Galactic open clusters}. \aap 440(1):403--408

\bibitem[{{Kharchenko} et~al.(2005{\natexlab{b}}){Kharchenko}, {Piskunov},
  {R{\"o}ser}, {Schilbach}, and {Scholz}}]{Kharchenko_05a}
{Kharchenko} NV, {Piskunov} AE, {R{\"o}ser} S, et~al. (2005{\natexlab{b}})
  {Astrophysical parameters of Galactic open clusters}. \aap 438(3):1163--1173

\bibitem[{{Kharchenko} et~al.(2012){Kharchenko}, {Piskunov}, {Schilbach},
  {R{\"o}ser}, and {Scholz}}]{Kharchenko_12}
{Kharchenko} NV, {Piskunov} AE, {Schilbach} E, et~al. (2012) {Global survey of
  star clusters in the Milky Way. I. The pipeline and fundamental parameters in
  the second quadrant}. \aap 543:A156

\bibitem[{{Kharchenko} et~al.(2013{\natexlab{a}}){Kharchenko}, {Piskunov},
  {Schilbach}, {R{\"o}ser}, and {Scholz}}]{Kharchenko_13}
{Kharchenko} NV, {Piskunov} AE, {Schilbach} E, et~al. (2013{\natexlab{a}})
  {Global survey of star clusters in the Milky Way. II. The catalogue of basic
  parameters}. \aap 558:A53

\bibitem[{{Kharchenko} et~al.(2013{\natexlab{b}}){Kharchenko}, {Piskunov},
  {Schilbach}, {R{\"o}ser}, and {Scholz}}]{2013A&A...558A..53K}
{Kharchenko} NV, {Piskunov} AE, {Schilbach} E, et~al. (2013{\natexlab{b}})
  {Global survey of star clusters in the Milky Way. II. The catalogue of basic
  parameters}. \aap 558:A53

\bibitem[{{Kharchenko} et~al.(2016){Kharchenko}, {Piskunov}, {Schilbach},
  {R{\"o}ser}, and {Scholz}}]{Kharchenko_16}
{Kharchenko} NV, {Piskunov} AE, {Schilbach} E, et~al. (2016) {Global survey of
  star clusters in the Milky Way. V. Integrated JHK$_{S}$ magnitudes and
  luminosity functions}. \aap 585:A101

\bibitem[{{Kim} et~al.(2017){Kim}, {Ostriker}, and {Raileanu}}]{kim2017}
{Kim} CG, {Ostriker} EC, and {Raileanu} R (2017) {Superbubbles in the
  Multiphase ISM and the Loading of Galactic Winds}. \apj 834(1):25

\bibitem[{{Kim} et~al.(2018{\natexlab{a}}){Kim}, {Kim}, and
  {Ostriker}}]{kim18a}
{Kim} JG, {Kim} WT, and {Ostriker} EC (2018{\natexlab{a}}) {Modeling UV
  Radiation Feedback from Massive Stars. II. Dispersal of Star-forming Giant
  Molecular Clouds by Photoionization and Radiation Pressure}. \apj 859(1):68

\bibitem[{{Kim} et~al.(2018{\natexlab{b}}){Kim}, {Ma}, {Grudi{\'c}}, {Hopkins},
  {Hayward}, {Wetzel}, {Faucher-Gigu{\`e}re}, {Kere{\v s}}, {Garrison-Kimmel},
  and {Murray}}]{kim18}
{Kim} Jh, {Ma} X, {Grudi{\'c}} MY, et~al. (2018{\natexlab{b}}) {Formation of
  globular cluster candidates in merging proto-galaxies at high redshift: a
  view from the FIRE cosmological simulations}. \mnras 474:4232--4244

\bibitem[{{Kimm} et~al.(2016){Kimm}, {Cen}, {Rosdahl}, and {Yi}}]{kimm16}
{Kimm} T, {Cen} R, {Rosdahl} J, et~al. (2016) {Formation of Globular Clusters
  in Atomic-cooling Halos Via Rapid Gas Condensation and Fragmentation during
  the Epoch of Reionization}. \apj 823(1):52

\bibitem[{{King}(1962)}]{1962AJ.....67..471K}
{King} I (1962) {The structure of star clusters. I. an empirical density law}.
  \aj 67:471

\bibitem[{{Koda} et~al.(2009){Koda}, {Scoville}, {Sawada}, {La Vigne}, {Vogel},
  {Potts}, {Carpenter}, {Corder}, {Wright}, {White}, {Zauderer}, {Patience},
  {Sargent}, {Bock}, {Hawkins}, {Hodges}, {Kemball}, {Lamb}, {Plambeck},
  {Pound}, {Scott}, {Teuben}, and {Woody}}]{Koda2009}
{Koda} J, {Scoville} N, {Sawada} T, et~al. (2009) {Dynamically Driven Evolution
  of the Interstellar Medium in M51}. \apjl 700(2):L132--L136

\bibitem[{{Krause} et~al.(2020){Krause}, {Offner}, {Charbonnel}, {Gieles},
  {Klessen}, {Vazquez-Semadeni}, {Ballesteros-Paredes}, {Grichidis},
  {Kruijssen}, {Ward}, and {Zinnecker}}]{Krause_20}
{Krause} GHM, {Offner} SRS, {Charbonnel} C, et~al. (2020) {The Physics of Star
  Cluster Formation and Evolution}

\bibitem[{{Krause} et~al.(2013){Krause}, {Fierlinger}, {Diehl}, {Burkert},
  {Voss}, and {Ziegler}}]{krause2013}
{Krause} M, {Fierlinger} K, {Diehl} R, et~al. (2013) {Feedback by massive stars
  and the emergence of superbubbles. I. Energy efficiency and Vishniac
  instabilities}. \aap 550:A49

\bibitem[{{Kruijssen}(2012)}]{kruijssen2012}
{Kruijssen} JMD (2012) {On the fraction of star formation occurring in bound
  stellar clusters}. \mnras 426(4):3008--3040

\bibitem[{{Kruijssen}(2014)}]{kruijssen14}
{Kruijssen} JMD (2014) {Globular cluster formation in the context of galaxy
  formation and evolution}. Classical and Quantum Gravity 31(24):244006

\bibitem[{{Kruijssen}(2015)}]{Kruijssen_15}
{Kruijssen} JMD (2015) {Globular clusters as the relics of regular star
  formation in `normal' high-redshift galaxies}. \mnras 454:1658--1686

\bibitem[{{Kruijssen}(2019)}]{Kruijssen2019e}
{Kruijssen} JMD (2019) {The minimum metallicity of globular clusters and its
  physical origin -- implications for the galaxy mass-metallicity relation and
  observations of proto-globular clusters at high redshift}. \mnras 486:L20

\bibitem[{{Kruijssen} et~al.(2011){Kruijssen}, {Pelupessy}, {Lamers},
  {Portegies Zwart}, and {Icke}}]{kruijssen11}
{Kruijssen} JMD, {Pelupessy} FI, {Lamers} HJGLM, et~al. (2011) {Modelling the
  formation and evolution of star cluster populations in galaxy simulations}.
  \mnras 414(2):1339--1364

\bibitem[{{Kruijssen} et~al.(2012{\natexlab{a}}){Kruijssen}, {Maschberger},
  {Moeckel}, {Clarke}, {Bastian}, and {Bonnell}}]{kruijssen12b}
{Kruijssen} JMD, {Maschberger} T, {Moeckel} N, et~al. (2012{\natexlab{a}}) {The
  dynamical state of stellar structure in star-forming regions}. \mnras
  419:841--853

\bibitem[{{Kruijssen} et~al.(2012{\natexlab{b}}){Kruijssen}, {Pelupessy},
  {Lamers}, {Portegies Zwart}, {Bastian}, and {Icke}}]{kruijssen12c}
{Kruijssen} JMD, {Pelupessy} FI, {Lamers} HJGLM, et~al. (2012{\natexlab{b}})
  {Formation versus destruction: the evolution of the star cluster population
  in galaxy mergers}. \mnras 421:1927--1941

\bibitem[{{Kruijssen} et~al.(2015){Kruijssen}, {Dale}, and
  {Longmore}}]{kruijssen15b}
{Kruijssen} JMD, {Dale} JE, and {Longmore} SN (2015) {The dynamical evolution
  of molecular clouds near the Galactic Centre - I. Orbital structure and
  evolutionary timeline}. \mnras 447:1059--1079

\bibitem[{{Kruijssen} et~al.(2019{\natexlab{a}}){Kruijssen}, {Dale},
  {Longmore}, {Walker}, {Henshaw}, {Jeffreson}, {Petkova}, {Ginsburg},
  {Barnes}, {Battersby}, {Immer}, {Jackson}, {Keto}, {Krieger}, {Mills},
  {S{\'a}nchez-Monge}, {Schmiedeke}, {Suri}, and {Zhang}}]{kruijssen2019c}
{Kruijssen} JMD, {Dale} JE, {Longmore} SN, et~al. (2019{\natexlab{a}}) {The
  dynamical evolution of molecular clouds near the Galactic Centre - II.
  Spatial structure and kinematics of simulated clouds}. \mnras
  484(4):5734--5754

\bibitem[{{Kruijssen} et~al.(2019{\natexlab{b}}){Kruijssen}, {Schruba},
  {Chevance}, {Longmore}, {Hygate}, {Haydon}, {McLeod}, {Dalcanton}, {Tacconi},
  and {van Dishoeck}}]{Kruijssen2019d}
{Kruijssen} JMD, {Schruba} A, {Chevance} M, et~al. (2019{\natexlab{b}}) {Fast
  and inefficient star formation due to short-lived molecular clouds and rapid
  feedback}. \nat 569:519

\bibitem[{{Kruijssen} et~al.(2019a){Kruijssen}, {Pfeffer}, {Crain}, and
  {Bastian}}]{kruijssen2019a}
{Kruijssen} JMD, {Pfeffer} JL, {Crain} RA, et~al. (2019a) {The E-MOSAICS
  project: tracing galaxy formation and assembly with the age-metallicity
  distribution of globular clusters}. \mnras 486(3):3134--3179

\bibitem[{{Kruijssen} et~al.(2019b){Kruijssen}, {Pfeffer}, {Reina-Campos},
  {Crain}, and {Bastian}}]{kruijssen2019b}
{Kruijssen} JMD, {Pfeffer} JL, {Reina-Campos} M, et~al. (2019b) {The formation
  and assembly history of the Milky Way revealed by its globular cluster
  population}. \mnras 486(3):3180--3202

\bibitem[{{Kruijssen} et~al.(2020){Kruijssen}, {Pfeffer}, {Chevance}, {Bonaca},
  {Trujillo-Gomez}, {Bastian}, {Reina-Campos}, {Crain}, and
  {Hughes}}]{kruijssen20}
{Kruijssen} JMD, {Pfeffer} JL, {Chevance} M, et~al. (2020) {Kraken reveals
  itself -- the merger history of the Milky Way reconstructed with the
  E-MOSAICS simulations}. \mnras submitted arXiv:2003.01119

\bibitem[{{Krumholz} and {McKee}(2019)}]{Krumholz_19}
{Krumholz} MR and {McKee} CF (2019) {How do bound star clusters form?} arXiv
  e-prints arXiv:1909.01565

\bibitem[{{Krumholz} et~al.(2019){Krumholz}, {McKee}, and {Bland
  -Hawthorn}}]{krumholz_araa19}
{Krumholz} MR, {McKee} CF, and {Bland -Hawthorn} J (2019) {Star Clusters Across
  Cosmic Time}. \araa 57:227--303

\bibitem[{{Kudritzki} and {Puls}(2000)}]{Kudritzki_00}
{Kudritzki} RP and {Puls} J (2000) {Winds from Hot Stars}. \araa 38:613--666

\bibitem[{{Kuhn} et~al.(2014){Kuhn}, {Feigelson}, {Getman}, {Baddeley},
  {Broos}, {Sills}, {Bate}, {Povich}, {Luhman}, {Busk}, {Naylor}, and
  {King}}]{Kuhn_14}
{Kuhn} MA, {Feigelson} ED, {Getman} KV, et~al. (2014) {The Spatial Structure of
  Young Stellar Clusters. I. Subclusters}. \apj 787(2):107

\bibitem[{{Lada} and {Lada}(2003)}]{Lada_03}
{Lada} CJ and {Lada} EA (2003) {Embedded Clusters in Molecular Clouds}. \araa
  41:57--115

\bibitem[{{Lada} et~al.(1984){Lada}, {Margulis}, and {Dearborn}}]{Lada_84a}
{Lada} CJ, {Margulis} M, and {Dearborn} D (1984) {The formation and early
  dynamical evolution of bound stellar systems}. \apj 285:141--152

\bibitem[{{Lagioia} et~al.(2019){Lagioia}, {Milone}, {Marino}, and
  {Dotter}}]{Lagioia_etal19}
{Lagioia} EP, {Milone} AP, {Marino} AF, et~al. (2019) {Helium Variation in Four
  Small Magellanic Cloud Globular Clusters}. \apj 871(2):140

\bibitem[{{Lah{\'e}n} et~al.(2019){Lah{\'e}n}, {Naab}, {Johansson},
  {Elmegreen}, {Hu}, {Walch}, {Steinwand el}, and {Moster}}]{lahen2019}
{Lah{\'e}n} N, {Naab} T, {Johansson} PH, et~al. (2019) {The GRIFFIN project --
  Formation of star clusters with individual massive stars in a simulated dwarf
  galaxy starburst}. arXiv e-prints arXiv:1911.05093

\bibitem[{{Lamers} et~al.(2005{\natexlab{a}}){Lamers}, {Gieles}, {Bastian},
  {Baumgardt}, {Kharchenko}, and {Portegies Zwart}}]{lamers05b}
{Lamers} HJGLM, {Gieles} M, {Bastian} N, et~al. (2005{\natexlab{a}}) {An
  analytical description of the disruption of star clusters in tidal fields
  with an application to Galactic open clusters}. \aap 441:117--129

\bibitem[{{Lamers} et~al.(2005{\natexlab{b}}){Lamers}, {Gieles}, and {Portegies
  Zwart}}]{lamers05}
{Lamers} HJGLM, {Gieles} M, and {Portegies Zwart} SF (2005{\natexlab{b}})
  {Disruption time scales of star clusters in different galaxies}. \aap
  429:173--179

\bibitem[{{Lamers} et~al.(2017){Lamers}, {Kruijssen}, {Bastian}, {Rejkuba},
  {Hilker}, and {Kissler-Patig}}]{lamers2017}
{Lamers} HJGLM, {Kruijssen} JMD, {Bastian} N, et~al. (2017) {The difference in
  metallicity distribution functions of halo stars and globular clusters as a
  function of galaxy type. A tracer of globular cluster formation and
  evolution}. \aap 606:A85

\bibitem[{{Larsen}(2002)}]{larsen2002}
{Larsen} SS (2002) {The Luminosity Function of Star Clusters in Spiral
  Galaxies}. \aj 124(3):1393--1409

\bibitem[{{Larsen}(2004)}]{2004A&A...416..537L}
{Larsen} SS (2004) {The structure and environment of young stellar clusters in
  spiral galaxies}. \aap 416:537--553

\bibitem[{{Larsen}(2006)}]{larsen2006}
{Larsen} SS (2006) {Young Massive Clusters - Formation Efficiencies and
  (Initial) Mass Functions}. arXiv e-prints astro-ph/0606625

\bibitem[{{Larsen}(2009)}]{larsen2009}
{Larsen} SS (2009) {The mass function of young star clusters in spiral
  galaxies}. \aap 494(2):539--551

\bibitem[{{Larson}(1981)}]{larson81}
{Larson} RB (1981) {Turbulence and star formation in molecular clouds.} \mnras
  194:809--826

\bibitem[{{Leaman} et~al.(2013){Leaman}, {VandenBerg}, and {Mendel}}]{leaman13}
{Leaman} R, {VandenBerg} DA, and {Mendel} JT (2013) {The bifurcated
  age-metallicity relation of Milky Way globular clusters and its implications
  for the accretion history of the galaxy}. \mnras 436:122--135

\bibitem[{{Lee} et~al.(2005){Lee}, {Rolleston}, {Dufton}, and {Ryans}}]{Lee_05}
{Lee} JK, {Rolleston} WRJ, {Dufton} PL, et~al. (2005) {Chemical compositions of
  four B-type supergiants in the SMC wing}. \aap 429:1025--1030

\bibitem[{{Lennon} et~al.(2018){Lennon}, {Evans}, {van der Marel}, {Anderson},
  {Platais}, {Herrero}, {de Mink}, {Sana}, {Sabbi}, {Bedin}, {Crowther},
  {Langer}, {Ramos Lerate}, {del Pino}, {Renzo}, {Sim{\'o}n-D{\'\i}az}, and
  {Schneider}}]{Lennon_18}
{Lennon} DJ, {Evans} CJ, {van der Marel} RP, et~al. (2018) {Gaia DR2 reveals a
  very massive runaway star ejected from R136}. \aap 619:A78

\bibitem[{Leroy et~al.(2013)Leroy, Walter, Sandstrom, Schruba, Munoz-Mateos,
  Bigiel, Bolatto, Brinks, de~Blok, Meidt, Rix, Rosolowsky, Schinnerer,
  Schuster, and Usero}]{leroy2013}
Leroy AK, Walter F, Sandstrom K, et~al. (2013) {Molecular Gas and Star
  Formation in Nearby Disk Galaxies}. \aj 146(2):19,
  \urlprefix\url{http://adsabs.harvard.edu/abs/2013AJ....146...19L{\%}5Cnhttp://stacks.iop.org/1538-3881/146/i=2/a=19}

\bibitem[{{Leroy} et~al.(2016){Leroy}, {Hughes}, {Schruba}, {Rosolowsky},
  {Blanc}, {Bolatto}, {Colombo}, {Escala}, {Kramer}, {Kruijssen}, {Meidt},
  {Pety}, {Querejeta}, {Sandstrom}, {Schinnerer}, {Sliwa}, and
  {Usero}}]{leroy2016}
{Leroy} AK, {Hughes} A, {Schruba} A, et~al. (2016) {A Portrait of Cold Gas in
  Galaxies at 60 pc Resolution and a Simple Method to Test Hypotheses That Link
  Small-scale ISM Structure to Galaxy-scale Processes}. \apj 831(1):16

\bibitem[{{Li} and {Gnedin}(2019)}]{li2019}
{Li} H and {Gnedin} OY (2019) {Star cluster formation in cosmological
  simulations - III. Dynamical and chemical evolution}. \mnras
  486(3):4030--4043

\bibitem[{{Li} et~al.(2017){Li}, {Gnedin}, {Gnedin}, {Meng}, {Semenov}, and
  {Kravtsov}}]{Li2017}
{Li} H, {Gnedin} OY, {Gnedin} NY, et~al. (2017) {Star Cluster Formation in
  Cosmological Simulations. I. Properties of Young Clusters}. \apj 834(1):69

\bibitem[{{Li} et~al.(2018){Li}, {Gnedin}, and {Gnedin}}]{Li2018}
{Li} H, {Gnedin} OY, and {Gnedin} NY (2018) {Star Cluster Formation in
  Cosmological Simulations. II. Effects of Star Formation Efficiency and
  Stellar Feedback}. \apj 861(2):107

\bibitem[{{Lim} and {Lee}(2015)}]{lim2015}
{Lim} S and {Lee} MG (2015) {The Star Cluster System in the Local Group
  Starburst Galaxy IC 10}. \apj 804(2):123

\bibitem[{{Linden} et~al.(2017){Linden}, {Evans}, {Rich}, {Larson}, {Armus},
  {D{\'\i}az-Santos}, {Privon}, {Howell}, {Inami}, {Kim}, {Chien}, {Vavilkin},
  {Mazzarella}, {Modica}, {Surace}, {Manning}, {Abdullah}, {Blake}, {Yarber},
  and {Lambert}}]{linden2017}
{Linden} ST, {Evans} AS, {Rich} J, et~al. (2017) {Massive Star Cluster
  Formation and Destruction in Luminous Infrared Galaxies in GOALS}. \apj
  843(2):91

\bibitem[{{Liu} and {Pang}(2019)}]{Liu_19}
{Liu} L and {Pang} X (2019) {A Catalog of Newly Identified Star Clusters in
  Gaia DR2}. \apjs 245(2):32

\bibitem[{{Lombardi} et~al.(2015){Lombardi}, {Alves}, and {Lada}}]{lombardi15}
{Lombardi} M, {Alves} J, and {Lada} CJ (2015) {Molecular clouds have power-law
  probability distribution functions}. \aap 576:L1

\bibitem[{{Longmore} et~al.(2014){Longmore}, {Kruijssen}, {Bastian}, {Bally},
  {Rathborne}, {Testi}, {Stolte}, {Dale}, {Bressert}, and
  {Alves}}]{Longmore_14}
{Longmore} SN, {Kruijssen} JMD, {Bastian} N, et~al. (2014) {The Formation and
  Early Evolution of Young Massive Clusters}. Protostars and Planets VI pp
  291--314

\bibitem[{{Ma} et~al.(2019){Ma}, {Grudi{\'c}}, {Quataert}, {Hopkins},
  {Faucher-Gigu{\`e}re}, {Boylan-Kolchin}, {Wetzel}, {Kim}, {Murray}, and
  {Kere{\v{s}}}}]{ma19}
{Ma} X, {Grudi{\'c}} MY, {Quataert} E, et~al. (2019) {Self-consistent
  proto-globular cluster formation in cosmological simulations of high-redshift
  galaxies}. \mnras\ submitted arXiv:1906.11261

\bibitem[{{Ma} et~al.(2020){Ma}, {Quataert}, {Wetzel}, {Hopkins},
  {Faucher-Gigu{\`e}re}, and {Kere{\v{s}}}}]{ma2020}
{Ma} X, {Quataert} E, {Wetzel} A, et~al. (2020) {No missing photons for
  reionization: moderate ionizing photon escape fractions from the FIRE-2
  simulations}. arXiv e-prints arXiv:2003.05945

\bibitem[{{Madau} et~al.(2019){Madau}, {Lupi}, {Diemand}, {Burkert}, and
  {Lin}}]{madau19}
{Madau} P, {Lupi} A, {Diemand} J, et~al. (2019) {Globular Cluster Formation
  from Colliding Substructure}. \apj\ submitted arXiv:1905.08951

\bibitem[{{Maeder} and {Meynet}(2006)}]{2006A&A...448L..37M}
{Maeder} A and {Meynet} G (2006) {On the origin of the high helium sequence in
  {$\omega$} Centauri}. \aap 448:L37--L41

\bibitem[{{Mar{\'{\i}}n-Franch} et~al.(2009){Mar{\'{\i}}n-Franch}, {Aparicio},
  {Piotto}, {Rosenberg}, {Chaboyer}, {Sarajedini}, {Siegel}, {Anderson},
  {Bedin}, {Dotter}, {Hempel}, {King}, {Majewski}, {Milone}, {Paust}, and
  {Reid}}]{marinfranch09}
{Mar{\'{\i}}n-Franch} A, {Aparicio} A, {Piotto} G, et~al. (2009) {The ACS
  Survey of Galactic Globular Clusters. VII. Relative Ages}. \apj
  694:1498--1516

\bibitem[{{Marino} et~al.(2015){Marino}, {Milone}, {Karakas}, {Casagrand e},
  {Yong}, {Shingles}, {Da Costa}, {Norris}, {Stetson}, {Lind}, {Asplund},
  {Collet}, {Jerjen}, {Sbordone}, {Aparicio}, and {Cassisi}}]{Marino_etal15}
{Marino} AF, {Milone} AP, {Karakas} AI, et~al. (2015) {Iron and s-elements
  abundance variations in NGC 5286: comparison with `anomalous' globular
  clusters and Milky Way satellites}. \mnras 450(1):815--845

\bibitem[{{Marino} et~al.(2018){Marino}, {Yong}, {Milone}, {Piotto},
  {Lundquist}, {Bedin}, {Chen{\'e}}, {Da Costa}, {Asplund}, and
  {Jerjen}}]{Marino_etal18}
{Marino} AF, {Yong} D, {Milone} AP, et~al. (2018) {Metallicity Variations in
  the Type II Globular Cluster NGC 6934}. \apj 859(2):81

\bibitem[{{Martins} et~al.(2020){Martins}, {Schaerer}, {Haemmerl{\'e}}, and
  {Charbonnel}}]{Martins_etal2020}
{Martins} F, {Schaerer} D, {Haemmerl{\'e}} L, et~al. (2020) {Spectral
  properties and detectability of supermassive stars in protoglobular clusters
  at high redshift}. \aap 633:A9

\bibitem[{{Martocchia} et~al.(2018{\natexlab{a}}){Martocchia}, {Cabrera-Ziri},
  {Lardo}, {Dalessandro}, {Bastian}, {Kozhurina-Platais}, {Usher},
  {Niederhofer}, {Cordero}, {Geisler}, {Hollyhead}, {Kacharov}, {Larsen}, {Li},
  {Mackey}, {Hilker}, {Mucciarelli}, {Platais}, and
  {Salaris}}]{2018MNRAS.473.2688M}
{Martocchia} S, {Cabrera-Ziri} I, {Lardo} C, et~al. (2018{\natexlab{a}}) {Age
  as a major factor in the onset of multiple populations in stellar clusters}.
  \mnras 473:2688--2700

\bibitem[{{Martocchia} et~al.(2018{\natexlab{b}}){Martocchia}, {Niederhofer},
  {Dalessandro}, {Bastian}, {Kacharov}, {Usher}, {Cabrera-Ziri}, {Lardo},
  {Cassisi}, {Geisler}, {Hilker}, {Hollyhead}, {Kozhurina-Platais}, {Larsen},
  {Mackey}, {Mucciarelli}, {Platais}, and {Salaris}}]{2018MNRAS.477.4696M}
{Martocchia} S, {Niederhofer} F, {Dalessandro} E, et~al. (2018{\natexlab{b}})
  {The search for multiple populations in Magellanic Cloud clusters - IV.
  Coeval multiple stellar populations in the young star cluster NGC 1978}.
  \mnras 477(4):4696--4705

\bibitem[{{Massari} et~al.(2019){Massari}, {Koppelman}, and
  {Helmi}}]{massari19}
{Massari} D, {Koppelman} HH, and {Helmi} A (2019) {Origin of the system of
  globular clusters in the Milky Way}. \aap 630:L4

\bibitem[{{Masseron} et~al.(2019){Masseron}, {Garc{\'\i}a-Hern{\'a}ndez},
  {M{\'e}sz{\'a}ros}, {Zamora}, {Dell'Agli}, {Allende Prieto}, {Edvardsson},
  {Shetrone}, {Plez}, {Fern{\'a}ndez-Trincado}, {Cunha}, {J{\"o}nsson},
  {Geisler}, {Beers}, and {Cohen}}]{Masseron_etal2019}
{Masseron} T, {Garc{\'\i}a-Hern{\'a}ndez} DA, {M{\'e}sz{\'a}ros} S, et~al.
  (2019) {Homogeneous analysis of globular clusters from the APOGEE survey with
  the BACCHUS code. I. The northern clusters}. \aap 622:A191

\bibitem[{{McCaughrean} and {O'dell}(1996)}]{McCaughrean96}
{McCaughrean} MJ and {O'dell} CR (1996) {Direct Imaging of Circumstellar Disks
  in the Orion Nebula}. \aj 111:1977

\bibitem[{{McLeod} et~al.(2015){McLeod}, {Dale}, {Ginsburg}, {Ercolano},
  {Gritschneder}, {Ramsay}, and {Testi}}]{McLeod_15}
{McLeod} AF, {Dale} JE, {Ginsburg} A, et~al. (2015) {The Pillars of Creation
  revisited with MUSE: gas kinematics and high-mass stellar feedback traced by
  optical spectroscopy}. \mnras 450(1):1057--1076

\bibitem[{{McLeod} et~al.(2016{\natexlab{a}}){McLeod}, {Gritschneder}, {Dale},
  {Ginsburg}, {Klaassen}, {Mottram}, {Preibisch}, {Ramsay}, {Reiter}, and
  {Testi}}]{McLeod_16b}
{McLeod} AF, {Gritschneder} M, {Dale} JE, et~al. (2016{\natexlab{a}})
  {Connecting the dots: a correlation between ionizing radiation and cloud
  mass-loss rate traced by optical integral field spectroscopy}. \mnras
  462(4):3537--3569

\bibitem[{{McLeod} et~al.(2016{\natexlab{b}}){McLeod}, {Weilbacher},
  {Ginsburg}, {Dale}, {Ramsay}, and {Testi}}]{McLeod_16a}
{McLeod} AF, {Weilbacher} PM, {Ginsburg} A, et~al. (2016{\natexlab{b}}) {A
  nebular analysis of the central Orion nebula with MUSE}. \mnras
  455(4):4057--4086

\bibitem[{{McLeod} et~al.(2018){McLeod}, {Reiter}, {Kuiper}, {Klaassen}, and
  {Evans}}]{McLeod_18}
{McLeod} AF, {Reiter} M, {Kuiper} R, et~al. (2018) {A parsec-scale optical jet
  from a massive young star in the Large Magellanic Cloud}. \nat
  554(7692):334--336

\bibitem[{{McMillan} et~al.(2007){McMillan}, {Vesperini}, and {Portegies
  Zwart}}]{McMillan_07}
{McMillan} SLW, {Vesperini} E, and {Portegies Zwart} SF (2007) {A Dynamical
  Origin for Early Mass Segregation in Young Star Clusters}. \apjl 655:L45--L49

\bibitem[{{Meingast} and {Alves}(2019)}]{meingast19}
{Meingast} S and {Alves} J (2019) {Extended stellar systems in the solar
  neighborhood. I. The tidal tails of the Hyades}. \aap 621:L3

\bibitem[{{Messa} et~al.(2018a){Messa}, {Adamo}, {{\"O}stlin}, {Calzetti},
  {Grasha}, {Grebel}, {Shabani}, {Chandar}, {Dale}, {Dobbs}, {Elmegreen},
  {Fumagalli}, {Gouliermis}, {Kim}, {Smith}, {Thilker}, {Tosi}, {Ubeda},
  {Walterbos}, {Whitmore}, {Fedorenko}, {Mahadevan}, {Andrews}, {Bright},
  {Cook}, {Kahre}, {Nair}, {Pellerin}, {Ryon}, {Ahmad}, {Beale}, {Brown},
  {Clarkson}, {Guidarelli}, {Parziale}, {Turner}, and {Weber}}]{messa2018a}
{Messa} M, {Adamo} A, {{\"O}stlin} G, et~al. (2018a) {The young star cluster
  population of M51 with LEGUS - I. A comprehensive study of cluster formation
  and evolution}. \mnras 473(1):996--1018

\bibitem[{{Messa} et~al.(2018b){Messa}, {Adamo}, {Calzetti}, {Reina-Campos},
  {Colombo}, {Schinnerer}, {Chand ar}, {Dale}, {Gouliermis}, {Grasha},
  {Grebel}, {Elmegreen}, {Fumagalli}, {Johnson}, {Kruijssen}, {{\"O}stlin},
  {Shabani}, {Smith}, and {Whitmore}}]{messa2018b}
{Messa} M, {Adamo} A, {Calzetti} D, et~al. (2018b) {The young star cluster
  population of M51 with LEGUS - II. Testing environmental dependences}. \mnras
  477(2):1683--1707

\bibitem[{{Messa} et~al.(2019){Messa}, {Adamo}, {{\"O}stlin}, {Melinder},
  {Hayes}, {Bridge}, and {Cannon}}]{messa2019}
{Messa} M, {Adamo} A, {{\"O}stlin} G, et~al. (2019) {Star-forming clumps in the
  Lyman Alpha Reference Sample of galaxies - I. Photometric analysis and
  clumpiness}. \mnras 487(3):4238--4260

\bibitem[{{M{\'e}sz{\'a}ros} et~al.(2019){M{\'e}sz{\'a}ros}, {Masseron},
  {Garc{\'\i}a-Hern{\'a}ndez}, {Allende Prieto}, {Beers}, {Bizyaev},
  {Chojnowski}, {Cohen}, {Cunha}, {Dell'Agli}, {Ebelke},
  {Fern{\'a}ndez-Trincado}, {Frinchaboy}, {Geisler}, {Hasselquist}, {Hearty},
  {Holtzman}, {Johnson}, {Lane}, {Lacerna}, {Longa-Pe{\~n}a}, {Majewski},
  {Martell}, {Minniti}, {Nataf}, {Nidever}, {Pan}, {Schiavon}, {Shetrone},
  {Smith}, {Sobeck}, {Stringfellow}, {Szigeti}, {Tang}, {Wilson}, and
  {Zamora}}]{2019MNRAS.tmp.3134M}
{M{\'e}sz{\'a}ros} S, {Masseron} T, {Garc{\'\i}a-Hern{\'a}ndez} DA, et~al.
  (2019) {Homogeneous analysis of globular clusters from the APOGEE survey with
  the BACCHUS code. II. The Southern clusters and overview}. \mnras p 3134

\bibitem[{{Meurer} et~al.(1995){Meurer}, {Heckman}, {Leitherer}, {Kinney},
  {Robert}, and {Garnett}}]{meurer1995}
{Meurer} GR, {Heckman} TM, {Leitherer} C, et~al. (1995) {Starbursts and Star
  Clusters in the Ultraviolet}. \aj 110:2665

\bibitem[{{Miholics} et~al.(2016){Miholics}, {Webb}, and {Sills}}]{miholics16}
{Miholics} M, {Webb} JJ, and {Sills} A (2016) {The dynamical evolution of
  accreted star clusters in the Milky Way}. \mnras 456(1):240--247

\bibitem[{{Miholics} et~al.(2017){Miholics}, {Kruijssen}, and
  {Sills}}]{miholics17}
{Miholics} M, {Kruijssen} JMD, and {Sills} A (2017) {A tight relation between
  the age distributions of stellar clusters and the properties of the
  interstellar medium in the host galaxy}. \mnras 470:1421--1435

\bibitem[{{Milone} et~al.(2014){Milone}, {Marino}, {Dotter}, {Norris},
  {Jerjen}, {Piotto}, {Cassisi}, {Bedin}, {Recio Blanco}, {Sarajedini},
  {Asplund}, {Monelli}, and {Aparicio}}]{2014ApJ...785...21M}
{Milone} AP, {Marino} AF, {Dotter} A, et~al. (2014) {Global and Nonglobal
  Parameters of Horizontal-branch Morphology of Globular Clusters}. \apj 785:21

\bibitem[{{Milone} et~al.(2015){Milone}, {Marino}, {Piotto}, {Bedin},
  {Anderson}, {Renzini}, {King}, {Bellini}, {Brown}, {Cassisi}, {D'Antona},
  {Jerjen}, {Nardiello}, {Salaris}, {Marel}, {Vesperini}, {Yong}, {Aparicio},
  {Sarajedini}, and {Zoccali}}]{Milone_etal15}
{Milone} AP, {Marino} AF, {Piotto} G, et~al. (2015) {The Hubble Space Telescope
  UV Legacy Survey of galactic globular clusters - II. The seven stellar
  populations of NGC 7089 (M2)$^{{\ensuremath{\star}}}$}. \mnras
  447(1):927--938

\bibitem[{{Milone} et~al.(2017){Milone}, {Piotto}, {Renzini}, {Marino},
  {Bedin}, {Vesperini}, {D'Antona}, {Nardiello}, {Anderson}, {King}, {Yong},
  {Bellini}, {Aparicio}, {Barbuy}, {Brown}, {Cassisi}, {Ortolani}, {Salaris},
  {Sarajedini}, and {van der Marel}}]{Milone_etal17}
{Milone} AP, {Piotto} G, {Renzini} A, et~al. (2017) {The Hubble Space Telescope
  UV Legacy Survey of Galactic globular clusters - IX. The Atlas of multiple
  stellar populations}. \mnras 464(3):3636--3656

\bibitem[{{Milone} et~al.(2018){Milone}, {Marino}, {Renzini}, {D'Antona},
  {Anderson}, {Barbuy}, {Bedin}, {Bellini}, {Brown}, {Cassisi}, {Cordoni},
  {Lagioia}, {Nardiello}, {Ortolani}, {Piotto}, {Sarajedini}, {Tailo}, {van der
  Marel}, and {Vesperini}}]{Milone_etal18}
{Milone} AP, {Marino} AF, {Renzini} A, et~al. (2018) {The Hubble Space
  Telescope UV legacy survey of galactic globular clusters - XVI. The helium
  abundance of multiple populations}. \mnras 481(4):5098--5122

\bibitem[{{Miville-Desch{\^e}nes} et~al.(2010){Miville-Desch{\^e}nes},
  {Martin}, {Abergel}, {Bernard}, {Boulanger}, {Lagache}, {Anderson},
  {Andr{\'e}}, {Arab}, {Baluteau}, {Blagrave}, {Bontemps}, {Cohen},
  {Compiegne}, {Cox}, {Dartois}, {Davis}, {Emery}, {Fulton}, {Gry}, {Habart},
  {Huang}, {Joblin}, {Jones}, {Kirk}, {Lim}, {Madden}, {Makiwa}, {Menshchikov},
  {Molinari}, {Moseley}, {Motte}, {Naylor}, {Okumura}, {Pinheiro
  Gon{\c{c}}alves}, {Polehampton}, {Rod{\'o}n}, {Russeil}, {Saraceno},
  {Schneider}, {Sidher}, {Spencer}, {Swinyard}, {Ward-Thompson}, {White}, and
  {Zavagno}}]{MivilleDeschenes2010}
{Miville-Desch{\^e}nes} MA, {Martin} PG, {Abergel} A, et~al. (2010)
  {Herschel-SPIRE observations of the Polaris flare: Structure of the diffuse
  interstellar medium at the sub-parsec scale}. \aap 518:L104

\bibitem[{{Mok} et~al.(2019){Mok}, {Chandar}, and {Fall}}]{mok2019}
{Mok} A, {Chandar} R, and {Fall} SM (2019) {Constraints on Upper Cutoffs in the
  Mass Functions of Young Star Clusters}. \apj 872(1):93

\bibitem[{{Mok} et~al.(2020){Mok}, {Chandar}, and {Fall}}]{mok2020}
{Mok} A, {Chandar} R, and {Fall} SM (2020) {Mass Functions of Giant Molecular
  Clouds and Young Star Clusters in Six Nearby Galaxies}. arXiv e-prints
  arXiv:2004.02698

\bibitem[{{Muench} et~al.(2008){Muench}, {Getman}, {Hillenbrand}, and
  {Preibisch}}]{Muench_08}
{Muench} A, {Getman} K, {Hillenbrand} L, et~al. (2008) {Star Formation in the
  Orion Nebula I: Stellar Content}, vol~4, p 483

\bibitem[{{Myeong} et~al.(2018){Myeong}, {Evans}, {Belokurov}, {Sand ers}, and
  {Koposov}}]{myeong18}
{Myeong} GC, {Evans} NW, {Belokurov} V, et~al. (2018) {The Sausage Globular
  Clusters}. \apjl 863(2):L28

\bibitem[{{Norris}(2004)}]{2004ApJ...612L..25N}
{Norris} JE (2004) {The Helium Abundances of {$\omega$} Centauri}. \apjl
  612:L25--L28

\bibitem[{{O'Dell} and {Henney}(2008)}]{O'Dell_08b}
{O'Dell} CR and {Henney} WJ (2008) {High Spatial Velocity Features in the Orion
  Nebula,}. \aj 136(4):1566--1586

\bibitem[{{O'Dell} et~al.(2008){O'Dell}, {Muench}, {Smith}, and
  {Zapata}}]{O'Dell_08a}
{O'Dell} CR, {Muench} A, {Smith} N, et~al. (2008) {Star Formation in the Orion
  Nebula II: Gas, Dust, Proplyds and Outflows}, vol~4, p 544

\bibitem[{{Offner} et~al.(2014){Offner}, {Clark}, {Hennebelle}, {Bastian},
  {Bate}, {Hopkins}, {Moraux}, and {Whitworth}}]{Offner_14}
{Offner} SSR, {Clark} PC, {Hennebelle} P, et~al. (2014) {The Origin and
  Universality of the Stellar Initial Mass Function}. In: {Beuther} H,
  {Klessen} RS, {Dullemond} CP, et~al. (eds) Protostars and Planets VI, p~53

\bibitem[{{Oka} et~al.(2001){Oka}, {Hasegawa}, {Sato}, {Tsuboi}, {Miyazaki},
  and {Sugimoto}}]{oka01}
{Oka} T, {Hasegawa} T, {Sato} F, et~al. (2001) {Statistical Properties of
  Molecular Clouds in the Galactic Center}. \apj 562:348--362

\bibitem[{{Pancino} et~al.(2017){Pancino}, {Bellazzini}, {Giuffrida}, and
  {Marinoni}}]{2017MNRAS.467..412P}
{Pancino} E, {Bellazzini} M, {Giuffrida} G, et~al. (2017) {Globular clusters
  with Gaia}. \mnras 467:412--427

\bibitem[{{Pang} et~al.(2013){Pang}, {Grebel}, {Allison}, {Goodwin}, {Altmann},
  {Harbeck}, {Moffat}, and {Drissen}}]{Pang_13}
{Pang} X, {Grebel} EK, {Allison} RJ, et~al. (2013) {On the Origin of Mass
  Segregation in NGC 3603}. \apj 764:73

\bibitem[{{Parker} et~al.(2014){Parker}, {Wright}, {Goodwin}, and
  {Meyer}}]{Parker_14}
{Parker} RJ, {Wright} NJ, {Goodwin} SP, et~al. (2014) {Dynamical evolution of
  star-forming regions}. \mnras 438:620--638

\bibitem[{{Parker} et~al.(2016){Parker}, {Goodwin}, {Wright}, {Meyer}, and
  {Quanz}}]{parker16}
{Parker} RJ, {Goodwin} SP, {Wright} NJ, et~al. (2016) {Mass segregation in star
  clusters is not energy equipartition}. \mnras 459(1):L119--L123

\bibitem[{{Pellegrini} et~al.(2012){Pellegrini}, {Oey}, {Winkler}, {Points},
  {Smith}, {Jaskot}, and {Zastrow}}]{pellegrini2012}
{Pellegrini} EW, {Oey} MS, {Winkler} PF, et~al. (2012) {The Optical Depth of H
  II Regions in the Magellanic Clouds}. \apj 755(1):40

\bibitem[{{Pfalzner}(2009)}]{2009A&A...498L..37P}
{Pfalzner} S (2009) {Universality of young cluster sequences}. \aap
  498:L37--L40

\bibitem[{{Pfeffer} et~al.(2018){Pfeffer}, {Kruijssen}, {Crain}, and
  {Bastian}}]{pfeffer2018}
{Pfeffer} J, {Kruijssen} JMD, {Crain} RA, et~al. (2018) {The E-MOSAICS project:
  simulating the formation and co-evolution of galaxies and their star cluster
  populations}. \mnras 475(4):4309--4346

\bibitem[{{Pfeffer} et~al.(2019a){Pfeffer}, {Bastian}, {Crain}, {Kruijssen},
  {Hughes}, and {Reina-Campos}}]{pfeffer2019a}
{Pfeffer} J, {Bastian} N, {Crain} RA, et~al. (2019a) {The evolution of the UV
  luminosity function of globular clusters in the E-MOSAICS simulations}.
  \mnras 487(4):4550--4564

\bibitem[{{Pfeffer} et~al.(2019b){Pfeffer}, {Bastian}, {Kruijssen},
  {Reina-Campos}, {Crain}, and {Usher}}]{pfeffer2019b}
{Pfeffer} J, {Bastian} N, {Kruijssen} JMD, et~al. (2019b) {Young star cluster
  populations in the E-MOSAICS simulations}. \mnras 490(2):1714--1733

\bibitem[{{Pfeffer} et~al.(2020){Pfeffer}, {Trujillo-Gomez}, {Kruijssen},
  {Crain}, {Hughes}, {Reina-Campos}, and {Bastian}}]{pfeffer20}
{Pfeffer} JL, {Trujillo-Gomez} S, {Kruijssen} JMD, et~al. (2020) {Predicting
  accreted satellite galaxy masses and accretion redshifts based on globular
  cluster orbits in the E-MOSAICS simulations}. \mnras submitted
  arXiv:2003.00076

\bibitem[{{Phipps} et~al.(2019){Phipps}, {Khochfar}, {Varri}, and {Dalla
  Vecchia}}]{phipps19}
{Phipps} F, {Khochfar} S, {Varri} AL, et~al. (2019) {The First Billion Years
  Project: Finding Infant Globular Clusters at z=6}. \mnras\ submitted
  arXiv:1910.09924

\bibitem[{{Piotto} et~al.(2007){Piotto}, {Bedin}, {Anderson}, {King},
  {Cassisi}, {Milone}, {Villanova}, {Pietrinferni}, and
  {Renzini}}]{Piotto_etal2007}
{Piotto} G, {Bedin} LR, {Anderson} J, et~al. (2007) {A Triple Main Sequence in
  the Globular Cluster NGC 2808}. \apjl 661(1):L53--L56

\bibitem[{{Piotto} et~al.(2012){Piotto}, {Milone}, {Anderson}, {Bedin},
  {Bellini}, {Cassisi}, {Marino}, {Aparicio}, and
  {Nascimbeni}}]{Piotto_etal2012}
{Piotto} G, {Milone} AP, {Anderson} J, et~al. (2012) {Hubble Space Telescope
  Reveals Multiple Sub-giant Branch in Eight Globular Clusters}. \apj 760(1):39

\bibitem[{{Piotto} et~al.(2015){Piotto}, {Milone}, {Bedin}, {Anderson}, {King},
  {Marino}, {Nardiello}, {Aparicio}, {Barbuy}, {Bellini}, {Brown}, {Cassisi},
  {Cool}, {Cunial}, {Dalessandro}, {D'Antona}, {Ferraro}, {Hidalgo}, {Lanzoni},
  {Monelli}, {Ortolani}, {Renzini}, {Salaris}, {Sarajedini}, {van der Marel},
  {Vesperini}, and {Zoccali}}]{Piotto_etal2015}
{Piotto} G, {Milone} AP, {Bedin} LR, et~al. (2015) {The Hubble Space Telescope
  UV Legacy Survey of Galactic Globular Clusters. I. Overview of the Project
  and Detection of Multiple Stellar Populations}. \aj 149(3):91

\bibitem[{{Piskunov} et~al.(2007){Piskunov}, {Schilbach}, {Kharchenko},
  {R{\"o}ser}, and {Scholz}}]{2007A&A...468..151P}
{Piskunov} AE, {Schilbach} E, {Kharchenko} NV, et~al. (2007) {Towards absolute
  scales for the radii and masses of open clusters}. \aap 468(1):151--161

\bibitem[{{Portegies Zwart} et~al.(2010){Portegies Zwart}, {McMillan}, and
  {Gieles}}]{portegies10}
{Portegies Zwart} SF, {McMillan} SLW, and {Gieles} M (2010) {Young Massive Star
  Clusters}. \araa 48:431--493

\bibitem[{{Pozzetti} et~al.(2019){Pozzetti}, {Maraston}, and
  {Renzini}}]{Pozzetti_etal19}
{Pozzetti} L, {Maraston} C, and {Renzini} A (2019) {Search instructions for
  globular clusters in formation at high redshifts}. \mnras 485(4):5861--5873

\bibitem[{{Prantzos} and {Charbonnel}(2006)}]{2006A&A...458..135P}
{Prantzos} N and {Charbonnel} C (2006) {On the self-enrichment scenario of
  galactic globular clusters: constraints on the IMF}. \aap 458:135--149

\bibitem[{{Prantzos} et~al.(2007){Prantzos}, {Charbonnel}, and
  {Iliadis}}]{2007A&A...470..179P}
{Prantzos} N, {Charbonnel} C, and {Iliadis} C (2007) {Light nuclei in galactic
  globular clusters: constraints on the self-enrichment scenario from
  nucleosynthesis}. \aap 470(1):179--190

\bibitem[{{Prantzos} et~al.(2017){Prantzos}, {Charbonnel}, and
  {Iliadis}}]{2017A&A...608A..28}
{Prantzos} N, {Charbonnel} C, and {Iliadis} C (2017) {Revisiting
  nucleosynthesis in globular clusters. The case of NGC 2808 and the role of He
  and K}. \aap 608:A28

\bibitem[{{Preibisch} and {Mamajek}(2008)}]{Preibisch_08}
{Preibisch} T and {Mamajek} E (2008) {The Nearest OB Association:
  Scorpius-Centaurus (Sco OB2)}, vol~5, p 235

\bibitem[{{Rahner} et~al.(2019){Rahner}, {Pellegrini}, {Glover}, and
  {Klessen}}]{rahner19}
{Rahner} D, {Pellegrini} EW, {Glover} SCO, et~al. (2019) {WARPFIELD 2.0:
  feedback-regulated minimum star formation efficiencies of giant molecular
  clouds}. \mnras 483(2):2547--2560

\bibitem[{{Randriamanakoto} et~al.(2013){Randriamanakoto}, {V{\"a}is{\"a}nen},
  {Ryder}, {Kankare}, {Kotilainen}, and {Mattila}}]{randria2013}
{Randriamanakoto} Z, {V{\"a}is{\"a}nen} P, {Ryder} S, et~al. (2013) {The K-band
  luminosity functions of super star clusters in luminous infrared galaxies,
  their slopes and the effects of blending}. \mnras 431(1):554--569

\bibitem[{{Reina-Campos} and {Kruijssen}(2017)}]{reinacampos2017}
{Reina-Campos} M and {Kruijssen} JMD (2017) {A unified model for the maximum
  mass scales of molecular clouds, stellar clusters and high-redshift clumps}.
  \mnras 469(2):1282--1298

\bibitem[{{Reina-Campos} et~al.(2018){Reina-Campos}, {Kruijssen}, {Pfeffer},
  {Bastian}, and {Crain}}]{reinacampos18}
{Reina-Campos} M, {Kruijssen} JMD, {Pfeffer} J, et~al. (2018) {Dynamical
  cluster disruption and its implications for multiple population models in the
  E-MOSAICS simulations}. \mnras 481:2851--2857

\bibitem[{{Reina-Campos} et~al.(2019{\natexlab{a}}){Reina-Campos}, {Hughes},
  {Kruijssen}, {Pfeffer}, {Bastian}, {Crain}, {Koch}, and
  {Grebel}}]{reinacampos19b}
{Reina-Campos} M, {Hughes} ME, {Kruijssen} JMD, et~al. (2019{\natexlab{a}})
  {The mass fraction of halo stars contributed by the disruption of globular
  clusters in the E-MOSAICS simulations}. \mnras\ submitted arXiv:1910.06973

\bibitem[{{Reina-Campos} et~al.(2019{\natexlab{b}}){Reina-Campos}, {Kruijssen},
  {Pfeffer}, {Bastian}, and {Crain}}]{reinacampos2019}
{Reina-Campos} M, {Kruijssen} JMD, {Pfeffer} JL, et~al. (2019{\natexlab{b}})
  {Formation histories of stars, clusters, and globular clusters in the
  E-MOSAICS simulations}. \mnras 486(4):5838--5852

\bibitem[{{Renaud}(2019)}]{renaud19}
{Renaud} F (2019) {3 things they don't tell you about star clusters}. arXiv
  e-prints arXiv:1908.02301

\bibitem[{{Renaud} and {Gieles}(2013)}]{renaud2013}
{Renaud} F and {Gieles} M (2013) {The role of galaxy mergers on the evolution
  of star clusters}. \mnras 431:L83--L87

\bibitem[{{Renaud} et~al.(2015){Renaud}, {Bournaud}, and {Duc}}]{renaud2015}
{Renaud} F, {Bournaud} F, and {Duc} PA (2015) {A parsec-resolution simulation
  of the Antennae galaxies: formation of star clusters during the merger}.
  \mnras 446(2):2038--2054

\bibitem[{{Renaud} et~al.(2017){Renaud}, {Agertz}, and {Gieles}}]{renaud17}
{Renaud} F, {Agertz} O, and {Gieles} M (2017) {The origin of the Milky Way
  globular clusters}. \mnras 465:3622--3636

\bibitem[{{Renzini}(2017)}]{Renzini2017}
{Renzini} A (2017) {Finding forming globular clusters at high redshifts}.
  \mnras 469(1):L63--L67

\bibitem[{{Renzini} et~al.(2015){Renzini}, {D'Antona}, {Cassisi}, {King},
  {Milone}, {Ventura}, {Anderson}, {Bedin}, {Bellini}, {Brown}, {Piotto}, {van
  der Marel}, {Barbuy}, {Dalessandro}, {Hidalgo}, {Marino}, {Ortolani},
  {Salaris}, and {Sarajedini}}]{Renzini_etal15}
{Renzini} A, {D'Antona} F, {Cassisi} S, et~al. (2015) {The Hubble Space
  Telescope UV Legacy Survey of Galactic Globular Clusters - V. Constraints on
  formation scenarios}. \mnras 454(4):4197--4207

\bibitem[{{Rey-Raposo} et~al.(2017){Rey-Raposo}, {Dobbs}, {Agertz}, and
  {Alig}}]{reyraposo2017}
{Rey-Raposo} R, {Dobbs} C, {Agertz} O, et~al. (2017) {The roles of stellar
  feedback and galactic environment in star-forming molecular clouds}. \mnras
  464(3):3536--3551

\bibitem[{{Ricotti}(2002)}]{ricotti02}
{Ricotti} M (2002) {Did globular clusters reionize the Universe?} \mnras
  336(2):L33--L37

\bibitem[{{Ricotti} et~al.(2016){Ricotti}, {Parry}, and {Gnedin}}]{ricotti16}
{Ricotti} M, {Parry} OH, and {Gnedin} NY (2016) {A Common Origin for Globular
  Clusters and Ultra-faint Dwarfs in Simulations of the First Galaxies}. \apj
  831(2):204

\bibitem[{{Rieder} et~al.(2013){Rieder}, {Ishiyama}, {Langelaan}, {Makino},
  {McMillan}, and {Portegies Zwart}}]{rieder13}
{Rieder} S, {Ishiyama} T, {Langelaan} P, et~al. (2013) {Evolution of star
  clusters in a cosmological tidal field}. \mnras 436:3695--3706

\bibitem[{{Rivera-Thorsen} et~al.(2019){Rivera-Thorsen}, {Dahle}, {Chisholm},
  {Florian}, {Gronke}, {Rigby}, {Gladders}, {Mahler}, {Sharon}, and
  {Bayliss}}]{Rivera2019}
{Rivera-Thorsen} TE, {Dahle} H, {Chisholm} J, et~al. (2019) {Gravitational
  lensing reveals ionizing ultraviolet photons escaping from a distant galaxy}.
  Science 366(6466):738--741

\bibitem[{{Robberto} et~al.(2013){Robberto}, {Soderblom}, {Bergeron},
  {Kozhurina-Platais}, {Makidon}, {McCullough}, {McMaster}, {Panagia}, {Reid},
  {Levay}, {Frattare}, {Da Rio}, {Andersen}, {O'Dell}, {Stassun}, {Simon},
  {Feigelson}, {Stauffer}, {Meyer}, {Reggiani}, {Krist}, {Manara},
  {Romaniello}, {Hillenbrand}, {Ricci}, {Palla}, {Najita}, {Ananna},
  {Scandariato}, and {Smith}}]{Robberto_13}
{Robberto} M, {Soderblom} DR, {Bergeron} E, et~al. (2013) {The Hubble Space
  Telescope Treasury Program on the Orion Nebula Cluster}. \apjs 207(1):10

\bibitem[{{Rolleston} et~al.(1999){Rolleston}, {Dufton}, {McErlean}, and
  {Venn}}]{Rolleston_99}
{Rolleston} WRJ, {Dufton} PL, {McErlean} ND, et~al. (1999) {The chemical
  composition of the young, Inter-Cloud population}. \aap 348:728--736

\bibitem[{{Roman-Duval} et~al.(2010){Roman-Duval}, {Jackson}, {Heyer},
  {Rathborne}, and {Simon}}]{romanduval10}
{Roman-Duval} J, {Jackson} JM, {Heyer} M, et~al. (2010) {Physical Properties
  and Galactic Distribution of Molecular Clouds Identified in the Galactic Ring
  Survey}. \apj 723(1):492--507

\bibitem[{{Roman-Duval} et~al.(2014){Roman-Duval}, {Gordon}, {Meixner}, {Bot},
  {Bolatto}, {Hughes}, {Wong}, {Babler}, {Bernard}, {Clayton}, {Fukui},
  {Galametz}, {Galliano}, {Glover}, {Hony}, {Israel}, {Jameson},
  {Lebouteiller}, {Lee}, {Li}, {Madden}, {Misselt}, {Montiel}, {Okumura},
  {Onishi}, {Panuzzo}, {Reach}, {Remy-Ruyer}, {Robitaille}, {Rubio}, {Sauvage},
  {Seale}, {Sewilo}, {Staveley-Smith}, and {Zhukovska}}]{Roman-Duval_14}
{Roman-Duval} J, {Gordon} KD, {Meixner} M, et~al. (2014) {Dust and Gas in the
  Magellanic Clouds from the HERITAGE Herschel Key Project. II. Gas-to-dust
  Ratio Variations across Interstellar Medium Phases}. \apj 797(2):86

\bibitem[{{R{\"o}ser} et~al.(2019){R{\"o}ser}, {Schilbach}, and
  {Goldman}}]{roeser19}
{R{\"o}ser} S, {Schilbach} E, and {Goldman} B (2019) {Hyades tidal tails
  revealed by Gaia DR2}. \aap 621:L2

\bibitem[{{Russell} and {Dopita}(1992)}]{Russell_92}
{Russell} SC and {Dopita} MA (1992) {Abundances of the Heavy Elements in the
  Magellanic Clouds. III. Interpretation of Results}. \apj 384:508

\bibitem[{{Ryon} et~al.(2014){Ryon}, {Adamo}, {Bastian}, {Smith}, {Gallagher},
  {Konstantopoulos}, {Larsen}, {Silva-Villa}, and {Zackrisson}}]{ryon2014}
{Ryon} JE, {Adamo} A, {Bastian} N, et~al. (2014) {The Snapshot Hubble U-Band
  Cluster Survey (SHUCS). II. The Star Cluster Population of NGC 2997}. \aj
  148(2):33

\bibitem[{{Ryon} et~al.(2015){Ryon}, {Bastian}, {Adamo}, {Konstantopoulos},
  {Gallagher}, {Larsen}, {Hollyhead}, {Silva-Villa}, and
  {Smith}}]{2015MNRAS.452..525R}
{Ryon} JE, {Bastian} N, {Adamo} A, et~al. (2015) {Sizes and shapes of young
  star cluster light profiles in M83}. \mnras 452(1):525--539

\bibitem[{{Ryon} et~al.(2017{\natexlab{a}}){Ryon}, {Gallagher}, {Smith},
  {Adamo}, {Calzetti}, {Bright}, {Cignoni}, {Cook}, {Dale}, {Elmegreen},
  {Fumagalli}, {Gouliermis}, {Grasha}, {Grebel}, {Kim}, {Messa}, {Thilker}, and
  {Ubeda}}]{ryon2017}
{Ryon} JE, {Gallagher} JS, {Smith} LJ, et~al. (2017{\natexlab{a}}) {Effective
  Radii of Young, Massive Star Clusters in Two LEGUS Galaxies}. \apj 841(2):92

\bibitem[{{Ryon} et~al.(2017{\natexlab{b}}){Ryon}, {Gallagher}, {Smith},
  {Adamo}, {Calzetti}, {Bright}, {Cignoni}, {Cook}, {Dale}, {Elmegreen},
  {Fumagalli}, {Gouliermis}, {Grasha}, {Grebel}, {Kim}, {Messa}, {Thilker}, and
  {Ubeda}}]{2017ApJ...841...92R}
{Ryon} JE, {Gallagher} JS, {Smith} LJ, et~al. (2017{\natexlab{b}}) {Effective
  Radii of Young, Massive Star Clusters in Two LEGUS Galaxies}. \apj 841(2):92

\bibitem[{{Sabbi} et~al.(2007){Sabbi}, {Sirianni}, {Nota}, {Tosi}, {Gallagher},
  {Meixner}, {Oey}, {Walterbos}, {Pasquali}, {Smith}, and
  {Angeretti}}]{Sabbi_07}
{Sabbi} E, {Sirianni} M, {Nota} A, et~al. (2007) {Past and Present Star
  Formation in the SMC: NGC 346 and its Neighborhood}. \aj 133:44--57

\bibitem[{{Sabbi} et~al.(2008){Sabbi}, {Sirianni}, {Nota}, {Tosi}, {Gallagher},
  {Smith}, {Angeretti}, {Meixner}, {Oey}, {Walterbos}, and
  {Pasquali}}]{sabbi2008}
{Sabbi} E, {Sirianni} M, {Nota} A, et~al. (2008) {The Stellar Mass Distribution
  in the Giant Star Forming Region NGC 346}. \aj 135(1):173--181

\bibitem[{{Sabbi} et~al.(2012){Sabbi}, {Lennon}, {Gieles}, {de Mink},
  {Walborn}, {Anderson}, {Bellini}, {Panagia}, {van der Marel}, and
  {Ma{\'{\i}}z Apell{\'a}niz}}]{Sabbi_12}
{Sabbi} E, {Lennon} DJ, {Gieles} M, et~al. (2012) {A Double Cluster at the Core
  of 30 Doradus}. \apjl 754:L37

\bibitem[{{Sabbi} et~al.(2018){Sabbi}, {Calzetti}, {Ubeda}, {Adamo}, {Cignoni},
  {Thilker}, {Aloisi}, {Elmegreen}, {Elmegreen}, {Gouliermis}, {Grebel},
  {Messa}, {Smith}, {Tosi}, {Dolphin}, {Andrews}, {Ashworth}, {Bright},
  {Brown}, {Chandar}, {Christian}, {Clayton}, {Cook}, {Dale}, {de Mink},
  {Dobbs}, {Evans}, {Fumagalli}, {Gallagher}, {Grasha}, {Herrero}, {Hunter},
  {Johnson}, {Kahre}, {Kennicutt}, {Kim}, {Krumholz}, {Lee}, {Lennon},
  {Martin}, {Nair}, {Nota}, {{\"O}stlin}, {Pellerin}, {Prieto}, {Regan},
  {Ryon}, {Sacchi}, {Schaerer}, {Schiminovich}, {Shabani}, {Van Dyk},
  {Walterbos}, {Whitmore}, and {Wofford}}]{Sabbi_18}
{Sabbi} E, {Calzetti} D, {Ubeda} L, et~al. (2018) {The Resolved Stellar
  Populations in the LEGUS Galaxies1}. \apjs 235(1):23

\bibitem[{{Sacchi} et~al.(2018){Sacchi}, {Cignoni}, {Aloisi}, {Tosi},
  {Calzetti}, {Lee}, {Adamo}, {Annibali}, {Dale}, {Elmegreen}, {Gouliermis},
  {Grasha}, {Grebel}, {Hunter}, {Sabbi}, {Smith}, {Thilker}, {Ubeda}, and
  {Whitmore}}]{Sacchi_18}
{Sacchi} E, {Cignoni} M, {Aloisi} A, et~al. (2018) {Star Formation Histories of
  the LEGUS Dwarf Galaxies. II. Spatially Resolved Star Formation History of
  the Magellanic Irregular NGC 4449}. \apj 857(1):63

\bibitem[{{Salpeter}(1955)}]{Salpeter_55}
{Salpeter} EE (1955) {The Luminosity Function and Stellar Evolution.} \apj
  121:161

\bibitem[{{Saracino} et~al.(2019){Saracino}, {Bastian}, {Kozhurina-Platais},
  {Cabrera-Ziri}, {Dalessandro}, {Kacharov}, {Lardo}, {Larsen}, {Mucciarelli},
  {Platais}, and {Salaris}}]{Saracino_etal19}
{Saracino} S, {Bastian} N, {Kozhurina-Platais} V, et~al. (2019) {An
  extragalactic chromosome map: the intermediate-age SMC cluster Lindsay 1}.
  \mnras 489(1):L97--L101

\bibitem[{{Schaefer}(2008)}]{Schaefer_08}
{Schaefer} BE (2008) {a Problem with the Clustering of Recent Measures of the
  Distance to the Large Magellanic Cloud}. \aj 135(1):112--119

\bibitem[{{Schaye} et~al.(2015){Schaye}, {Crain}, {Bower}, {Furlong},
  {Schaller}, {Theuns}, {Dalla Vecchia}, {Frenk}, {McCarthy}, {Helly},
  {Jenkins}, {Rosas-Guevara}, {White}, {Baes}, {Booth}, {Camps}, {Navarro},
  {Qu}, {Rahmati}, {Sawala}, {Thomas}, and {Trayford}}]{schaye2015}
{Schaye} J, {Crain} RA, {Bower} RG, et~al. (2015) {The EAGLE project:
  simulating the evolution and assembly of galaxies and their environments}.
  \mnras 446(1):521--554

\bibitem[{{Schechter}(1976)}]{schechter76}
{Schechter} P (1976) {An analytic expression for the luminosity function for
  galaxies.} \apj 203:297--306

\bibitem[{{Scheepmaker} et~al.(2007){Scheepmaker}, {Haas}, {Gieles}, {Bastian},
  {Larsen}, and {Lamers}}]{2007A&A...469..925S}
{Scheepmaker} RA, {Haas} MR, {Gieles} M, et~al. (2007) {ACS imaging of star
  clusters in M 51. I. Identification and radius distribution}. \aap
  469(3):925--940

\bibitem[{{Schinnerer} et~al.(2019){Schinnerer}, {Leroy}, {Blanc}, {Emsellem},
  {Hughes}, {Rosolowsky}, {Schruba}, {Bigiel}, {Escala}, {Groves}, {Kreckel},
  {Kruijssen}, {Lee}, {Meidt}, {Pety}, {Sanchez-Blazquez}, {Sandstrom},
  {Usero}, {Barnes}, {Belfiore}, {Be{\v{s}}li{\'c}}, {Chandar},
  {Chatzigiannakis}, {Chevance}, {Congiu}, {Dale}, {Faesi}, {Gallagher},
  {Garcia-Rodriguez}, {Glover}, {Grasha}, {Henshaw}, {Herrera}, {Ho}, {Hygate},
  {Jimenez-Donaire}, {Kessler}, {Kim}, {Klessen}, {Koch}, {Lang}, {Larson}, {Le
  Reste}, {Liu}, {McElroy}, {Nofech}, {Ostriker}, {Pessa Gutierrez},
  {Puschnig}, {Querejeta}, {Razza}, {Saito}, {Santoro}, {Stuber}, {Sun},
  {Thilker}, {Turner}, {Ubeda}, {Utreras}, {Utomo}, {van Dyk}, {Ward}, and
  {Whitmore}}]{schinnerer19}
{Schinnerer} E, {Leroy} A, {Blanc} G, et~al. (2019) {The Physics at High
  Angular resolution in Nearby GalaxieS (PHANGS) Surveys}. The Messenger
  177:36--41

\bibitem[{{Schmeja} et~al.(2008){Schmeja}, {Kumar}, and
  {Ferreira}}]{Schmeja2008}
{Schmeja} S, {Kumar} MSN, and {Ferreira} B (2008) {The structures of embedded
  clusters in the Perseus, Serpens and Ophiuchus molecular clouds}. \mnras
  389(3):1209--1217

\bibitem[{{Schmeja} et~al.(2014){Schmeja}, {Kharchenko}, {Piskunov},
  {R{\"o}ser}, {Schilbach}, {Froebrich}, and {Scholz}}]{Schmeja_14}
{Schmeja} S, {Kharchenko} NV, {Piskunov} AE, et~al. (2014) {Global survey of
  star clusters in the Milky Way. III. 139 new open clusters at high Galactic
  latitudes}. \aap 568:A51

\bibitem[{{Schneider} et~al.(2018){Schneider}, {Sana}, {Evans}, {Bestenlehner},
  {Castro}, {Fossati}, {Gr{\"a}fener}, {Langer}, {Ram{\'\i}rez-Agudelo},
  {Sab{\'\i}n-Sanjuli{\'a}n}, {Sim{\'o}n-D{\'\i}az}, {Tramper}, {Crowther}, {de
  Koter}, {de Mink}, {Dufton}, {Garcia}, {Gieles}, {H{\'e}nault-Brunet},
  {Herrero}, {Izzard}, {Kalari}, {Lennon}, {Ma{\'\i}z Apell{\'a}niz},
  {Markova}, {Najarro}, {Podsiadlowski}, {Puls}, {Taylor}, {van Loon}, {Vink},
  and {Norman}}]{Schneider_18}
{Schneider} FRN, {Sana} H, {Evans} CJ, et~al. (2018) {An excess of massive
  stars in the local 30 Doradus starburst}. Science 359(6371):69--71

\bibitem[{{Scholz} et~al.(2015){Scholz}, {Kharchenko}, {Piskunov}, {R{\"o}ser},
  and {Schilbach}}]{Scholz_15}
{Scholz} RD, {Kharchenko} NV, {Piskunov} AE, et~al. (2015) {Global survey of
  star clusters in the Milky Way. IV. 63 new open clusters detected by proper
  motions}. \aap 581:A39

\bibitem[{{Schruba} et~al.(2019){Schruba}, {Kruijssen}, and
  {Leroy}}]{schruba2019}
{Schruba} A, {Kruijssen} JMD, and {Leroy} AK (2019) {How Galactic Environment
  Affects the Dynamical State of Molecular Clouds and Their Star Formation
  Efficiency}. \apj 883(1):2

\bibitem[{{Seth} et~al.(2019){Seth}, {Neumayer}, and {Boeker}}]{seth2019}
{Seth} AC, {Neumayer} N, and {Boeker} T (2019) {The Properties of Nuclear Star
  Clusters and their Host Galaxies}. arXiv e-prints arXiv:1908.00022

\bibitem[{{Shapiro} et~al.(2010){Shapiro}, {Genzel}, and {F{\"o}rster
  Schreiber}}]{shapiro10}
{Shapiro} KL, {Genzel} R, and {F{\"o}rster Schreiber} NM (2010) {Star-forming
  galaxies at z \~{} 2 and the formation of the metal-rich globular cluster
  population}. \mnras 403:L36--L40

\bibitem[{{Shibuya} et~al.(2016){Shibuya}, {Ouchi}, {Kubo}, and
  {Harikane}}]{shibuya2016}
{Shibuya} T, {Ouchi} M, {Kubo} M, et~al. (2016) {Morphologies of
  \raisebox{-0.5ex}\textasciitilde190,000 Galaxies at z = 0-10 Revealed with
  HST Legacy Data. II. Evolution of Clumpy Galaxies}. \apj 821(2):72

\bibitem[{{Sim} et~al.(2019){Sim}, {Lee}, {Ann}, and {Kim}}]{Sim_19}
{Sim} G, {Lee} SH, {Ann} HB, et~al. (2019) {207 New Open Star Clusters within 1
  kpc from Gaia Data Release 2}. Journal of Korean Astronomical Society
  52:145--158

\bibitem[{{Simmerer} et~al.(2013){Simmerer}, {Ivans}, {Filler}, {Francois},
  {Charbonnel}, {Monier}, and {James}}]{Simmerer_etal13}
{Simmerer} J, {Ivans} II, {Filler} D, et~al. (2013) {Star-to-star Iron
  Abundance Variations in Red Giant Branch Stars in the Galactic Globular
  Cluster NGC 3201}. \apjl 764(1):L7

\bibitem[{{Skrutskie} et~al.(2006){Skrutskie}, {Cutri}, {Stiening}, {Weinberg},
  {Schneider}, {Carpenter}, {Beichman}, {Capps}, {Chester}, {Elias}, {Huchra},
  {Liebert}, {Lonsdale}, {Monet}, {Price}, {Seitzer}, {Jarrett}, {Kirkpatrick},
  {Gizis}, {Howard}, {Evans}, {Fowler}, {Fullmer}, {Hurt}, {Light}, {Kopan},
  {Marsh}, {McCallon}, {Tam}, {Van Dyk}, and {Wheelock}}]{Skrutskie_06}
{Skrutskie} MF, {Cutri} RM, {Stiening} R, et~al. (2006) {The Two Micron All Sky
  Survey (2MASS)}. \aj 131(2):1163--1183

\bibitem[{{Smith} et~al.(2016){Smith}, {Crowther}, {Calzetti}, and
  {Sidoli}}]{smith2016}
{Smith} LJ, {Crowther} PA, {Calzetti} D, et~al. (2016) {The Very Massive Star
  Content of the Nuclear Star Clusters in NGC 5253}. \apj 823(1):38

\bibitem[{{Smith} and {Brooks}(2008)}]{Smith_08}
{Smith} N and {Brooks} KJ (2008) {The Carina Nebula: A Laboratory for Feedback
  and Triggered Star Formation}, vol~8, Astronomical Society of the Pacific, p
  138

\bibitem[{Solomon et~al.(1987)Solomon, Rivolo, Barrett, and
  Yahil}]{solomon1987}
Solomon PM, Rivolo aR, Barrett J, et~al. (1987) {Mass, luminosity, and line
  width relations of Galactic molecular clouds}. \apj 319:730

\bibitem[{{Spera} et~al.(2016){Spera}, {Mapelli}, and {Jeffries}}]{spera16}
{Spera} M, {Mapelli} M, and {Jeffries} RD (2016) {Do open star clusters evolve
  towards energy equipartition?} \mnras 460(1):317--328

\bibitem[{{Spitzer}(1958)}]{spitzer58}
{Spitzer} L Jr (1958) {Disruption of Galactic Clusters.} \apj 127:17--+

\bibitem[{{Stanke} et~al.(2006){Stanke}, {Smith}, {Gredel}, and
  {Khanzadyan}}]{Stanke2006}
{Stanke} T, {Smith} MD, {Gredel} R, et~al. (2006) {An unbiased search for the
  signatures of protostars in the {\ensuremath{\rho}} Ophiuchi molecular cloud
  . II. Millimetre continuum observations}. \aap 447(2):609--622

\bibitem[{{Stewart} et~al.(2008){Stewart}, {Bullock}, {Wechsler}, {Maller}, and
  {Zentner}}]{stewart08}
{Stewart} KR, {Bullock} JS, {Wechsler} RH, et~al. (2008) {Merger Histories of
  Galaxy Halos and Implications for Disk Survival}. \apj 683:597-610

\bibitem[{{Stolte} et~al.(2006){Stolte}, {Brandner}, {Brandl}, and
  {Zinnecker}}]{Stolte_06}
{Stolte} A, {Brandner} W, {Brandl} B, et~al. (2006) {The Secrets of the Nearest
  Starburst Cluster. II. The Present-Day Mass Function in NGC 3603}. \aj
  132:253--270

\bibitem[{{Stolte} et~al.(2010){Stolte}, {Morris}, {Ghez}, {Do}, {Lu},
  {Wright}, {Ballard}, {Mills}, and {Matthews}}]{Stolte_10}
{Stolte} A, {Morris} MR, {Ghez} AM, et~al. (2010) {Disks in the Arches Cluster
  - Survival in a Starburst Environment}. \apj 718:810--831

\bibitem[{{Stolte} et~al.(2015){Stolte}, {Hu{\ss}mann}, {Olczak}, {Brandner},
  {Habibi}, {Ghez}, {Morris}, {Lu}, {Clarkson}, and {Anderson}}]{Stolte_15}
{Stolte} A, {Hu{\ss}mann} B, {Olczak} C, et~al. (2015) {Circumstellar discs in
  Galactic centre clusters: Disc-bearing B-type stars in the Quintuplet and
  Arches clusters}. \aap 578:A4

\bibitem[{{Sun} et~al.(2018){Sun}, {Leroy}, {Schruba}, {Rosolowsky}, {Hughes},
  {Kruijssen}, {Meidt}, {Schinnerer}, {Blanc}, {Bigiel}, {Bolatto}, {Chevance},
  {Groves}, {Herrera}, {Hygate}, {Pety}, {Querejeta}, {Usero}, and
  {Utomo}}]{sun18}
{Sun} J, {Leroy} AK, {Schruba} A, et~al. (2018) {Cloud-scale Molecular Gas
  Properties in 15 Nearby Galaxies}. \apj 860:172

\bibitem[{{Sung} and {Bessell}(2004)}]{Sung_04}
{Sung} H and {Bessell} MS (2004) {The Initial Mass Function and Stellar Content
  of NGC 3603}. \aj 127:1014--1028

\bibitem[{{Thilker} et~al.(2002){Thilker}, {Walterbos}, {Braun}, and
  {Hoopes}}]{thilker2002}
{Thilker} DA, {Walterbos} RAM, {Braun} R, et~al. (2002) {H II Regions and
  Diffuse Ionized Gas in 11 Nearby Spiral Galaxies}. \aj 124(6):3118--3134

\bibitem[{{Toomre}(1964)}]{toomre1964}
{Toomre} A (1964) {On the gravitational stability of a disk of stars.} \apj
  139:1217--1238

\bibitem[{{Trenti} and {van der Marel}(2013)}]{trenti13}
{Trenti} M and {van der Marel} R (2013) {No energy equipartition in globular
  clusters}. \mnras 435(4):3272--3282

\bibitem[{{Trenti} et~al.(2015){Trenti}, {Padoan}, and {Jimenez}}]{trenti15}
{Trenti} M, {Padoan} P, and {Jimenez} R (2015) {The Relative and Absolute Ages
  of Old Globular Clusters in the LCDM Framework}. \apjl 808(2):L35

\bibitem[{{Trujillo-Gomez} et~al.(2019){Trujillo-Gomez}, {Reina-Campos}, and
  {Kruijssen}}]{trujillogomez19}
{Trujillo-Gomez} S, {Reina-Campos} M, and {Kruijssen} JMD (2019) {A model for
  the minimum mass of bound stellar clusters and its dependence on the galactic
  environment}. \mnras 488(3):3972--3994

\bibitem[{{Trujillo-Gomez} et~al.(2020){Trujillo-Gomez}, {Kruijssen},
  {Reina-Campos}, {Pfeffer}, {Keller}, {Crain}, {Bastian}, and
  {Hughes}}]{trujillo20}
{Trujillo-Gomez} S, {Kruijssen} JMD, {Reina-Campos} M, et~al. (2020) {The
  kinematics of globular cluster populations in the E-MOSAICS simulations and
  their implications for the assembly history of the Milky Way}.
  \mnras~submitted

\bibitem[{{Tsuge} et~al.(2019){Tsuge}, {Sano}, {Tachihara}, {Yozin}, {Bekki},
  {Inoue}, {Mizuno}, {Kawamura}, {Onishi}, and {Fukui}}]{Tsuge_19}
{Tsuge} K, {Sano} H, {Tachihara} K, et~al. (2019) {Formation of the Active
  Star-forming Region LHA 120-N 44 Triggered by Tidally Driven Colliding H I
  Flows}. \apj 871(1):44

\bibitem[{{Usher} et~al.(2019){Usher}, {Brodie}, {Forbes}, {Romanowsky},
  {Strader}, {Pfeffer}, and {Bastian}}]{usher19}
{Usher} C, {Brodie} JP, {Forbes} DA, et~al. (2019) {The SLUGGS survey:
  measuring globular cluster ages using both photometry and spectroscopy}.
  \mnras 490(1):491--501

\bibitem[{{van der Marel} et~al.(2018){van der Marel}, {Williams}, and
  {Bruderer}}]{van_der_Marel_18}
{van der Marel} N, {Williams} JP, and {Bruderer} S (2018) {Rings and Gaps in
  Protoplanetary Disks: Planets or Snowlines?} \apjl 867(1):L14

\bibitem[{{van Dokkum} and {Conroy}(2010)}]{vanDokkum_10}
{van Dokkum} PG and {Conroy} C (2010) {A substantial population of low-mass
  stars in luminous elliptical galaxies}. \nat 468(7326):940--942

\bibitem[{{Vanzella} et~al.(2017){Vanzella}, {Calura}, {Meneghetti},
  {Mercurio}, {Castellano}, {Caminha}, {Balestra}, {Rosati}, {Tozzi}, {De
  Barros}, {Grazian}, {D'Ercole}, {Ciotti}, {Caputi}, {Grillo}, {Merlin},
  {Pentericci}, {Fontana}, {Cristiani}, and {Coe}}]{Vanzella_etal17}
{Vanzella} E, {Calura} F, {Meneghetti} M, et~al. (2017) {Paving the way for the
  JWST: witnessing globular cluster formation at z \&gt; 3}. \mnras
  467(4):4304--4321

\bibitem[{{Vanzella} et~al.(2019){Vanzella}, {Calura}, {Meneghetti},
  {Castellano}, {Caminha}, {Mercurio}, {Cupani}, {Rosati}, {Grillo}, {Gilli},
  {Mignoli}, {Fiorentino}, {Arcidiacono}, {Lombini}, and
  {Cortecchia}}]{Vanzella_etal19}
{Vanzella} E, {Calura} F, {Meneghetti} M, et~al. (2019) {Massive star cluster
  formation under the microscope at z = 6}. \mnras 483(3):3618--3635

\bibitem[{{Vanzella} et~al.(2020){Vanzella}, {Caminha}, {Calura}, {Cupani},
  {Meneghetti}, {Castellano}, {Rosati}, {Mercurio}, {Sani}, {Grillo}, {Gilli},
  {Mignoli}, {Comastri}, {Nonino}, {Cristiani}, {Giavalisco}, and
  {Caputi}}]{vanzella2020}
{Vanzella} E, {Caminha} GB, {Calura} F, et~al. (2020) {Ionizing the
  intergalactic medium by star clusters: the first empirical evidence}. \mnras
  491(1):1093--1103

\bibitem[{{Ventura} et~al.(2001){Ventura}, {D'Antona}, {Mazzitelli}, and
  {Gratton}}]{2001ApJ...550L..65V}
{Ventura} P, {D'Antona} F, {Mazzitelli} I, et~al. (2001) {Predictions for
  Self-Pollution in Globular Cluster Stars}. \apjl 550:L65--L69

\bibitem[{{Walborn} and {Blades}(1997)}]{Walborn_97}
{Walborn} NR and {Blades} JC (1997) {Spectral Classification of the 30 Doradus
  Stellar Populations}. \apjs 112(2):457--485

\bibitem[{{Walker} et~al.(2015){Walker}, {Longmore}, {Bastian}, {Kruijssen},
  {Rathborne}, {Jackson}, {Foster}, and {Contreras}}]{walker15}
{Walker} DL, {Longmore} SN, {Bastian} N, et~al. (2015) {Tracing the conversion
  of gas into stars in Young Massive Cluster Progenitors}. \mnras 449:715--725

\bibitem[{{Walker} et~al.(2016){Walker}, {Longmore}, {Bastian}, {Kruijssen},
  {Rathborne}, {Galv{\'a}n-Madrid}, and {Liu}}]{walker16}
{Walker} DL, {Longmore} SN, {Bastian} N, et~al. (2016) {Comparing young massive
  clusters and their progenitor clouds in the Milky Way}. \mnras 457:4536--4545

\bibitem[{{Ward} and {Kruijssen}(2018)}]{ward18}
{Ward} JL and {Kruijssen} JMD (2018) {Not all stars form in clusters -
  measuring the kinematics of OB associations with Gaia}. \mnras
  475(4):5659--5676

\bibitem[{{Ward} et~al.(2019){Ward}, {Kruijssen}, and {Rix}}]{ward19}
{Ward} JL, {Kruijssen} JMD, and {Rix} HW (2019) {Not all stars form in clusters
  -- $Gaia$-DR2 uncovers the origin of OB associations}. arXiv e-prints
  arXiv:1910.06974

\bibitem[{{Webb} et~al.(2019){Webb}, {Reina-Campos}, and {Kruijssen}}]{webb19}
{Webb} JJ, {Reina-Campos} M, and {Kruijssen} JMD (2019) {A systematic analysis
  of star cluster disruption by tidal shocks - I. Controlled N-body simulations
  and a new theoretical model}. \mnras 486(4):5879--5894

\bibitem[{{Wei} et~al.(2012){Wei}, {Keto}, and {Ho}}]{wei12}
{Wei} LH, {Keto} E, and {Ho} LC (2012) {Two Populations of Molecular Clouds in
  the Antennae Galaxies}. \apj 750:136

\bibitem[{{Wei} et~al.(2019){Wei}, {Huerta}, {Whitmore}, {Lee}, {Hannon},
  {Chandar}, {Dale}, {Larson}, {Thilker}, {Ubeda}, {Boquien}, {Chevance},
  {Kruijssen}, {Schruba}, {Blanc}, and {Congiu}}]{wei19}
{Wei} W, {Huerta} EA, {Whitmore} BC, et~al. (2019) {Deep Transfer Learning for
  Star Cluster Classification: I. Application to the PHANGS-HST Survey}. arXiv
  e-prints arXiv:1909.02024

\bibitem[{{Westerlund}(1961)}]{Westerlund_61}
{Westerlund} B (1961) {On the identification of a radio source in Carina}.
  Arkiv for Astronomi 2:419--425

\bibitem[{{Whitmore}(2000)}]{whitmore2000}
{Whitmore} BC (2000) {The Formation of Star Clusters}. arXiv e-prints
  astro-ph/0012546

\bibitem[{{Whitmore} and {Schweizer}(1995)}]{whitmore95}
{Whitmore} BC and {Schweizer} F (1995) {Hubble Space Telescope Observations of
  Young Star Clusters in NGC 4038/4039, ``The Antennae'' Galaxies}. \aj 109:960

\bibitem[{{Whitmore} et~al.(2007){Whitmore}, {Chandar}, and
  {Fall}}]{whitmore07}
{Whitmore} BC, {Chandar} R, and {Fall} SM (2007) {Star Cluster Demographics. I.
  A General Framework and Application to the Antennae Galaxies}. \aj
  133:1067--1084

\bibitem[{{Whitmore} et~al.(2014){Whitmore}, {Brogan}, {Chandar}, {Evans},
  {Hibbard}, {Johnson}, {Leroy}, {Privon}, {Remijan}, and
  {Sheth}}]{whitmore2014}
{Whitmore} BC, {Brogan} C, {Chandar} R, et~al. (2014) {ALMA Observations of the
  Antennae Galaxies. I. A New Window on a Prototypical Merger}. \apj 795:156

\bibitem[{{Wilson} et~al.(2003){Wilson}, {Scoville}, {Madden}, and
  {Charmandaris}}]{wilson2003}
{Wilson} CD, {Scoville} N, {Madden} SC, et~al. (2003) {The Mass Function of
  Supergiant Molecular Complexes and Implications for Forming Young Massive
  Star Clusters in the Antennae (NGC 4038/4039)}. \apj 599(2):1049--1066

\bibitem[{{Winter} et~al.(2020){Winter}, {Kruijssen}, {Chevance}, {Keller}, and
  {Longmore}}]{winter20}
{Winter} AJ, {Kruijssen} JMD, {Chevance} M, et~al. (2020) {Prevalent externally
  driven protoplanetary disc dispersal as a function of the galactic
  environment}. \mnras 491(1):903--922

\bibitem[{{Wright} and {Mamajek}(2018)}]{wright18}
{Wright} NJ and {Mamajek} EE (2018) {The kinematics of the Scorpius-Centaurus
  OB association from Gaia DR1}. \mnras 476(1):381--398

\bibitem[{{Wyse}(2001)}]{wyse01}
{Wyse} RFG (2001) {The Merging History of the Milky Way Disk}. In: {Funes} JG
  and {Corsini} EM (eds) Galaxy Disks and Disk Galaxies, Astronomical Society
  of the Pacific Conference Series, vol 230, pp 71--80

\bibitem[{{Yonekura} et~al.(2005){Yonekura}, {Asayama}, {Kimura}, {Ogawa},
  {Kanai}, {Yamaguchi}, {Barnes}, and {Fukui}}]{Yonekura_05}
{Yonekura} Y, {Asayama} S, {Kimura} K, et~al. (2005) {High-Mass Cloud Cores in
  the {$\eta$} Carinae Giant Molecular Cloud}. \apj 634:476--494

\bibitem[{{Zari} et~al.(2019){Zari}, {Brown}, and {de Zeeuw}}]{Zari_19}
{Zari} E, {Brown} AGA, and {de Zeeuw} PT (2019) {Structure, kinematics, and
  ages of the young stellar populations in the Orion region}. \aap 628:A123

\bibitem[{{Zeidler} et~al.(2015){Zeidler}, {Sabbi}, {Nota}, {Grebel}, {Tosi},
  {Bonanos}, {Pasquali}, {Christian}, {de Mink}, and {Ubeda}}]{Zeidler_15}
{Zeidler} P, {Sabbi} E, {Nota} A, et~al. (2015) {A High-resolution Multiband
  Survey of Westerlund 2 with the Hubble Space Telescope. I. Is the Massive
  Star Cluster Double?} \aj 150:78

\bibitem[{{Zeidler} et~al.(2016{\natexlab{a}}){Zeidler}, {Grebel}, {Nota},
  {Sabbi}, {Pasquali}, {Tosi}, {Bonanos}, and {Christian}}]{Zeidler_16b}
{Zeidler} P, {Grebel} EK, {Nota} A, et~al. (2016{\natexlab{a}}) {A
  High-resolution Multiband Survey of Westerlund 2 with the Hubble Space
  Telescope. II. Mass Accretion in the Pre-main-sequence Population}. \aj
  152:84

\bibitem[{{Zeidler} et~al.(2016{\natexlab{b}}){Zeidler}, {Preibisch}, {Ratzka},
  {Roccatagliata}, and {Petr-Gotzens}}]{Zeidler_16a}
{Zeidler} P, {Preibisch} T, {Ratzka} T, et~al. (2016{\natexlab{b}}) {The VISTA
  Carina Nebula Survey. II. Spatial distribution of the
  infrared-excess-selected young stellar population}. \aap 585:A49

\bibitem[{{Zeidler} et~al.(2017){Zeidler}, {Nota}, {Grebel}, {Sabbi},
  {Pasquali}, {Tosi}, and {Christian}}]{Zeidler_17}
{Zeidler} P, {Nota} A, {Grebel} EK, et~al. (2017) {A High-resolution Multiband
  Survey of Westerlund 2 with the Hubble Space Telescope. III. The Present-day
  Stellar Mass Function}. \aj 153:122

\bibitem[{{Zeidler} et~al.(2018){Zeidler}, {Sabbi}, {Nota}, {Pasquali},
  {Grebel}, {McLeod}, {Kamann}, {Tosi}, {Cignoni}, and {Ramsay}}]{Zeidler_18}
{Zeidler} P, {Sabbi} E, {Nota} A, et~al. (2018) {The Young Massive Star Cluster
  Westerlund 2 Observed with MUSE. I. First Results on the Cluster Internal
  Motion from Stellar Radial Velocities}. \aj 156(5):211

\bibitem[{{Zennaro} et~al.(2019){Zennaro}, {Milone}, {Marino}, {Cordoni},
  {Lagioia}, and {Tailo}}]{Zennaro_etal19}
{Zennaro} M, {Milone} AP, {Marino} AF, et~al. (2019) {Four stellar populations
  and extreme helium variation in the massive outer-halo globular cluster NGC
  2419}. \mnras 487(3):3239--3251

\bibitem[{{Zepf} et~al.(1999){Zepf}, {Ashman}, {English}, {Freeman}, and
  {Sharples}}]{1999AJ....118..752Z}
{Zepf} SE, {Ashman} KM, {English} J, et~al. (1999) {The Formation and Evolution
  of Candidate Young Globular Clusters in NGC 3256}. \aj 118:752--764

\bibitem[{{Zhang} et~al.(2001){Zhang}, {Fall}, and {Whitmore}}]{zhang2001}
{Zhang} Q, {Fall} SM, and {Whitmore} BC (2001) {A Multiwavelength Study of the
  Young Star Clusters and Interstellar Medium in the Antennae Galaxies}. \apj
  561(2):727--750

\bibitem[{{Zick} et~al.(2018){Zick}, {Weisz}, and {Boylan-Kolchin}}]{zick2018}
{Zick} TO, {Weisz} DR, and {Boylan-Kolchin} M (2018) {Globular clusters in
  high-redshift dwarf galaxies: a case study from the Local Group}. \mnras
  477(1):480--490

\bibitem[{{Zinnecker} and {Yorke}(2007)}]{zinnecker2007}
{Zinnecker} H and {Yorke} HW (2007) {Toward Understanding Massive Star
  Formation}. \araa 45(1):481--563

\end{thebibliography}

\end{document}